\documentclass[
  twocolumn,english,aps,prb,
  superscriptaddress,amsmath,amssymb,floatfix,nofootinbib,longbibliography
]{revtex4-2}

\usepackage{amsthm}
\usepackage{amsfonts}
\usepackage{siunitx}
\usepackage{amsmath}
\usepackage{amssymb}
\usepackage{graphicx}
\usepackage{verbatim}
\usepackage[colorlinks]{hyperref}
\usepackage{tikz}
\usepackage{pgfplots}
\usepackage{adjustbox}
\usepackage{braket}
\usepackage{xcolor}
\usepackage{physics}
\usepackage{amssymb} 
\usepackage{graphicx}
\usepackage{dcolumn}
\usepackage{bm}
\usepackage{mathtools}
\usepackage{hyperref}
\usepackage{mathrsfs}
\usepackage{dashrule}
\usepackage{caption}
\usepackage{subcaption}
\usepackage[version=4]{mhchem}

\usepackage[font=small,labelfont=bf,
   justification=justified,
   format=plain]{caption} 
   
\definecolor{linkcolor}{RGB}{0,83,166}
\hypersetup{
  colorlinks = true,
  allcolors = {linkcolor}
}

\begin{document}

\title{Probing Antiferromagnetic Hysteresis on Programmable Quantum Annealers}

\author{Elijah Pelofske}
\email[]{epelofske@lanl.gov}
\affiliation{Information Systems \& Modeling, Los Alamos National Laboratory}
\affiliation{Theoretical Division, Quantum \& Condensed Matter Physics, Los Alamos National Laboratory}
\affiliation{Center for Quantum Computing, Los Alamos National Laboratory}

\author{Pratik Sathe}
\affiliation{Theoretical Division, Quantum \& Condensed Matter Physics, Los Alamos National Laboratory}
\affiliation{Information Science and Technology Institute, Los Alamos National Laboratory}

\author{Cristiano Nisoli}
\affiliation{Theoretical Division, Quantum \& Condensed Matter Physics, Los Alamos National Laboratory}
\affiliation{Center for Nonlinear Studies, Los Alamos National Laboratory}
\affiliation{Information Science and Technology Institute, Los Alamos National Laboratory}
\affiliation{Center for Quantum Computing, Los Alamos National Laboratory}

\author{Frank Barrows}
\affiliation{Theoretical Division, Quantum \& Condensed Matter Physics, Los Alamos National Laboratory}
\affiliation{Center for Nonlinear Studies, Los Alamos National Laboratory}
\affiliation{Center for Quantum Computing, Los Alamos National Laboratory}

\begin{abstract}

Using programmable analog quantum annealing processors, we implement a sampling-based magnetic hysteresis protocol to probe the counterintuitive notion of magnetic memory in antiferromagnetic models. 
A key component of this protocol responsible for the hysteresis is a transverse field, which enables state transitions, while the longitudinal magnetic field sweep is done via a longitudinal control field. We observe intriguing non-monotonic magnetization, alongside full saturation and reversal of the hysteresis curve, Barkhausen-like noise, as well as emergent magnetic domains mediated by quantum fluctuations that give rise to the magnetic memory effect in antiferromagnets.

\end{abstract}

\maketitle

\section{Introduction}\label{section:Introduction}

Quantum annealing is a type of analog quantum computation that is based on the adiabatic theorem of quantum mechanics~\cite{Morita_2008, farhi2000quantumcomputationadiabaticevolution, Santoro_2002, Kadowaki_1998}. 
While originally designed to solve optimization problems, quantum annealers are also emerging as programmable laboratories for physical phenomena which are otherwise inaccessible using either bulk material observables in magnetic laboratories, or classical simulation techniques~\cite{PRXQuantum.2.030317, harris2018phase, PhysRevB.105.195101, King_2018, King_2021, king2024computationalsupremacyquantumsimulation,PhysRevB.104.L081107,PhysRevB.110.054432,lopez2023kagome,lopez2024quantum,Kairys_2020}. 

In recent work, we expanded the phenomenological range of these experiments by introducing a protocol to emulate magnetic hysteresis experiments using D-Wave's quantum annealers, which are conceptually identical to measurements performed in a laboratory~\cite{pelofske2025magnetichysteresisexperimentsperformed}. 
In Refs.~\cite{pelofske2025magnetichysteresisexperimentsperformed, barrows2025magneticmemoryhysteresisquantum}, we applied this method to  ferromagnetic and $\pm J$ disordered Ising models.
In this study, we apply our method to simulations of a counterintuitive type: hysteresis in antiferromagnetically interacting spins.
We perform these experiments on four different D-Wave quantum annealing quantum processing units (QPUs). We find several interesting phenomena, including non-monotonicities similar to those observed in low-dimensional ferromagnetic models~\cite{barrows2025magneticmemoryhysteresisquantum}. Where meaningful, we extract both spin structure factor and the antiferromagnetic (N\'eel) order parameter from the $2D$ antiferromagnetic models to probe the magnetic order throughout the hysteresis cycles.

Historically, magnetic hysteresis has most commonly been observed and studied in ferromagnetic compounds, where it arises from the breakdown of ergodicity associated with the spontaneous symmetry breaking around a specific magnetization. While a sufficiently strong field can polarize an antiferromagnet, one would not easily expect hysteretic magnetization, because the local field acts to destroy the net moment. 

Antiferromagnetic hysteresis is a rare but known phenomenon explained with diverse mechanisms in various systems~\cite{PhysRevB.55.R14717, gilles2002magnetic, PhysRevB.111.064304, PhysRevB.107.L060403, PhysRevB.43.11107,opherden2018inverted}, though typically not in \emph{purely} antiferromagnetic models, where it remains poorly understood. Indeed, even the textbook explanation for standard, ferromagnetic hysteresis in terms of sticky metastabilities alone proves deceptively simple: glossing over, in the classical case, the effects of kinetics in dynamical hysteresis~\cite{chakrabarti1999dynamic,broner1997dynamical}; and neglecting, in quantum systems, the role of quantum fluctuations in driving state transitions via tunneling. This motivates a study of hysteresis precisely where such metastability is absent~\cite{broner1997dynamical}: the pure antiferromagnet.

In this work, we explore an exotic form of magnetic hysteresis in various systems of antiferromagnetically interacting qubits exposed to quantum fluctuations induced by transverse fields. Of the three varieties of models we consider, two exhibit frustration~\cite{PhysRevB.63.224401}.
There exist many magnetic compounds that can be effectively modeled as antiferromagnetic quantum Hamiltonians with various quasi-one-dimensional or triangular 2-dimensional lattices, such as \ce{BaCo_2V_2O_8} \cite{PhysRevLett.127.077201}, \ce{SrCo_2V_2O_8} \cite{PhysRevLett.123.067203}, and \ce{TmMgGaO_4} \cite{PhysRevResearch.2.043013}. In this study, we focus on simple Transverse-Field Ising Model (TFIM) open quantum systems that are 1D and 2D as canonical baseline geometries on which to probe antiferromagnetic hysteresis. We also experimentally show results from D-Wave processor lattices whose connectivity does not correspond to simple low-dimensional quantum Hamiltonian models. However, this protocol can be applied to many other lattices and other quantum Hamiltonians besides what we report here, in particular for D-Wave quantum processors any Ising model that can be expressed as a sub-graph of the hardware graph. Because magnetic hysteresis is a fundamental emergent phenomenon of interacting magnetic spins, the possibility of analog quantum experiments of antiferromagnetic hysteresis is a notable computational capability of contemporary, noisy, analog quantum annealing processors.

\section{Methods}\label{section:methods}
\subsection{The Hysteresis Protocol} 
We perform analog simulations on programmable quantum annealers manufactured by D-Wave Systems~\cite{johnson2011quantum, Lanting_2014}. These devices use superconducting flux qubits~\cite{Bunyk_2014, dickson2013thermally, PhysRevB.82.024511, Harris_2007, Johnson_2010} connected in a graph comprising repeating subgraph units, and operate at approximately 15 millikelvin (15 mK). 
D-Wave hardware realizes time-dependent transverse field Ising models of the form
\begin{align}
\begin{split}
    {\mathcal H} &= - \frac{A(s)}{2} \sum_i \hat{\sigma}_{x}^{(i)}  + \\
    & \frac{B(s)} {2} \left[ g(t) \sum_i h_i \hat{\sigma_z}^{(i)} + \sum_{i\ne j} J_{ij} \hat{\sigma_z}^{(i)} \hat{\sigma_z}^{(j)} \right],
    \end{split}
    \label{equation:QA_Hamiltonian_h_gain}
\end{align}
where $s\in[0,1]$ is a programmable time-dependent annealing parameter, and $A(s)$ and $B(s)$ are hardware-dependent energy scales (see Supplementary Information~\ref{section:A_s_B_s_functions}). 
The dimensionless coupler interaction strengths $\{J_{ij}\}$ and longitudinal fields $\{h_i\}$ can be programmed by the user. The $\sigma_{x}$ Pauli term is the non-commuting driving Hamiltonian. Crucially, the function $g(t)$, which multiplies the longitudinal field, can be controlled independently of $s$, and is thus central to our methods, as described in Ref.~\cite{pelofske2025magnetichysteresisexperimentsperformed}. 
In all our experiments, we set each $h_i$ to the maximum allowed normalized-hardware programmable value. 

The ability to vary $g$ in time, central to our experiments, is currently available only in the total-simulation-time regime of on the order of microseconds. 
While quantum annealers behave coherently in the fast annealing regime (with anneal times of the order of tens of nanoseconds)~\cite{king2022coherent, king2023quantum}, in the slow anneal regime, the devices are described more accurately by an open quantum system~\cite{Amin_2015}.
In this regime, coupling to the environment leading to thermalization~\cite{PhysRevApplied.19.034053, PhysRevApplied.8.064025, buffoni2020thermodynamics, Chancellor_2016}, analog control errors and cross talk~\cite{nelson2021singlequbitfidelityassessmentquantum, tüysüz2025learningresponsefunctionsanalog, PRXQuantum.3.020317, Zaborniak_2021}, time-dependent noise characteristics~\cite{Pelofske_2023_noise}, and spin bath polarization~\cite{lanting2020probingenvironmentalspinpolarization} can all cause deviations from a closed quantum evolution.

\begin{figure*}[ht!]
    \centering
    \includegraphics[width=0.73\linewidth]{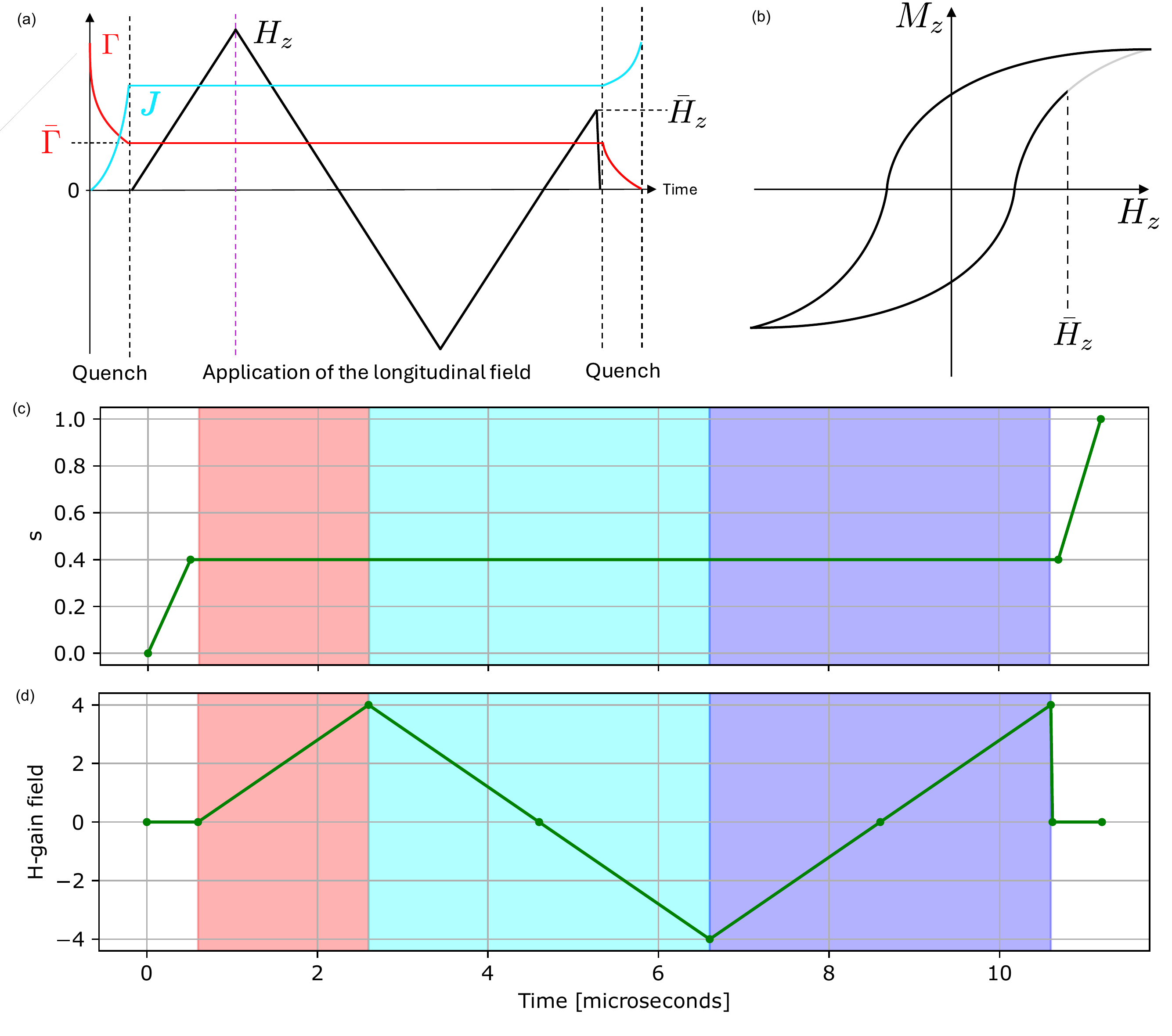}
    \caption{\textbf{Schematic of the magnetic hysteresis protocol}. Panel a shows the field ramps over time, as schematics (not actual energy scales or timescales) where we begin recording magnetization at the maximum-polarization longitudinal field denoted by the dashed purple vertical line, and panel b shows the conceptual tracing of the average magnetization response to the applied longitudinal field $H_z$. Panels c and d show the programmed D-Wave hardware waveforms for the \emph{final} simulation time of $11.2 \mu$s, where $s$ is the ``anneal-schedule'' control parameter, and the h-gain field is $H_z$. Adapted from ref.~\cite{pelofske2025magnetichysteresisexperimentsperformed}. }
    \label{fig:protocol_diagram}
\end{figure*}

\begin{table*}[ht!]
    \begin{center}
        \begin{tabular}{|l||l|l|l|l|}
            \hline
            D-Wave QPU Chip & Graph name & Qubits & Couplers & Maximum h-gain field strength \\
            \hline
            \hline
            \texttt{Advantage\_system4.1} & Pegasus $P_{16}$ & 5627 & 40279 & $\pm 1.75$ \\
            \hline
            \texttt{Advantage\_system6.4} & Pegasus $P_{16}$ & 5612 & 40088 & $\pm 4$ \\
            \hline
            \texttt{Advantage\_system7.1} & Pegasus $P_{16}$ & 5554 & 39238 & $\pm 3.5$ \\
            \hline
            \texttt{Advantage2\_prototype2.6} & Zephyr $Z_{6, 4}$ & 1248 & 10827 & $\pm 3$ \\
            \hline
        \end{tabular}
    \end{center}
    \caption{D-Wave QPU summary, where the maximum possible h-gain field that can be applied is in hardware-normalized units (relative to the specific Hamiltonian on each QPU). }
    \label{table:hardware_summary}
\end{table*}

For all measurements we use an $11.2$ microsecond total annealing time.
We implement ``forward anneals'' (i.e., those where $s$ is a non-decreasing function of time, which starts and ends at 0 and 1 respectively) with a pause in the annealing parameter $s$ during which we activate the longitudinal field. 
As in Ref.~\cite{pelofske2025magnetichysteresisexperimentsperformed}, we report average magnetization as a function of applied longitudinal field, with a reversed magnetization sign to compensate for the sign of the D-Wave QPU Hamiltonian Eq.~\eqref{equation:QA_Hamiltonian_h_gain}.

This hysteresis protocol is a probabilistic, sampling based protocol, and it requires both the transverse field (denoted as $\Gamma$) on the D-Wave QPUs, which facilitates state transitions, and a sufficiently strong lattice coupling $J$, without which no hysteresis can be measured. 
Our protocol has several important tunable parameters. 
The physical energy scale ratio $\Gamma/J$ at which the hysteresis cycle is swept has distinct behaviors at the two extremes. 
If it is too small, the state does not change (not even due to thermal fluctuations).
On the other hand, if $\Gamma/J$ is too large, then we are left with featureless unpolarized states with memory-less trajectories. 
Additionally, the longitudinal field must be sufficiently strong to force the system to become fully polarized against the antiferromagnetic bonds: therefore if the D-Wave hardware controls do not allow significantly strong longitudinal fields, only minor hysteresis loops are obtained. In principle one could always obtain saturation by using smaller coupling values, but that would amplify the role of thermal fluctuations and further destroy memory in an undesirable manner. Each QPU also has a different maximum longitudinal field that can be applied; the maximum field values of the ``h-gain'' schedule, i.e., of $g(t)$ in Eq.~\eqref{equation:QA_Hamiltonian_h_gain} are provided in Table~\ref{table:hardware_summary}. With the exception of the quasi-3D QPU hardware graph experiments, for all the reported experiments, we use default hardware settings, with no additional analog tuning; for instance, the flux bias offsets were left to default values, and no gauge-averaging was applied.

\subsection{Models and Embeddings}

We consider three different antiferromagnetic models: one-dimensional with periodic boundary conditions, two-dimensional with open boundary conditions, and a specific type of ``pseudo-3-dimensional'' model that involves the full  hardware graph. 
For all models, we set the coupler $J$ values to $1$, the maximum coupler strength that can be specified on the hardware.

The 2D square-grid antiferromagnetic model case corresponds to entirely unfrustrated magnetic systems because they have open boundary conditions. The pseudo-3D model is magnetically frustrated; it is not bipartite. The 1D antiferromagnetic systems we study have an odd number of spins and are hence frustrated on one bond.

For embedding or mapping the one and two-dimensional models onto the hardware graph, we used the Glasgow solver~\cite{mccreesh2020glasgow}, a graph isomorphism finder, which is a part of the \texttt{minorminer} package~\cite{cai2014practicalheuristicfindinggraph}. The maximum system sizes for the 1d model we were able to embed are: $4905$ qubits on \texttt{Advantage\_system4.1}; $4885$ qubits on \texttt{Advantage\_system6.4}, $4791$ qubits on \texttt{Advantage\_system7.1}; and $1131$ qubits on \texttt{Advantage2\_prototype2.6}. 
For the two dimensional model, the maximum sizes of the models we realize on the QPU are: $32\times 32$ lattice on \texttt{Advantage\_system4.1} and \texttt{Advantage\_system6.4}, $33\times 33$ on \texttt{Advantage\_system7.1}, and $26 \times 26$ on \texttt{Advantage2\_prototype2.6}. See Supplementary Information~\ref{section:appendix_embeddings} for visual renderings of the embeddings.

In the hardware-defined antiferromagnetic models, we set every programmable coupler on the QPU to $J=1$, which is the maximum coupler strength that can be specified on the hardware (in normalized programmable units) and whose energy scale is given in Supplementary Information~\ref{section:A_s_B_s_functions}. The number of qubits and couplers for the various QPUs are listed in Table~\ref{table:hardware_summary}. The QPUs have either Pegasus~\cite{dattani2019pegasussecondconnectivitygraph, boothby2020nextgenerationtopologydwavequantum} or Zephyr~\cite{zephyr} connectivity. 

\begin{figure*}[ht!]
    \centering
    \includegraphics[width=0.496\linewidth]{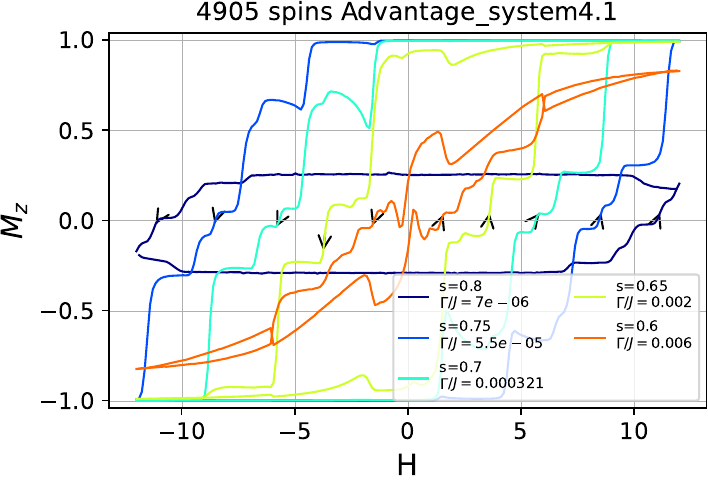}
    \includegraphics[width=0.496\linewidth]{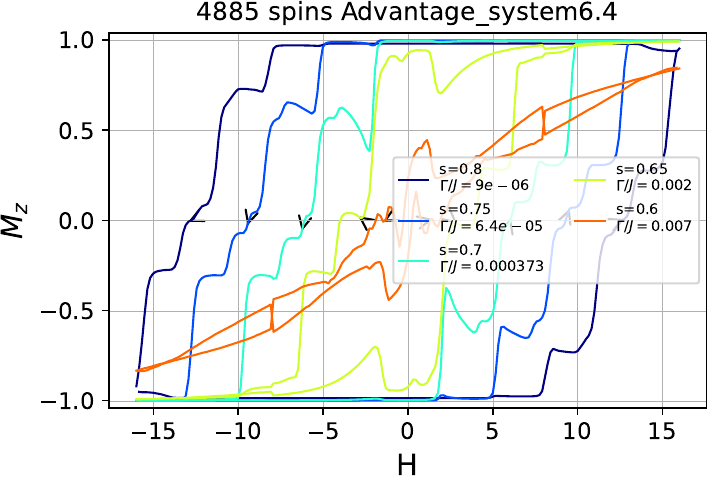}
    \includegraphics[width=0.496\linewidth]{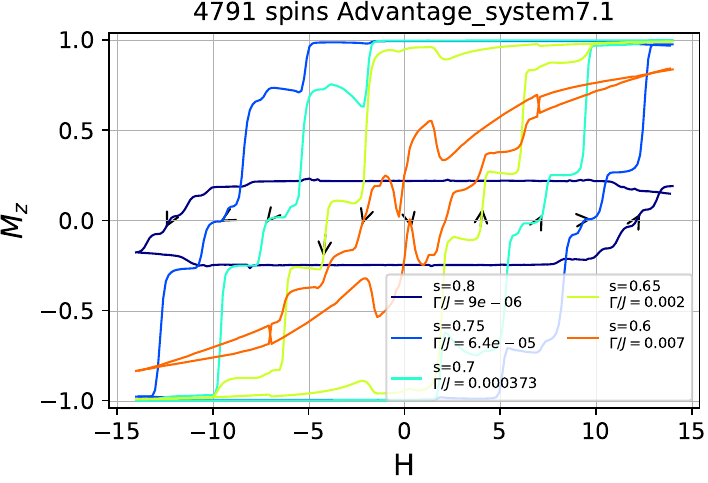}
    \includegraphics[width=0.496\linewidth]{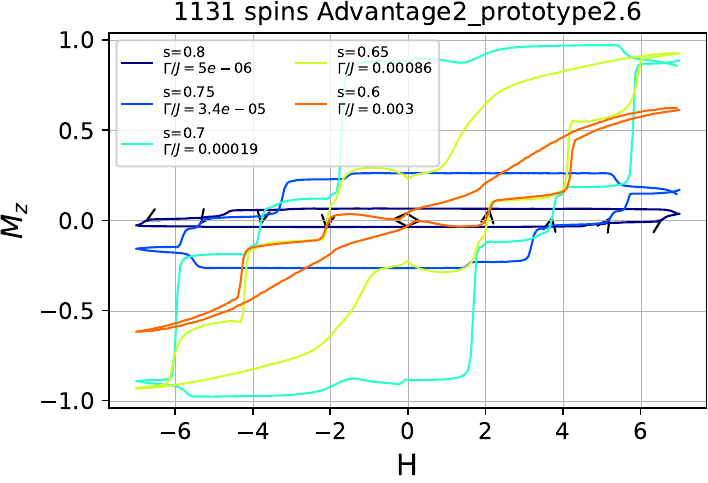}
    \caption{{\bf Magnetic hysteresis on 1D odd antiferromagnetic rings}. Average magnetization as a function of the applied longitudinal field $H$. The overlaid black arrows denote the time-progression of the experiment as well as the direction of the longitudinal field sweeps. The different lines correspond to D-Wave experiments performed at different anneal-schedule-pauses, defined by $s$, which corresponds to a particular physical field ratio of $\Gamma/J$.  }
    \label{fig:1D_AFM}
\end{figure*}

\subsection{Calibration Refinement by Statistical Balancing}\label{section:methods_calibration}
Non-idealities in the quantum annealers, even small fluctuations in the magnetic environment, can result in non-uniform response even in models that are uniform.
These effects can be mitigated by leveraging the symmetries of the model being studied via an iterative statistical-balancing calibration refinement technique~\cite{Chern_2023}, which has been implemented in a broad range of experiments~\cite{Chern_2023, King_2023_critical, Kairys_2020, king2022coherent, ali2024quantum, PhysRevB.110.054432, sathe2025classicalcriticalityquantumannealing, king2024computationalsupremacyquantumsimulation, Pelofske_2021_graph_partitioning, PhysRevA.102.042403}. 
In some of our ``quasi-3D'' hardware-defined experiments, we implement a flux bias offset (FBO) calibration refinement procedure.
In particular, before implementing any hysteresis experiment, we implement trial experiments with zero longitudinal field.
Due to $\mathbb Z_2$ symmetry, the average magnetization of each qubit should be 0. 
Due to experimental non-idealities, we instead find a spread in the values of qubit magnetizations.
By iteratively modifying the FBOs, the spread of qubit magnetizations can be reduced, achieving more uniformity.
(See Supplementary Information~\ref{section:flux_bias_offset_calibration} for more details.) 
The FBO values obtained at the end of the calibration refinement procedure are then used for the hysteresis experiments to improve accuracy of the results. 
While more advanced calibration refinement techniques could be utilized, such as modifying coupler strengths to make coupler frustration uniform, or refining individual $h_i$ values in each ``time-slice'' of the hysteresis protocol, such a procedure would use significantly more QPU time particularly for models of the size we study here. 
Hence, we restrict to FBO refinement in this work.
For completeness we report both calibrated and uncalibrated data for the quasi-3D models.

\begin{figure}[ht!]
    \centering
    \includegraphics[width=0.494\linewidth]{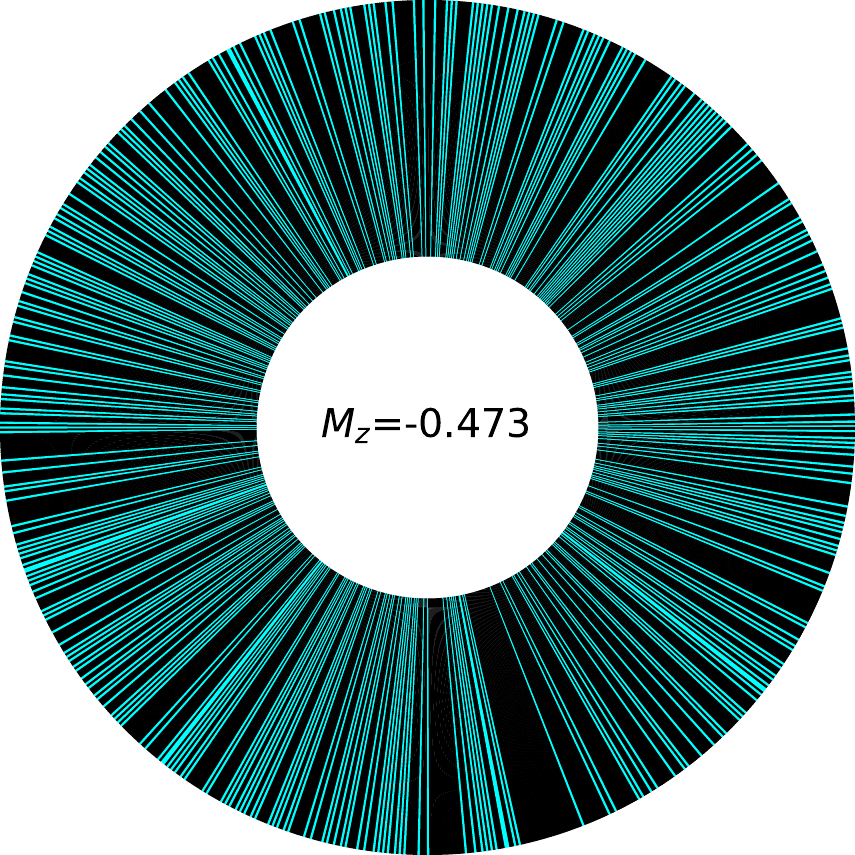}
    \includegraphics[width=0.494\linewidth]{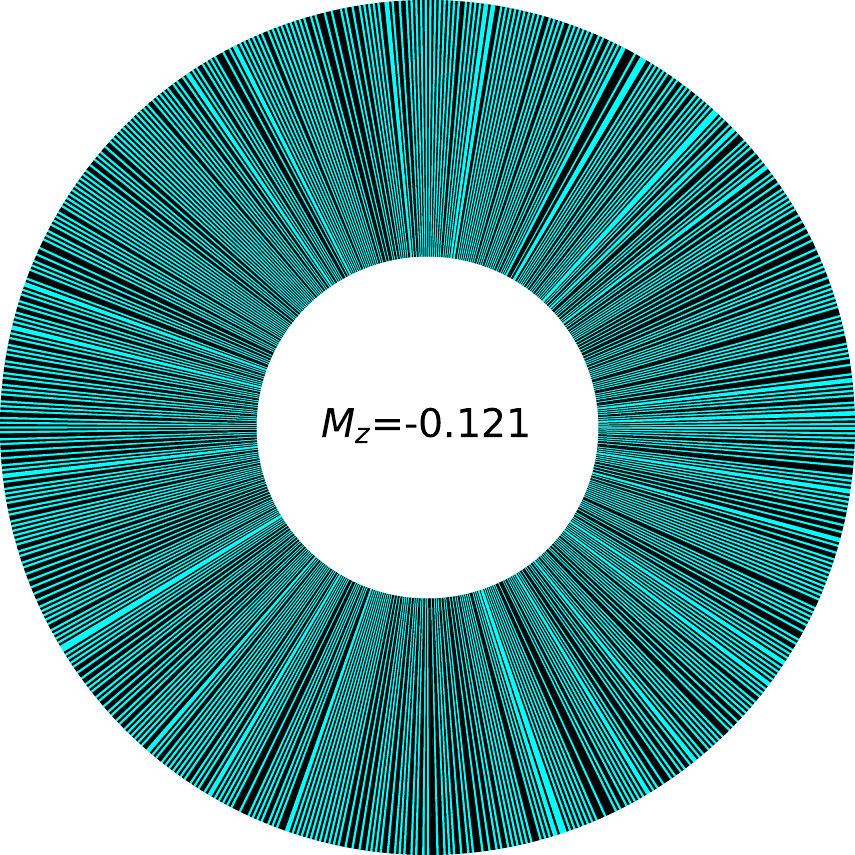}
    \caption{Selected representative example 1D antiferromagnet spin configurations measured during the hysteresis cycle run on \texttt{Advantage2\_prototype2.6} (at $s=0.7$), comprised of $1131$ spins. Each spin configuration is from a different applied longitudinal field value in the non-saturated region of the hysteresis cycle, both near $H \approx -2$. The spin configurations are shown as a circular annular wedge plot where black slices denote spin-down qubit measurements and cyan slices denote spin up qubit measurements. The net magnetization for each sample is shown in the center of each plot. These selected samples show different cases of fragmented antiferromagnetic ordering, which is to be expected for these experiments.  }
    \label{fig:1D_spin_configs_Advantage2_system2.6}
\end{figure}

\subsection{Observables}\label{section:methods_structure_factor}
From the experimental data, we extract the average longitudinal magnetization
\begin{equation}
M^z = \frac{1}{N} \sum_i \langle \sigma^z_i \rangle,
\end{equation}
where $\langle \dots \rangle$ denotes an average over $2{,}000$ samples ($8{,}000$ samples in the case of the quasi-3D models) measured on the hardware sampler.
The experimental average over a relatively large number of samples mitigates the possibility of apparent ordering due to finite sampling effects, which may occur if the number of samples is especially small.

It is natural to also consider the antiferromagnetic order parameter, which is defined as
\begin{equation}
M_s = \frac{1}{N} \sum_i (-1)^i \langle \sigma^z_i \rangle.
\label{eq:Neel}
\end{equation}
This quantity is well-defined only for a bipartite graph, and hence can be extracted only for the 2d models and not the odd-length one dimensional chains or the quasi-3D models.
Interestingly, unlike magnetization, the \emph{staggered} antiferromagnetic order parameter is inaccessible to experiments in a magnetic laboratory without a scattering experiment such as neutron scattering, but is computationally easily accessible to experiments on quantum annealers because they provide information on the elementary degrees of freedom.

Our access to the individual degrees of freedom allows for the direct extraction  of other interesting observables, including the real space snapshots of spin configurations, which we will show in section~\ref{section:results}. 
This also allows us to extract the magnetic structure factor, which gives clear visual evidence for different types of magnetic ordering during hysteresis cycles. While in experiments it is obtained by neutron scattering, we can reconstruct it from individual spins, as 
\begin{equation}
S(\mathbf{q}) = \left\langle \sum_{i, j} e^{i \mathbf{q} \cdot (\mathbf{r}_i - \mathbf{r}_j)}  \sigma_z^i \sigma_z^j \right\rangle
\end{equation}
where $\mathbf{r}$ denotes spin lattice locations. For the 2D case on a square lattice, we discretize the momentum vector $\mathbf{q}$ on a $200\times 200$ uniformly spaced grid spanning $(-2\pi,2\pi)$ along both dimensions. 

The pseudo-3D models lack a sensible definition for a Brillouin zone, and hence we compute the structure factor only for the one and two dimensional models.

For the 1D models (which always have odd lengths in our study), we examine domain wall counts which ranges from $1$ (in the ground state) to $L$ for the fully polarized state (where $L$ is the number of sites in the 1D model). 

\begin{figure*}[ht!]
    \centering
    \includegraphics[width=0.49\linewidth]{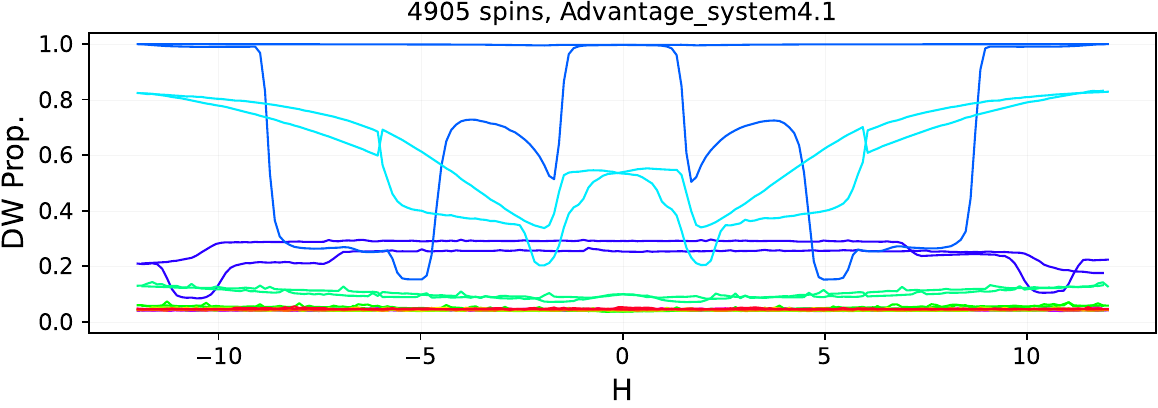}
    \includegraphics[width=0.49\linewidth]{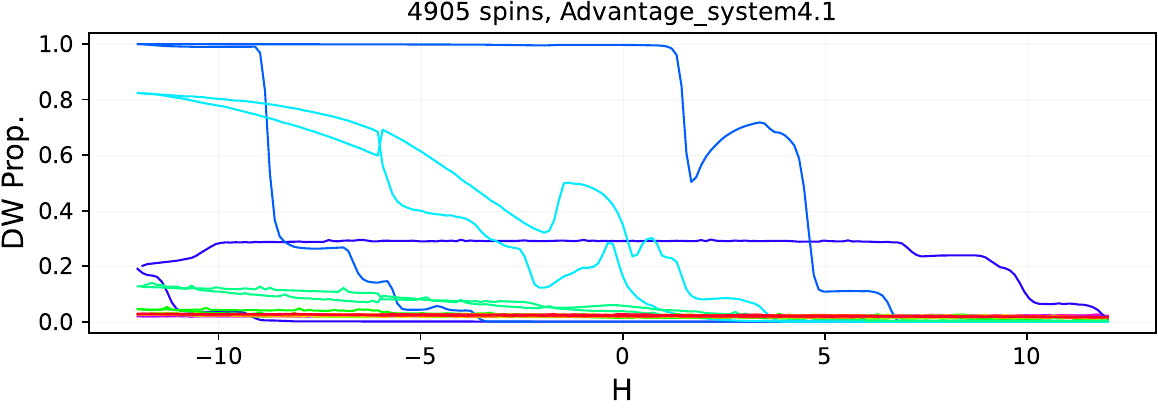}
    \includegraphics[width=0.6\linewidth]{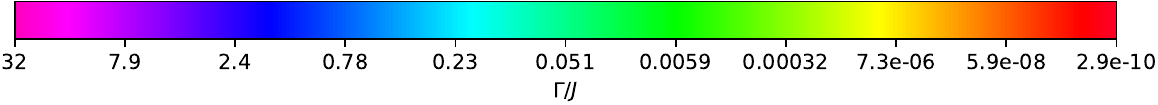}
    \caption{\textbf{Domain wall density on a 1D antiferromagnetic ring during the hysteresis cycles as a function of applied longitudinal field $H$}. Left hand plot shows the total, average, domain wall density, regardless of the sign of the domain wall. Right hand plot shows the spin-down ($\downarrow \downarrow$) domain wall density. Each closed curve corresponds to a different $\Gamma/J$ value on the D-Wave processor. Note that because these rings have an odd number of spins, we are always guaranteed to have at least one pinned domain wall somewhere on the ring. Color encodes the $\Gamma/J$ energy scale ratio which corresponds to an ``anneal''-pause value $s$. Both at very high $\Gamma/J$ and small $\Gamma/J$ the antiferromagnetic domain wall proportion drops to very close to zero, which is due to state transitions being energetically favorable in one case and unfavorable in the other case. However, the lack of domain walls means that for most of the plotted color-spectrum, the data lines are nearly identical, and therefore visually only a certain range of $\Gamma/J$ show differences. }
    \label{fig:1D_domain_wall_proportions}
\end{figure*}

\section{Results}\label{section:results}

We report antiferromagnetic hysteresis for the 1D, 2D and quasi-3D models in sections~\ref{section:results_1D},~\ref{section:results_2D} and \ref{section:results_3D} respectively. All reported results for the 1D and 2D (quasi-3D) models are computed from $2{,}000$ ($8{,}000$) measurements per data point, for each change in experimental configuration. All reported net magnetization hysteresis results are statistical averages. The hysteresis cycle plots are presented as a function of $H$, where $H$ field units are in hardware-specific programmable units.

\subsection{1-Dimensional Antiferromagnet Experiments}\label{section:results_1D}
Fig.~\ref{fig:1D_AFM} presents magnetic hysteresis experiments for 1D antiferromagnetic models with periodic boundary conditions, performed at different values of $\Gamma/J$, which we control by varying the value of $s$. It shows interesting non-monotonic features. In particular, a ``pinch-point'' occurs at $s=0.6$ on three of the QPUs, except \texttt{Advantage2\_prototype2.6}. When $\Gamma/J$ is sufficiently small, we achieve full polarization, unlike for larger values of $\Gamma$. 

Fig.~\ref{fig:1D_spin_configs_Advantage2_system2.6} shows examples of real-space spin configurations sampled during one of the hysteresis cycles run on the D-Wave QPU \texttt{Advantage2\_prototype2.6}, specifically in the non-saturated regions of the cycle. These spin configurations show very small antiferromagnetic domains at small $M_z$; the distribution of alternating sign spins is not uniform -- which could be in part due to small J coupling energy scales differences on the analog hardware. One sees formation of magnetic domains for larger $M_z$, as expected.

Fig.~\ref{fig:1D_domain_wall_proportions} reports 1D domain wall density in the 1D hysteresis cycles, including the proportion of $\downarrow \downarrow$ domain walls (note that, like magnetization, the sign of this domain wall has been reversed in order to compensate for the sign of the D-Wave processor Hamiltonian~\cite{pelofske2025magnetichysteresisexperimentsperformed}). The domain wall density on the other three QPUs is qualitatively similar, which can be found in Supplementary Information~\ref{section:appendix_Domain_wall_density}.

\subsubsection{Odd Rings, Domain-Wall Sectors, and Their Use as a Defect Transport Probe}
To provide more insight on these experiments, consider the transverse-field Ising antiferromagnet on a ring,
\begin{equation}
    H = J\sum_{i=1}^{N}\sigma_i^z\sigma_{i+1}^z \;-\; h\sum_{i=1}^{N}\sigma_i^z \;-\; \Gamma\sum_{i=1}^{N}\sigma_i^x,
\end{equation}
\label{eq:TFI}
with  $\sigma_{N+1}^z\equiv\sigma_1^z$ and  $J>0$. Because $N$ is odd, there are an odd number of domain walls for any computational basis ($\sigma^z$ basis) configuration. When $\abs{h}, \Gamma\ll J$, the ground state subspace, comprising configurations with exactly one domain wall ($D=1$), is $N$-fold degenerate.

Local spin flips generated by the transverse part of the Hamiltonian, proportional to $\Gamma$, toggle the two bonds adjacent to the flipped site, so the total number of domain walls changes by $0$ or $\pm 2$, thus conserving the parity of domain wall count. 
Thus, when $h=0$, domain wall parity is a conserved quantity.
Additionally, the domain walls come in two flavors, corresponding to both the spins pointing up, or down. A longitudinal field $h\neq0$ breaks the global $\mathbb{Z}_2$ symmetry and splits these two local configurations in energy. These are not distinct dynamical sectors once $\Gamma>0$, because a single spin flip at the wall converts a spin-up domain wall to a spin-down domain wall (and vice-versa) while shifting the wall by one bond. Thus, odd rings naturally realize a single-wall system in which $\Gamma$ drives coherent domain-wall motion and $h$ biases its drift.

This built-in, mobile defect makes odd rings a probe of quantum-assisted defect dynamics and antiferromagnetic hysteresis. In contrast, an even-$N$ ring possesses a wall-free N\'eel lowest energy configuration; reversal at small $|h|$ must first nucleate a kink–antikink pair (energy cost $\approx 4J$ at $\Gamma=0$), which strongly suppresses response and complicates interpretation. Because even rings require pair creation, their low-field loops are comparatively featureless and are not informative about defect-transport dynamics, we therefore concentrate on odd rings for quantitative analysis. For this reason, and because our experiments target quantum assisted transport and pinning phenomena, we restrict to odd rings in all the one-dimensional experiments in this study. 

Weak static inhomogeneity (bond or field offsets) pins the wall at preferred bonds. At a pinned location the two local sign states form a driven two-level subsystem; the transverse field $\Gamma$ mixes them, while $h$ sets their energy difference. As detailed below, this competition produces an avoided crossing and underlies the stick–slip dynamics and step-like magnetization changes characteristic of the antiferromagnetic hysteresis we observe.

In summary, the 1-dimensional antiferromagnet model experiments show non-monotonic magnetization features and interesting switching that occurs at high longitudinal field. These features have been observed before in prior quantum annealer experiments on different types of models~\cite{barrows2025magneticmemoryhysteresisquantum, pelofske2025magnetichysteresisexperimentsperformed}. In these models however, there is more ``corrugation'' compared to ferromagnet models~\cite{barrows2025magneticmemoryhysteresisquantum}. Examining domain wall statistics and individual measurements reveals fragmented antiferromagnetic ordering in the approximately de-polarized region of the hysteresis cycles.

\begin{figure*}[ht!]
    \centering
    \includegraphics[width=0.496\linewidth]{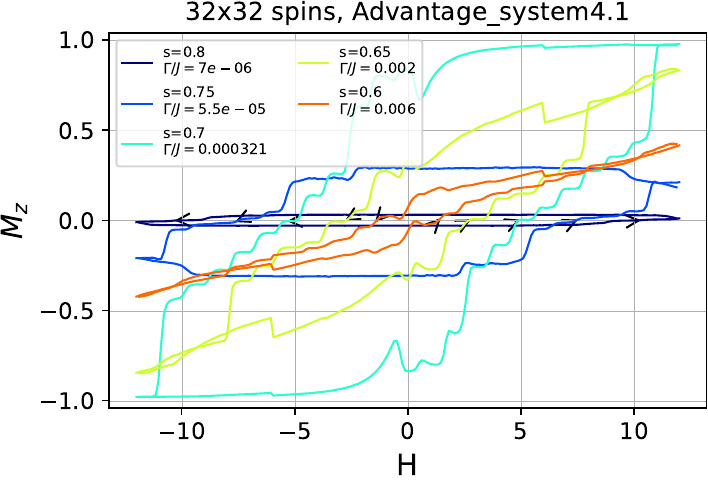}
    \includegraphics[width=0.496\linewidth]{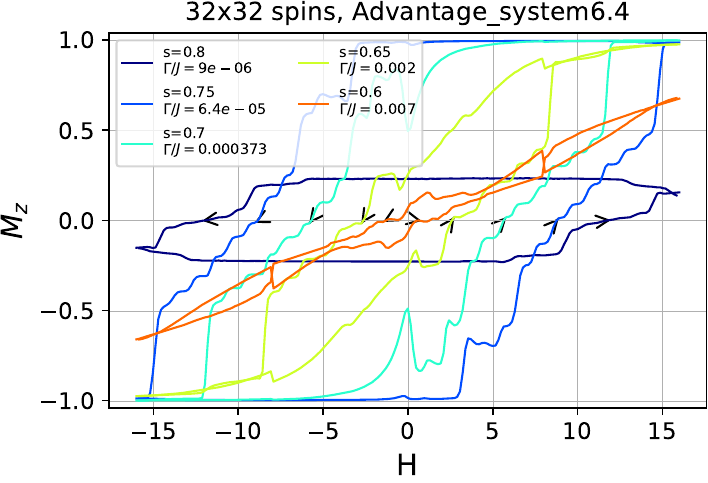}
    \includegraphics[width=0.496\linewidth]{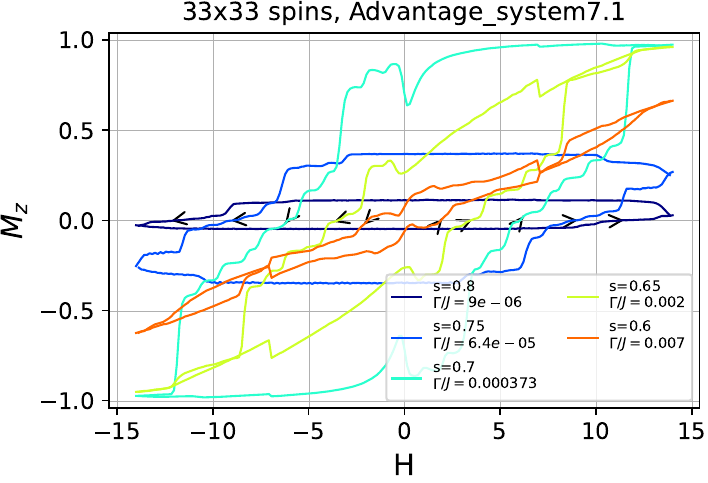}
    \includegraphics[width=0.496\linewidth]{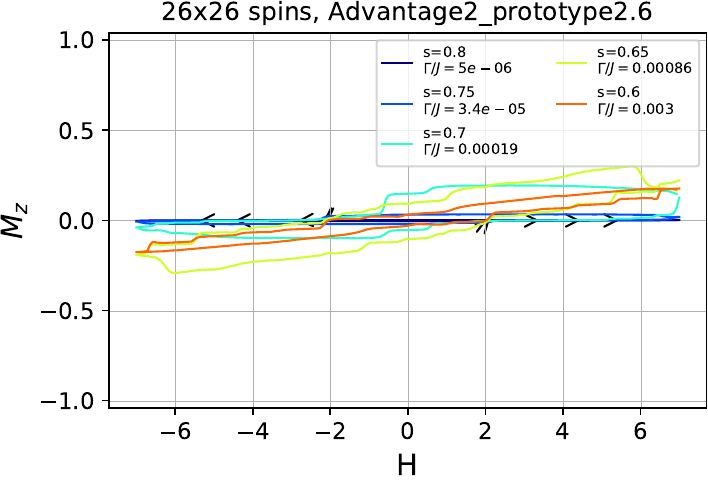}
    \caption{Magnetic hysteresis on 2D antiferromagnetic square lattices, on four different D-Wave quantum annealers. Average magnetization (y-axis) as a function of the applied longitudinal field (x-axis). The overlaid black arrows on the average magnetization lines sampled from the QPUs simultaneously denote the time-progression of the analog simulations, as well as the direction of the longitudinal field sweeps. The lower-right subplot results do not get close to saturation for any $\Gamma/J$ energy scale ratio causing the magnetization lines to overlap. The reason for this is because the maximum longitudinal field hardware energy scales are not as large as for the other processors.   }
    \label{fig:2D_AFM}
\end{figure*}

\subsubsection{Domain-wall Theory: Effective Hamiltonian, Dispersion, and Many-Wall Processes}

In the single-wall regime we label basis states $\ket{j,w}$, where $j\in\{1,\dots,N\}$ denotes the location of the domain wall and $w=\pm 1$ denotes its flavor (both up or both down). 
For these states, the exchange energy contribution is given by
\begin{equation}
E_J = -J(N-2),
\end{equation}
because one bond is ferromagnetic $(+J)$ and the remaining $N-1$ bonds are antiferromagnetic $(-J)$. The longitudinal field contributes a term that depends only on the wall sign: for odd $N$ the total magnetization in a one-wall state is $M=\sum_i\sigma_i^z=w$, so
\begin{equation}
H_h \to -h w \quad \text{on}\quad \ket{j,w},
\end{equation}
where the longitudinal-field term is denoted by \(H_h=-h\sum_i\sigma_i^z\). 
A single $\sigma^x$ acting on either spin adjacent to the wall reverses that spin, toggles the wall sign, and shifts the wall by one bond. To leading order in $\Gamma/J$, degenerate perturbation theory yields an effective Hamiltonian given by
\begin{widetext}
    \begin{equation}
\label{eq:Heff}
H_{\rm eff} = E_J  + \sum_{w=\pm 1} \sum_{j=1}^N\Big[-h  w\ket{j,w} \bra{j,w}- \Gamma\big(\ket{j+1,-w}\bra{j,w}+\ket{j-1,-w}\bra{j,w}\big)\Big],
\end{equation}
\end{widetext}
with periodic indexing $j\equiv j{+}N$. Equation~\eqref{eq:Heff} is a tight-binding model on a two-sublattice (sign) chain.

For a homogeneous ring we diagonalize with Bloch states $\ket{k,w}=\tfrac{1}{\sqrt{N}}\sum_j e^{ikj}\ket{j,w}$, $k=2\pi m/N$. In the $\{\ket{k,+},\ket{k,-}\}$ basis the Bloch Hamiltonian reads
\begin{equation}
\mathcal{H}(k) = -h\sigma^z - 2\Gamma\cos\left(k \right)\sigma^x,
\end{equation}
with eigenenergies
\begin{equation}
\label{eq:bands}
E_\pm(k) = E_J \pm \sqrt{h^2 + 4\Gamma^2\cos^2\left(k\right)}.
\end{equation}
Thus the transverse field endows the wall with a hopping amplitude $\sim \Gamma$ and, at $h \approx 0$, a single-domain-wall spectrum spanning an energy width $4\Gamma$; the time dependent longitudinal field can populate excited one-wall states. A finite longitudinal field splits the two bands by at least $2|h|$ (at $k=\pi$), producing a drift bias for wall motion. The corresponding group velocity is
\begin{equation}
v_\pm(k) = \pm \frac{2\Gamma^2 \sin(k)}{\sqrt{h^2 + 4\Gamma^2\cos^2(k)}}.
\end{equation}

\begin{figure*}[ht!]
    \centering
    \includegraphics[width=0.999\linewidth]{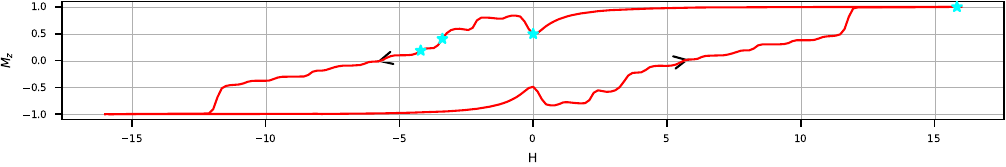}
    \includegraphics[width=0.245\linewidth]{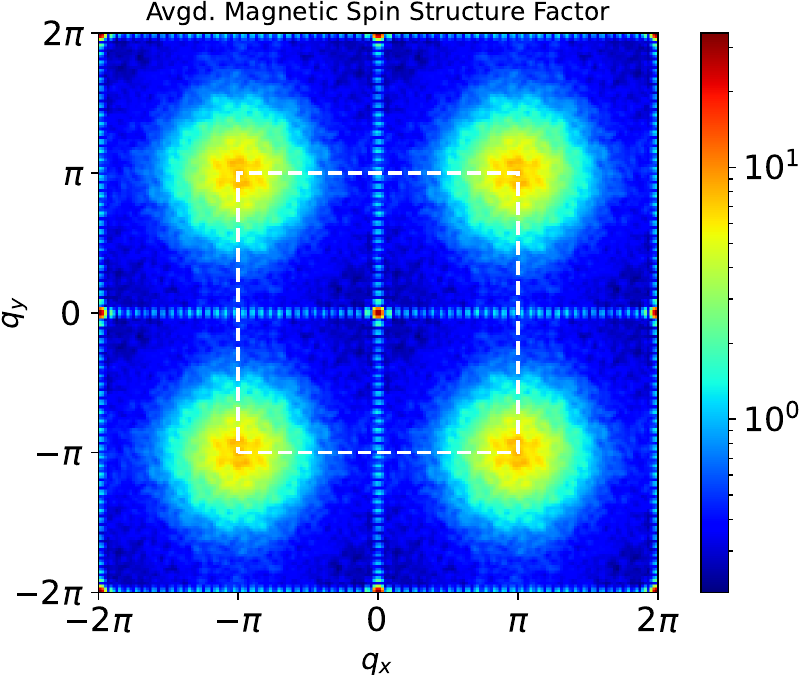}
    \includegraphics[width=0.245\linewidth]{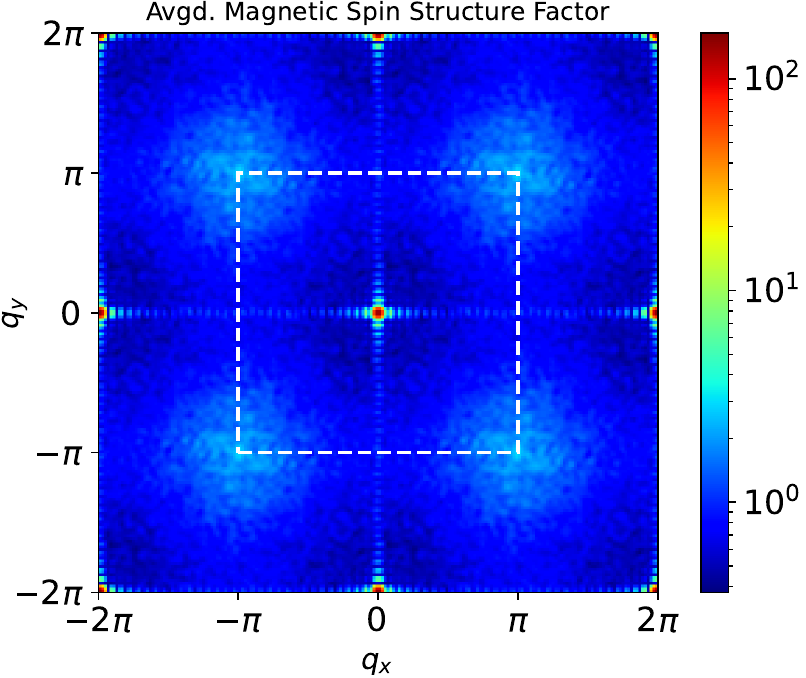}
    \includegraphics[width=0.245\linewidth]{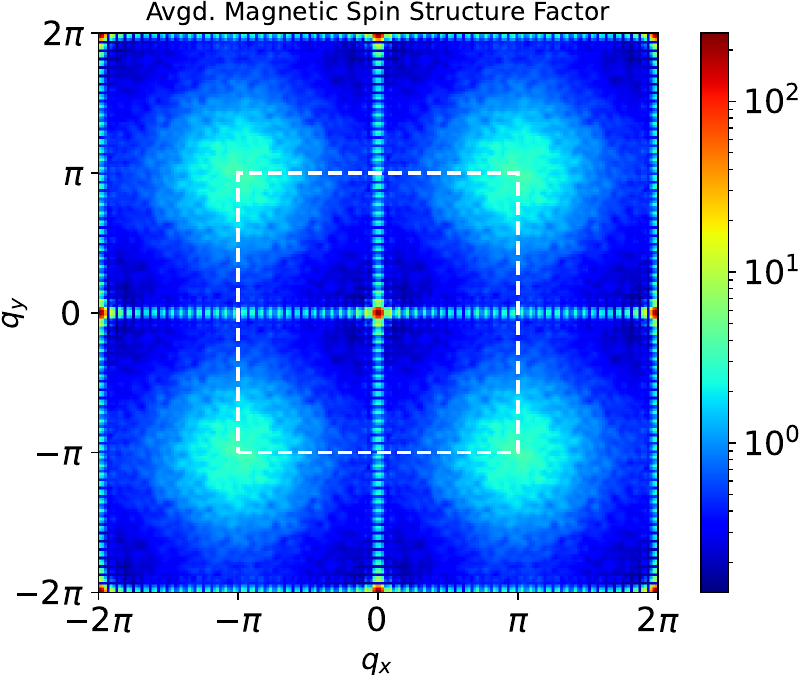}
    \includegraphics[width=0.245\linewidth]{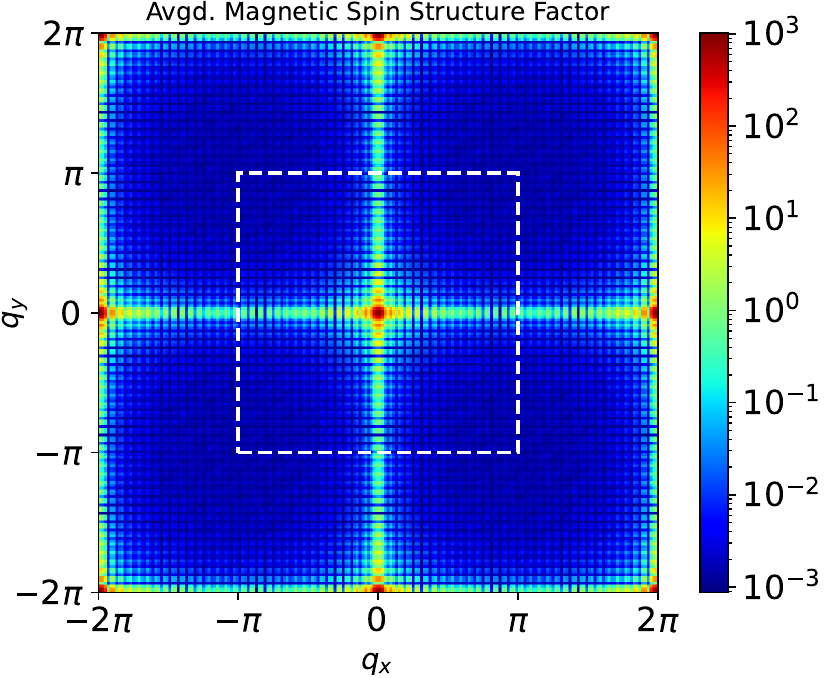}
    \caption{Magnetic structure factors $\abs{S(q)}$ heatmaps (bottom row) within the 2D antiferromagnetic hysteresis cycles (averaged over 100 independent samples), run on \texttt{Advantage\_system6.4} at $s=0.7$. The Ising model being simulated is a $32\times 32$ grid of (antiferromagnetically coupled) spins, open boundary conditions. The top row shows the single hysteresis cycle in terms of $M_z$ average magnetization, from which we extracted $\abs{S(q)}$ heatmaps for 4 points along the first longitudinal field sweep, denoted by cyan asterisks. The dashed white line on the structure factor heat maps outlines the first Brillouin zone. The heat maps are log-scaled and scales  are specific to each heatmap subplot. 
    The order of the $\abs{S(q)}$ heatmaps follows the same sweep direction denoted by the top hysteresis cycle plot (the static structure factor from the fully saturated regime is shown in the right hand subplot). Note the peaks at the corner of the Brillouin zone denoting large antiferromagnetic clusters, and coexisting with a growing peak at the center denoting magnetization. }
    \label{fig:SSF1}
\end{figure*}

While odd $N$ fixes the wall-number parity, additional kink–antikink pairs can be nucleated at finite drive. At $\Gamma=0$ and small $|h|$ a single local spin flip in a N\'eel region creates two additional domain walls and costs
\begin{equation}
\Delta E_{\rm pair} \approx 4J - 2h\delta M,
\end{equation}
with $\delta M=\pm 1$ in our normalization ($\sum_i\sigma_i^z=M$). Quantum mechanically, pair creation arises from higher-order virtual processes in $\Gamma/J$, while annihilation occurs when a wall meets an antiwall. On our platform thermal activation is negligible; the dynamics is therefore dominated by wall hopping, field-biased drift, quantum-assisted pair nucleation, and annihilation.

The interplay of hopping ($\sim\Gamma$) and field bias ($h$),  pinning, and rare nucleation/annihilation events produces the antiferromagnetic hysteresis observed in odd rings. In the single-wall-dominated regime, magnetization evolves via sequences of hops punctuated by depinning events at local avoided crossings, generating discrete, Barkhausen-like steps. As the drive amplitude increases, occasional nucleation events seed additional walls which drift and annihilate; this ``dilute domain-wall gas'' smooths the loop while preserving a strong odd-ring response near low fields.  A simple estimate gives the pair-creation energy $\Delta E_{\mathrm{pair}} \approx 4J-2h \delta M$, so pair nucleation becomes likely once $|h| \sim 2J$, beyond this “nucleation threshold” the loop evolves in a dilute domain-wall gas rather than by single-wall motion. In the data the rise of total domain wall density and magnetization reversal set in around this scale.

For odd-$N$ antiferromagnetic rings the domain-wall number $D$ is constrained to be odd, in the regime of small $|h|$ with $\Gamma\ll J$ the lowest-energy states have a single domain wall. In Fig.~\ref{fig:1D_domain_wall_proportions}, in the same low-field window the sign-resolved wall proportion exhibits a strong, hysteretic bias between $\uparrow\uparrow$ and $\downarrow\downarrow$ walls, with forward–reverse asymmetry. This pattern—weak change in total domain wall proportion but strong, history-dependent redistribution between the two wall signs—is exactly what the effective single-wall Hamiltonian predicts: $h$ splits the two wall-sign states $w=\pm1$ while $\Gamma$ mixes them, yielding a biased two-level population that traces a hysteresis loop.  In equilibrium this bias would reduce to a sigmoidal form $\tfrac12[1+\tanh(\beta_{\mathrm{eff}}h)]$. 

As $|h|$ grows the total domain wall proportion becomes large and the loop area is dominated by many-wall dynamics. The conspicuous non-monotonicity near $|h| \approx 1$ at $s\!\approx\!0.6$ lies in this regime. Such large non-monotonicities can be understood as a result of synchronized depinning at reproducible nucleation fields set by hardware pinning, i.e., pinning sites and hopping barriers introduced by hardware disorder and inhomogeneities in device parameters. The non-monotonicities and resulting Barkhausen steps survive ensemble averaging because those nucleation fields recur shot-to-shot.

\begin{figure*}[ht!]
    \centering
    \includegraphics[width=0.15\linewidth]{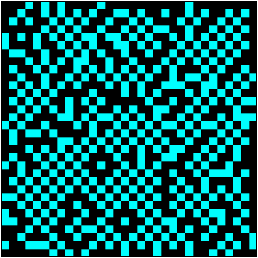}
    \includegraphics[width=0.15\linewidth]{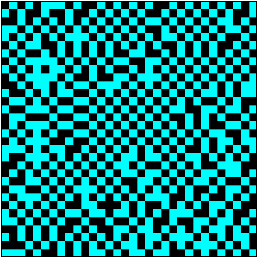}
    \includegraphics[width=0.15\linewidth]{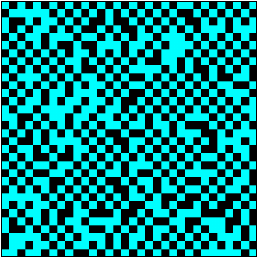}
    \includegraphics[width=0.15\linewidth]{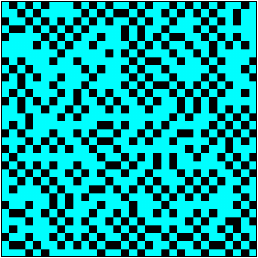}
    \includegraphics[width=0.15\linewidth]{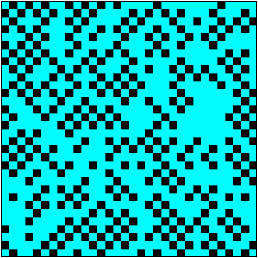}
    \includegraphics[width=0.15\linewidth]{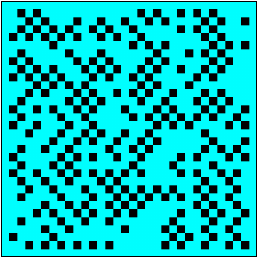}
    \includegraphics[width=0.15\linewidth]{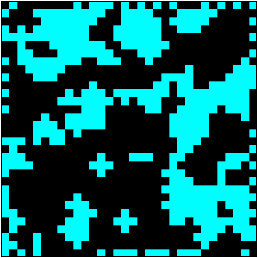}
    \includegraphics[width=0.15\linewidth]{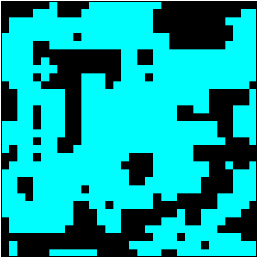}
    \includegraphics[width=0.15\linewidth]{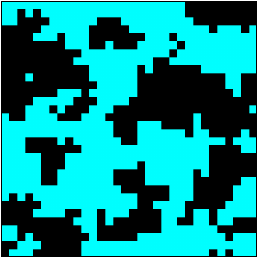}
    \includegraphics[width=0.15\linewidth]{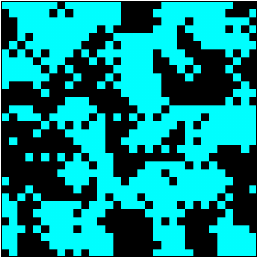}
    \includegraphics[width=0.15\linewidth]{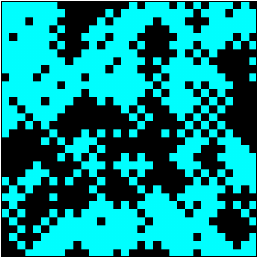}
    \includegraphics[width=0.15\linewidth]{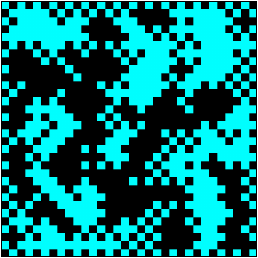}
    \caption{\textbf{Selected real-space spin configurations sampled on the \texttt{Advantage\_system6.4} QPU, } illustrated as pixel plots where black pixels denote spin-down qubit measurements and cyan pixels denote spin-up qubit measurements (top row), from the middle of the magnetization reversal at various longitudinal field strengths of the first magnetic hysteresis sweep of a $32\times 32$ 2D square grid antiferromagnet, at $s=0.7$. Notice that the checkerboard antiferromagnetic ground-state emerges in sections of the lattices. Also notice that the boundary effects from the open boundary conditions are noticeable in many of the measured spin configurations. Bottom row: staggered domain height function $\tau$ from Eq.~(\ref{eq:tau}), corresponding to the same spin configurations on top row, making the antiferromagnetic domains apparent.}
    \label{fig:2D_spin_configs_Advantage_system6.4}
\end{figure*}

\subsection{2-Dimensional Antiferromagnet Experiments}\label{section:results_2D}

Fig.~\ref{fig:2D_AFM} presents 2-dimensional antiferromagnetic hysteresis simulations. Again, we observe full saturation for some values of $\Gamma/J$, on most QPUs. On the \texttt{Advantage2\_prototype2.6} processor, the maximum longitudinal field that can be applied is comparatively weaker, resulting in the net magnetization being lower than $\pm 0.5$ for any $\Gamma/J$ value. In some regimes, such as $s=0.7$ and $s=0.75$ where the system is fully polarized, we observe Barkhausen-like noise (e.g., crackling) in these experiments.

Here we can also extract the spin structure factor in order to examine the types of magnetic ordering that occur during the hysteresis cycles. Fig.~\ref{fig:SSF1} shows averaged magnetic structure factors\footnote{Also referred to as Static Structure Factor (SSF)} from various points along the antiferromagnetic hysteresis cycle. We see clear boundary effects (due to the open boundary conditions) that occur closer to the demagnetized regions, seen as lines along $x=y=0$ axes. In the saturated regimes, we see either bright peaks along $q_x=\pi$ and $q_y=\pi$ when $\Gamma/J$ is high (meaning, the magnetization is not fully saturated), or closer to ferromagnetic-like ordering (only due to the system being polarized) when the ratio $\Gamma/J$ is smaller and also the longitudinal field is strong enough. The clearest signal we see is strong antiferromagnetic ordering, with peaks at the four corners of the first Brillouin zone, as the systems become more demagnetized. Across the different D-Wave QPUs and physical device settings, there are slight differences in the magnetic ordering captured by the SSF plots, but qualitatively they are all similar to each other with the primary difference being to what extent full saturation can be achieved; additional SSF plots under other settings are shown in Supplementary Information~\ref{section:additional_SSF_plots}.

Fig.~\ref{fig:2D_spin_configs_Advantage_system6.4} shows examples of real space single-spin configurations measured during one of the 2-dimensional antiferromagnet hysteresis cycles. These samples show fragmentation into domain walls of the expected checkerboard ordering in the ground-state of the 2D square antiferromagnet, and moreover we see clear boundary effects (due to the open boundary conditions) as the system becomes more magnetized. Qualitatively, the measured spin configurations during the hysteresis cycles are similar across the different devices and different physical energy scales, Supplementary Information~\ref{section:appendix_additional_2D_samples} reports similar measurements on a different D-Wave QPU. To better illustrate the antiferromagnetic domains in Fig.~\ref{fig:2D_spin_configs_Advantage_system6.4} we plot also on the bipartite lattice the domain height function
\begin{equation}
    \tau_{i_x i_y} =(-1)^{i_x+i_y}\,\sigma_{i_xi_y}^z,
    \label{eq:tau}
\end{equation}
which is uniform and valued $\pm1 $ in each domain. Domain walls are then the contours separating $\tau\!=\!+1$ and $\tau\!=\!-1$. Clearly, it is related to the antiferromagnetic order parameter via $M_s=\sum_{i_x i_y}\tau_{i_x, i_y}/N$.  

\begin{figure*}[ht]
    \centering
    \includegraphics[width=0.4965\linewidth]{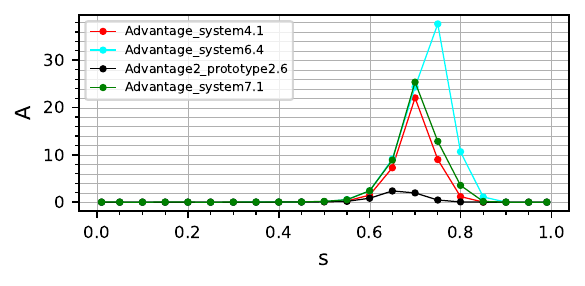}
    \includegraphics[width=0.4965\linewidth]{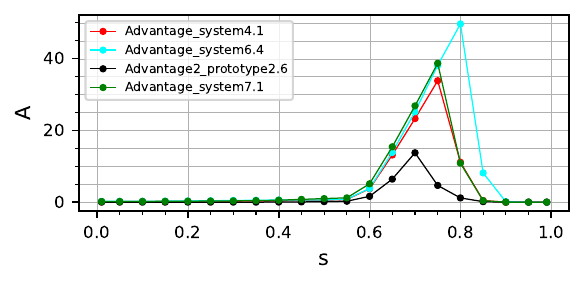}
    \caption{Magnetic hysteresis cycle area (y-axis) versus the annealing parameter $s$ at which the hysteresis is performed, for the 2D square antiferromagnetic lattice with open boundaries, (left) and 1D antiferromagnetic rings, with periodic boundary conditions (right).  }
    \label{fig:hysteresis_area}
\end{figure*}

In our finite 2D samples the reversal appears to proceed through a competition between nucleation of favorably oriented antiferromagnetic droplets (e.g., antiferromagnetic domains, seen as contiguous cyan or black regions in the lower row of Fig.~\ref{fig:2D_spin_configs_Advantage_system6.4}) and propagation and roughening of pre-existing interfaces, associated with antiferromagnetic symmetry breaking, as shown by the staggered boundaries (cyan-black pixels) between domains in the lower row of Fig.~\ref{fig:2D_spin_configs_Advantage_system6.4}. The antiferromagnetic domains, and the boundaries between those domains, can be seen visually by the checkerboard ordering in the spin configurations in the top row of Fig.~\ref{fig:2D_spin_configs_Advantage_system6.4}, but are better visually highlighted by the staggered height function in the bottom row of Fig.~\ref{fig:2D_spin_configs_Advantage_system6.4}. Open boundaries play a decisive role: edges and, in particular, corners reduce the interfacial cost, so the first irreversible events predominantly originate at the boundary. Once a supercritical droplet appears, growth continues mainly by interfacial translation (with occasional coalescence when neighboring droplets meet), producing the step-like changes in magnetization that are characteristic of pinned, driven interfaces. At larger field magnitude the rate of new-droplet formation increases and the loop correspondingly smooths as propagation competes with fresh nucleation. In the raw-spin plots of Fig.~\ref{fig:2D_spin_configs_Advantage_system6.4}-(top), the open-boundary effect appears as extended runs of similarly aligned spins along the outer edges, where spins have fewer antiferromagnetic neighbors than in the bulk. In the height-function plots of Fig.~\ref{fig:2D_spin_configs_Advantage_system6.4}, changes across the sampled field values highlight antiferromagnetic domains (cyan and black pixel regions) and in the shapes of their contours are consistent with propagation and roughening of the corresponding domain walls.

These processes leave clear fingerprints in the spin-structure factor. Near saturation the structure factor weight concentrates around $\mathbf{q}\!=\!0$, reflecting the large polarized regions. As the field is swept toward the demagnetized regime, intensity migrates to antiferromagnetic wavevectors at the corners of the Brillouin zone, consistent with the build-up of checkerboard order driven by local field, inside expanding droplets and along translating interfaces. As the field sweeps, the hysteresis emerges from the polarized domain walls. 

Because of the finite window and open boundaries, we also observe streaks of intensity aligned with the axes of the Brillouin zone at $q_x=q_y=0$, which track the prevalence of straight boundary-pinned segments. The streaks of intensity at $q_x=q_y=0$ can be seen in all four structure factor panels of Fig.~\ref{fig:SSF1}, but most clearly in the lower right panel which shows ordering when the model is entirely saturated. On the return sweep, the same features recur with a field offset set by the pinning landscape, yielding a reproducible hysteretic trajectory in both the net magnetization and the structure factor. 

While both reciprocal space and real space data show formation of large antiferromagnetic domains and their coexistence with net magnetization, a relevant question is whether the data show order that can be properly interpreted as spontaneous antiferromagnetic symmetry breaking. To explore that, we move to the N\'eel order parameter [Eq.~\eqref{eq:Neel}] in Supplementary Information~\ref{section:appendix_2D_Neel_order_param}, which reveals that there is some latent bias in these 2D experiments that should be mitigated in future work using analog control statistics balancing methods.

In summary, these 2-dimensional experiments show interesting non-monotonic magnetization features, the models can be fully saturated given appropriate tuning of $\Gamma/J$ and the longitudinal field is sufficiently strong. Moreover, the extracted static structure factors show ordering that is consistent with an antiferromagnet model undergoing reversal and saturation at high longitudinal field. By extracting individual spin configurations we can visually see antiferromagnetic domains, as well as boundary effects due to the open boundary conditions of the model embedded on the hardware.

\begin{figure*}[ht!]
    \centering
    \includegraphics[width=0.496\linewidth]{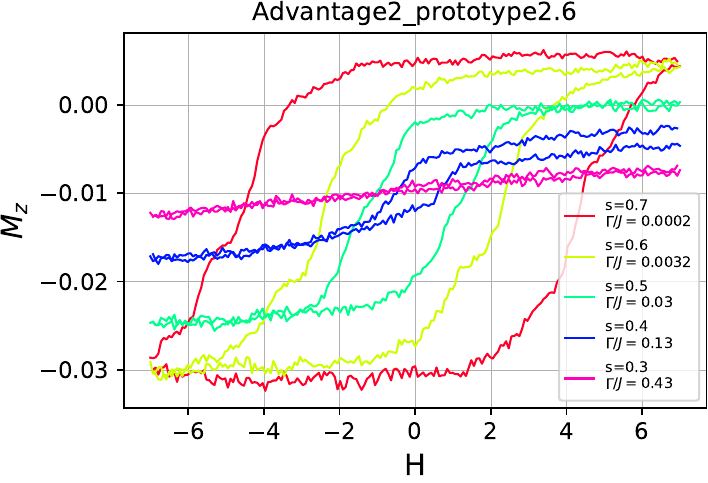}
    \includegraphics[width=0.496\linewidth]{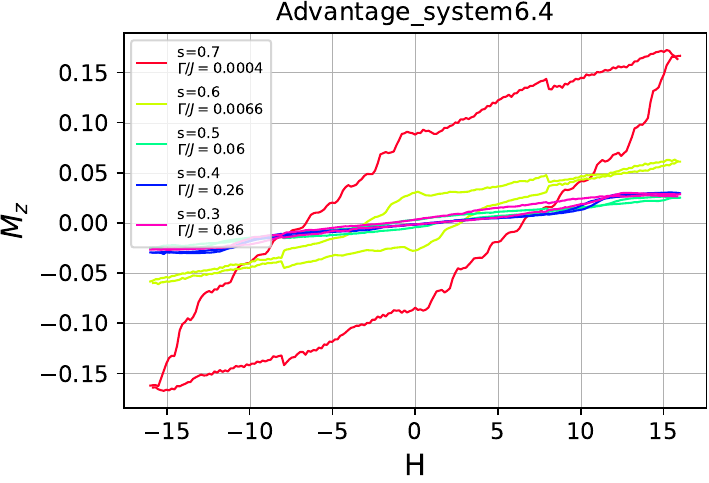}
    \includegraphics[width=0.496\linewidth]{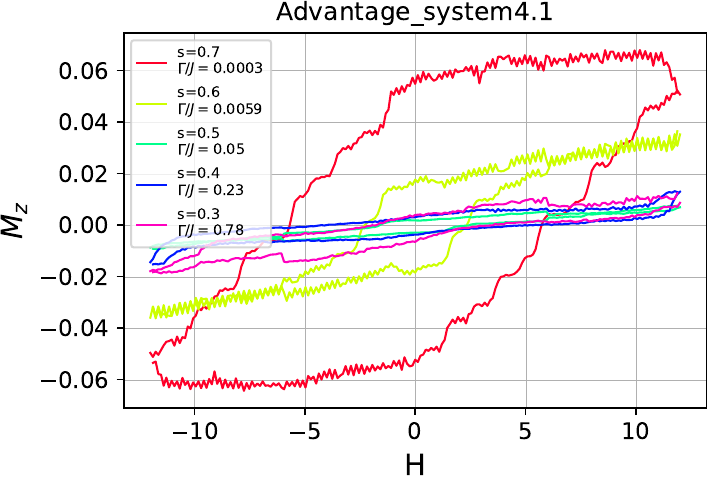}
    \includegraphics[width=0.496\linewidth]{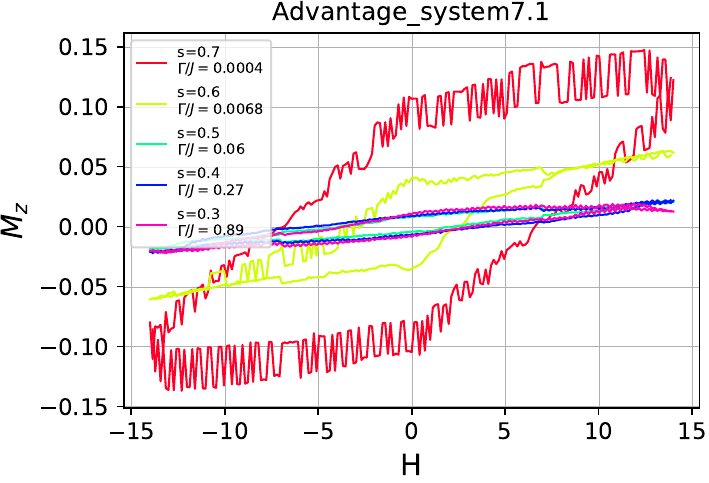}
    \caption{\textbf{Magnetic hysteresis of 3D hardware-defined antiferromagnets, with no calibration refinement}. Average magnetization (y-axis) as a function of the applied longitudinal field (x-axis). Because the absolute range of magnetization (y-axis) is much smaller than the other Ising models, here we increase the shot count (i.e., the total number of samples obtained by a complete anneal-readout cycle) to $8000$ for each data point. This increase in the number of samples reduces local jitter of the magnetization curves. No flux bias offset calibration was used for these simulations (e.g., before calibration).   }
    \label{fig:hardware_3D_AFM_hysteresis}
\end{figure*}

\begin{figure*}[ht!]
    \centering
    \includegraphics[width=0.496\linewidth]{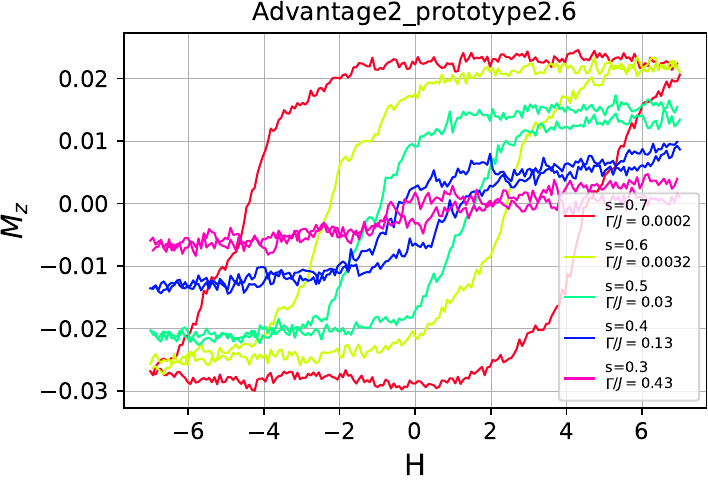}
    \includegraphics[width=0.496\linewidth]{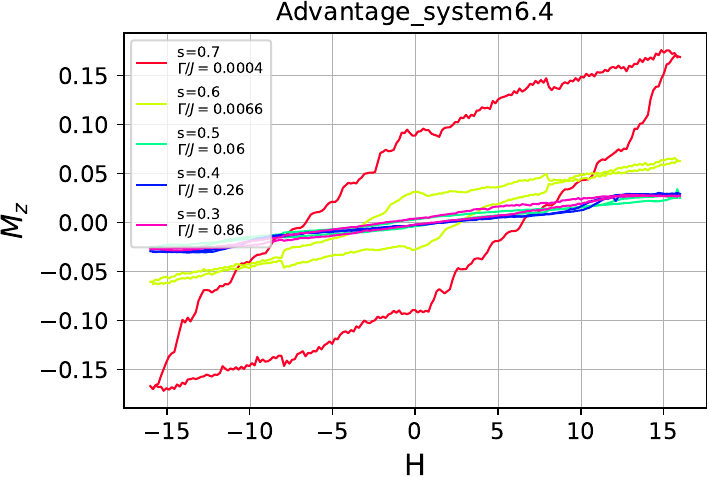}
    \includegraphics[width=0.496\linewidth]{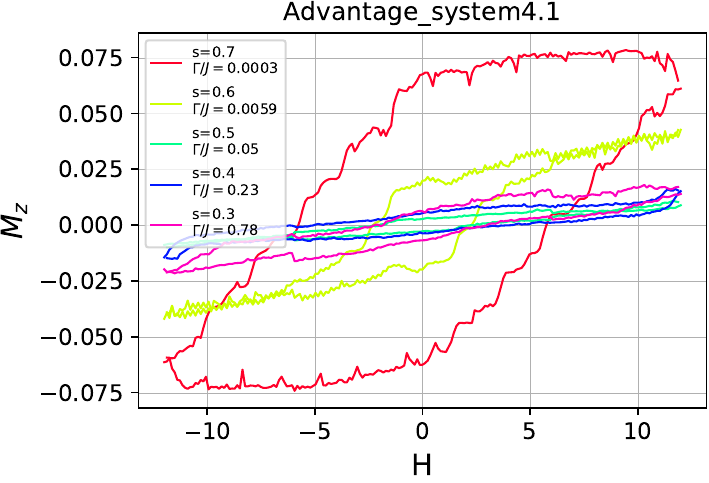}
    \includegraphics[width=0.496\linewidth]{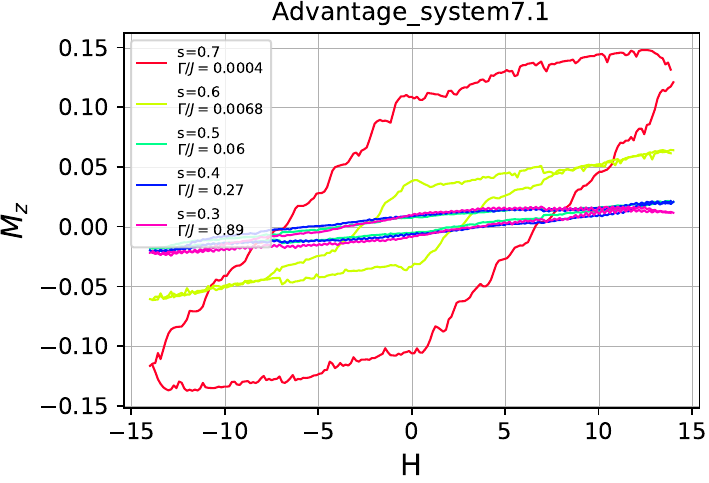}
    \caption{\textbf{Magnetic hysteresis on hardware-defined antiferromagnets, with extensive whole-lattice FBO calibration refinement}. Average magnetization (y-axis) as a function of the applied longitudinal field (x-axis). Each data point on the hysteresis curves are averaged single site magnetization from $8000$ samples. In comparison to Fig.~\ref{fig:hardware_3D_AFM_hysteresis} where no statistical balancing calibration was used, here we see in general smoother magnetization curves, in particular on the \texttt{Advantage\_system7.1} processor.  }
    \label{fig:FBO_calibrated_hardware_3D_AFM_hysteresis}
\end{figure*}

\subsection{Hysteresis Curve Areas From 1D and 2D Antiferromagnetic Hysteresis Cycles}
\label{section:results_hysteresis_areas}

Fig.~\ref{fig:hysteresis_area} plots the numerically integrated enclosed area of the hysteresis cycles as a function of $s$ (equivalently, $\Gamma/J$), from the 1D and 2D antiferromagnet simulations. This shows that the region of the annealing parameter ($s$, equivalently $\Gamma/J$) where the hysteresis loop area is non-zero is smaller compared to the ferromagnetic models or disordered frustrated models studied in Ref.~\cite{pelofske2025magnetichysteresisexperimentsperformed}. 

Across 1D rings and 2D grids the hysteresis loop area, as a function of $s$, is non-monotonic in $s$: it turns on from near zero at small $s$, peaks in an intermediate window, and decreases again at larger $s$. This reflects a competition between pinning-limited reversal at low $\Gamma/J$ and strong transverse-field mixing  at high $\Gamma/J$, with device-to-device amplitude set primarily by the usable longitudinal field range, seen in Table~\ref{table:hardware_summary}. The reduced longitudinal field produces minor loops and systematically smaller areas at the same $s$. In 2D, areas are further suppressed relative to 1D at matched $s$, consistent with reduced magnetic saturation in the 2D case.

Statistical uncertainty quantification, measured by magnetization standard deviation, is reported in Supplementary Information~\ref{section:appendix_standard_deviation_plots}.

\subsection{Quasi 3-Dimensional Antiferromagnet Experiments}\label{section:results_3D}

Fig.~\ref{fig:hardware_3D_AFM_hysteresis} and \ref{fig:FBO_calibrated_hardware_3D_AFM_hysteresis} show magnetic hysteresis simulations performed on 3D antiferromagnets -- specifically these use the entire hardware graph (every coupler being set to an antiferromagnetic coupling).

Fig.~\ref{fig:hardware_3D_AFM_hysteresis} shows hysteresis curve data from Ising models that are \emph{entirely antiferromagnetic}. These simulations are notable because while full magnetization is not reached on any of the antiferromagnetic lattices, the non-zero transverse field does allow state transitions to take place and we recover a minor-loop hysteresis curve that resembles classical ferromagnetic hysteresis. Moreover, like in the prior antiferromagnetic models we observe Barkhausen-like steps in the magnetization cycles (Fig.~\ref{fig:hardware_3D_AFM_hysteresis} and \ref{fig:FBO_calibrated_hardware_3D_AFM_hysteresis}), however, the apparent strength of these signals is weak because of the high shot noise and low-magnetization values. Being able to probe antiferromagnetic hysteresis using this technique is a notable computational capability given the relatively high coordination number of these graphs.

The general trend that we observe here is that as the coordination number of the antiferromagnetic Ising models increases, it becomes harder to fully magnetize the systems in the analog QPU simulation. This is to be expected, because a higher coordination number means more energy is required, in this case a stronger longitudinal field, in order to flip the spins against the strong $J$ lattice coupling. 

In summary, these quasi-3D antiferromagnet hysteresis experiments show that sufficiently high coordination number substantially suppresses the hysteresis effects. Despite this, some hysteresis can still be observed. Moreover, after applying observable bias reduction techniques the hysteresis effects remained effectively unchanged.

\subsubsection{Suppressed hysteresis in 3D Hardware Lattice}

Our 3D results exhibit only small, often near–minor loops. In highly connected antiferromagnetic graphs due to the large interfacial cost set by the coordination number and $J$, there is limited mobility when the transverse field $\Gamma$ is either too small (frozen) or too large (paramagnetic).

On an average $c$-coordinated AF lattice/graph, any domain surface cuts a number of bonds proportional to $c$ per unit area. Each cut bond contributes an energy scale $\sim J$, so the effective surface tension obeys
\begin{equation}
\lambda \propto cJ, \nonumber
\end{equation}
up to geometric factors set by the lattice or embedding. In three dimensions, the excess energy of a droplet, assumed nearly spherical, of reversed N\'eel order with radius $R$ is well approximated by a surface term that scales as
\begin{equation}
E_{\text{surf}}(R) \simeq 4\pi \lambda R^2. \nonumber
\end{equation}
Higher coordination therefore stiffens domain surfaces and penalizes both surface translation and the creation of new droplets.

A longitudinal field $h$ favors spins aligned with its sign. Flipping a spin gains Zeeman energy $2|h|$ (in our $\sigma^z=\pm1$ normalization), so a droplet containing a volume $\tfrac{4}{3}\pi R^3$ gains
\begin{equation}
E_{\text{field}}(R) \simeq - \tfrac{4}{3}\pi R^3  (2|h| m_0),
\end{equation}
where $m_0$ is the spin magnitude per site (for perfectly staggered order $m_0 = 1$). Setting the derivative of the sum of the surface and field terms to zero, $d(E_{\mathrm{surf}}+E_{\mathrm{field}})/dR=0$, yields the critical radius
\begin{equation}
R_c = \frac{2\lambda}{2|h| m_0} = \frac{\lambda}{|h| m_0},
\end{equation}
and an activation barrier
\begin{align}
E_c = E_{\text{surf}}(R_c)+E_{\text{field}}(R_c)
 = \frac{4\pi}{3}\frac{\lambda^3}{|h|^2 m_0^2}.
\end{align}
Thus in 3D the surface-area cost increases both $R_c$ and $E_c$ sharply, in addition the high hardware-coordination $c$ raises $\lambda$ and therefore pushes $R_c$ and $E_c$ up, suppressing bulk nucleation across the sweep range we employ.

On finite samples, droplets can nucleate as caps on faces (roughly halving both surface and volume terms) or at edges/corners (reducing cost further). This boundary-assisted nucleation produces small droplets and corresponding small hysteresis. Most field steps only rattle pinned facets without producing large, hysteretic changes in $M_z$.

The interfacial picture also clarifies how the control knobs set the loop morphology. First, the sweep amplitude (envelope) determines whether droplets can ever become supercritical: if the field never reaches values for which the critical radius $R_c$ is microscopic, nucleation remains subcritical and the response collapses to minor, low-area cycles. Second, the sweep rate and dwell time control the opportunity for rare, multi-spin processes that assist nucleation and for marginal droplets to grow: faster ramps and shorter dwells reduce this opportunity and thus diminish loop area. Third, the transverse field fixes the matrix element for local surface moves and pair creation; for $\Gamma/J\ll 1$ surfaces are effectively frozen, while for very large $\Gamma/J$, strong quantum mixing erases memory of the sweep path so that hysteresis collapses. A pronounced, reproducible loop therefore occurs only in an intermediate window of $\Gamma/J$ that is large enough to enable surface translation across pinning sites but not so large as to randomize domains during the protocol.

\section{Discussion and Conclusion}\label{section:conclusion}
We have demonstrated for the first time large-scale simulations of antiferromagnetic hysteresis using quantum computers. This includes observation of unexpected non-monotonic magnetization curves as has been seen before in these types of analog QPU simulations on 1D ferromagnets~\cite{barrows2025magneticmemoryhysteresisquantum}, along with extraction of the spin structure factor, which shows clear transitions of magnetic ordering during the hysteresis cycles. The structure factor plots show a clear antiferromagnetic ordering transition from ferromagnetic like ordering when the systems are fully polarized, as well as boundary effects. We have also demonstrated that the use of statistics-balancing calibration on the analog quantum hardware can improve the quality of the hysteresis simulations, as demonstrated on full-hardware-graph antiferromagnets. One of the primary results that we demonstrate is that for low-coordination antiferromagnets (1D and 2D Ising models in particular), under an appropriately tuned ratio of transverse field ($\Gamma$) to lattice coupling ($J$), we can achieve full polarization of the antiferromagnets. Furthermore, the subsequent reversal resembles many of the key features of standard classical hysteresis in disordered, and even ferromagnetic, spin systems.

Our measurements establish robust dynamical hysteresis in programmable antiferromagnets across geometries. In 1D odd rings, a built-in domain wall yields reproducible, step-like reversal under a transverse field. In 2D finite samples, boundary-assisted switching and droplet growth are consistent with the observed evolution of the spin-structure factor. The highly coordinated hardware graphs suppress the ferromagnetic-like ordering, producing weak, near-minor loops. Overall, loop morphology is set by a small set of tunable parameters—geometry and coordination, the sweep protocol, and the ratio $\Gamma/J$—highlighting quantum annealers as controllable platforms for exploring nonequilibrium antiferromagnetic dynamics. The phenomena we quantify are reproduced on four different D-Wave quantum processors. There are intrinsic noise differences and timescale differences on the different analog quantum hardware, as well as different Hamiltonian energy scales. Despite these differences, the hysteretic characteristics we identify are robust across the different quantum processors.

Interestingly, some of the hysteresis experiments we present in this study share some qualities with measured magnetic hysteresis cycles in certain molecular magnets~\cite{JIANG20102227}. Similarities between these analog open-quantum system hysteresis experiments and molecular magnetic hysteresis deserves further investigation. 

Future work can explore the dynamical origin of this hysteresis, its potential relationship with the qubit quantum dynamics, and also establish a closer connection with real magnetic materials and compare results from quantum annealers to those from laboratory experiments. Future studies should also investigate additional calibrated antiferromagnetic hysteresis cycle simulations (in particular of the couplers and of the flux bias offsets), in particular on the 2-dimensional systems where the antiferromagnetic order parameter data shows much larger statistical fluctuations.

This computational capability of near term analog quantum computers highlights that the full programmability of these devices inherently gives us the capability to examine phenomena which are hard to systematically study in magnetic laboratories -- in this case, antiferromagnetic hysteresis.

\section*{Acknowledgments}\label{sec:acknowledgments}
We thank Vivien Zapf for productive discussions. This work was supported by the U.S. Department of Energy through the Los Alamos National Laboratory. Los Alamos National Laboratory is operated by Triad National Security, LLC, for the National Nuclear Security Administration of U.S. Department of Energy (Contract No. 89233218CNA000001). The research presented in this article was supported by the Laboratory Directed Research and Development program of Los Alamos National Laboratory under project number 20240032DR. This research used resources provided by the Los Alamos National Laboratory Institutional Computing Program. The authors would also like to thank the New Mexico Consortium, under subcontract C2778, the Quantum Cloud Access Project (QCAP) for providing quantum computing resources. LANL report LA-UR-25-28275.

\appendix

\begin{figure}[ht!]
    \centering
    \includegraphics[width=0.999\linewidth]{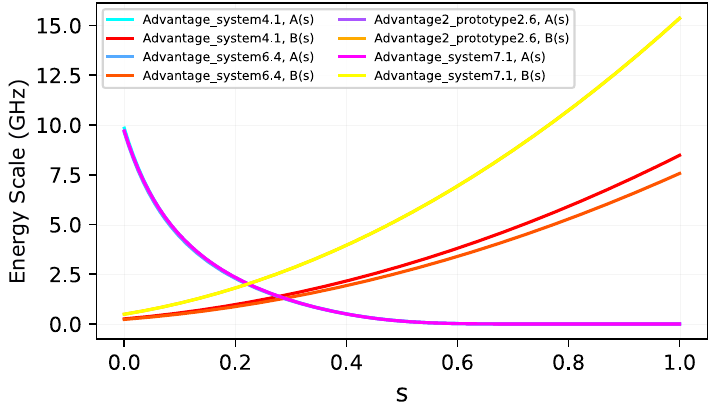}
    \caption{Overlaid $A(s)$ and $B(s)$ hardware specific functions for the $4$ D-Wave QPUs used in this study. Note that the $A(s)$ functions on all four processors are effectively identical, which makes the lines visually indistinguishable. }
    \label{fig:hardware_A_B_functions}
\end{figure}

\begin{figure*}[ht!]
    \centering
    \includegraphics[width=0.496\linewidth]{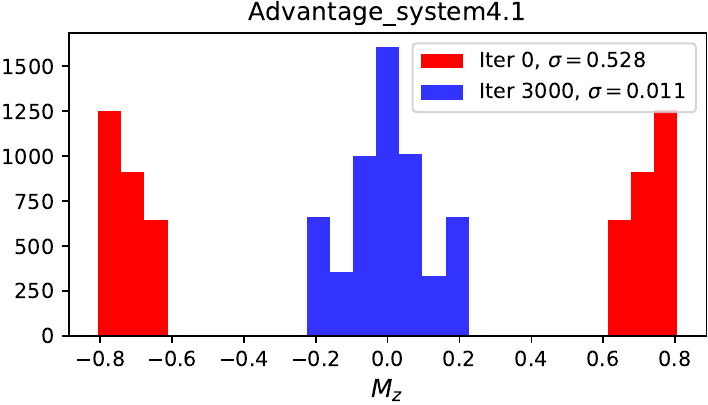}
    \includegraphics[width=0.496\linewidth]{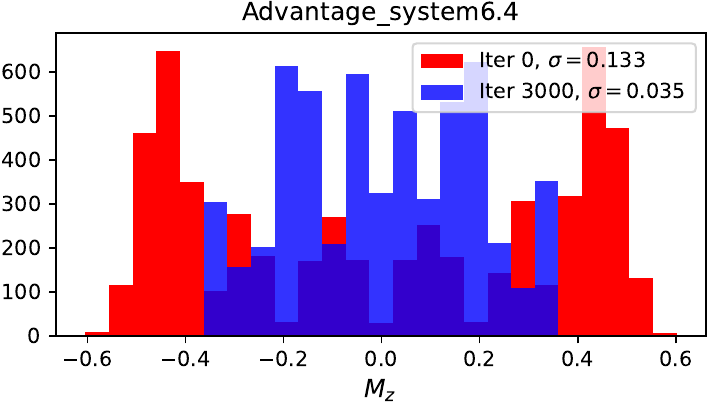}
    \includegraphics[width=0.496\linewidth]{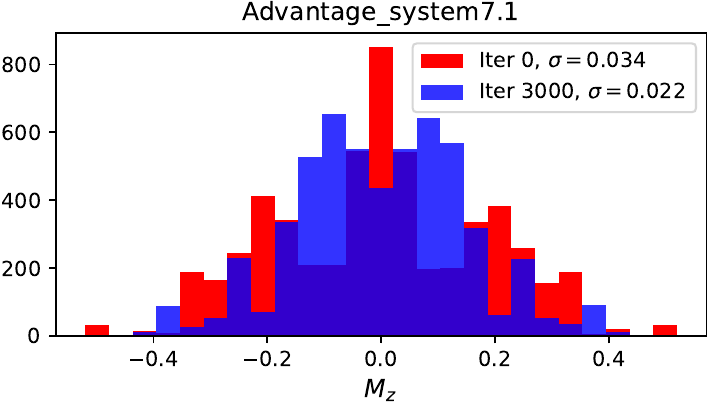}
    \includegraphics[width=0.496\linewidth]{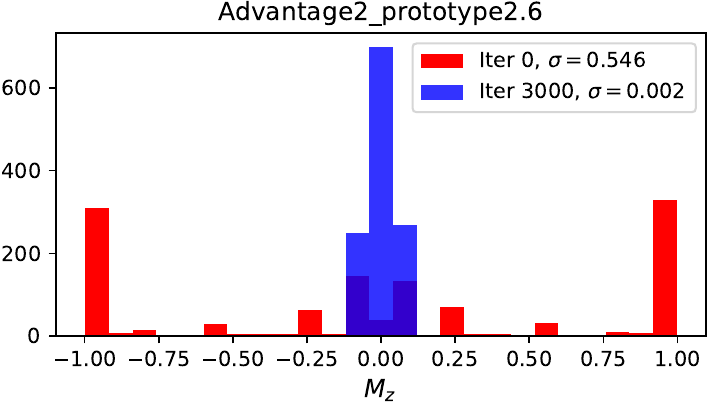}
    \caption{Single site magnetization distributions, for each qubit in the hardware lattice, before flux bias offset calibration and after calibration for a full antiferromagnetic lattice defined on the entire hardware graph. $3000$ gradient descent steps were used in total, and the flux bias offset stepsize is $2e^{-6}$ (with initial flux bias offset being uniformly zero). The standard deviation of the distribution of average single-site magnetization before (iteration 0) and after (iteration 3000) the FBO calibration is noted in the legend -- in all cases we see a decrease in the variance of the single-site magnetization distribution, which is the goal of the statistical balancing process. Each iteration of this calibration process, on each QPU, uses exactly $3000$ samples. The flux bias offset quantity calibration is shown in Fig.~\ref{fig:FBO_updates_shimming_learning}.  }
    \label{fig:magnetization_FBO_calibration}
\end{figure*}

\begin{figure*}[ht!]
    \centering
    \includegraphics[width=0.496\linewidth]{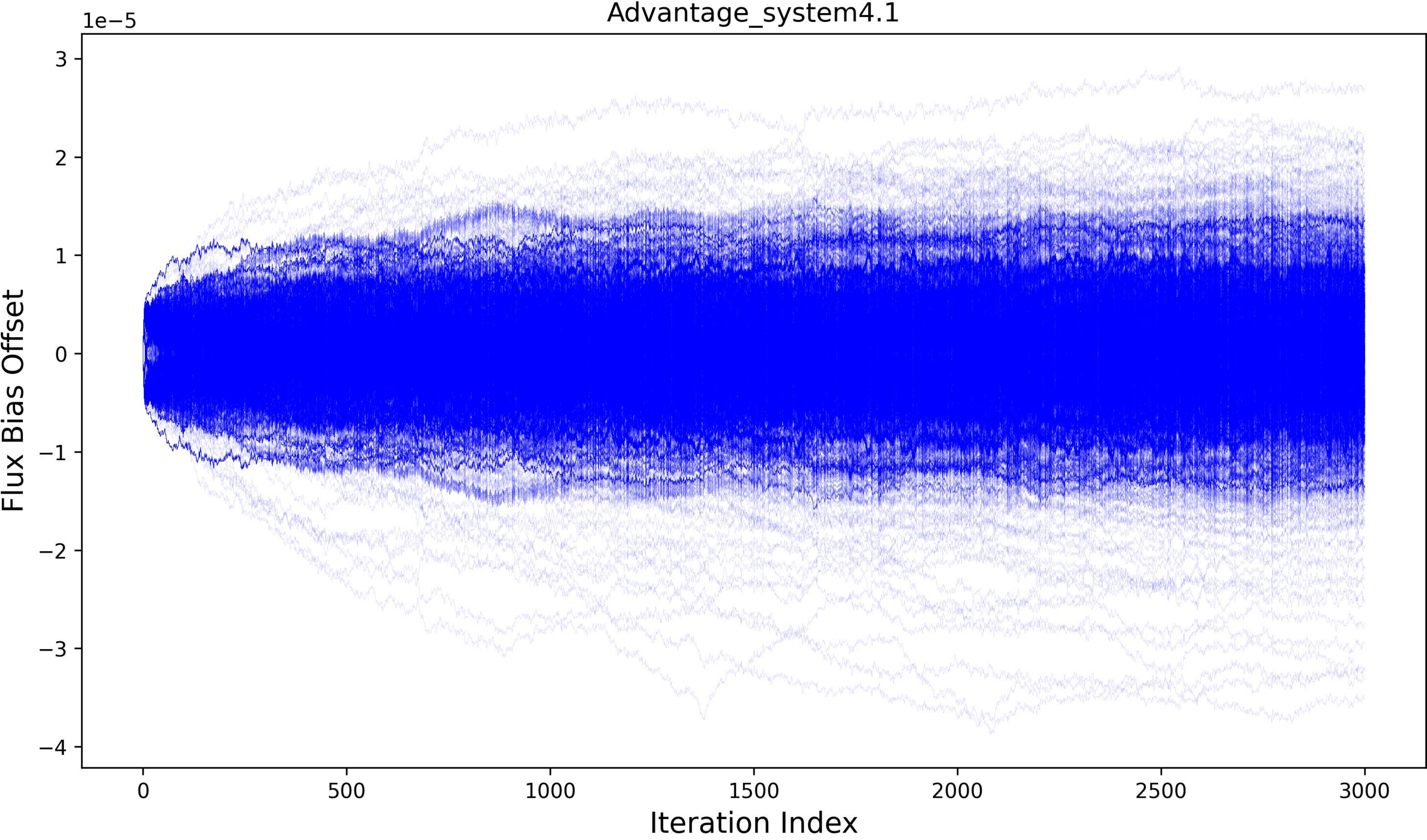}
    \includegraphics[width=0.496\linewidth]{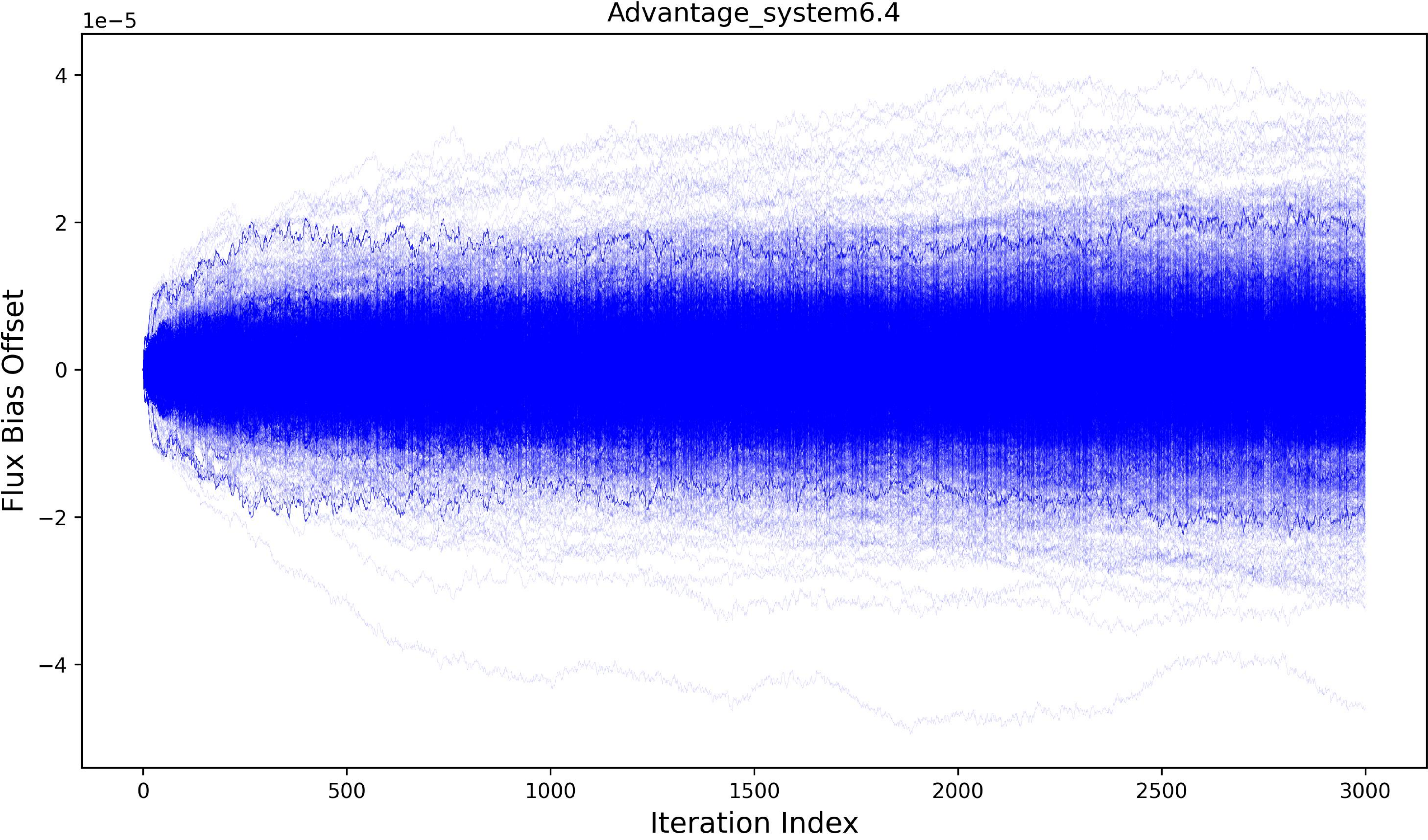}
    \includegraphics[width=0.496\linewidth]{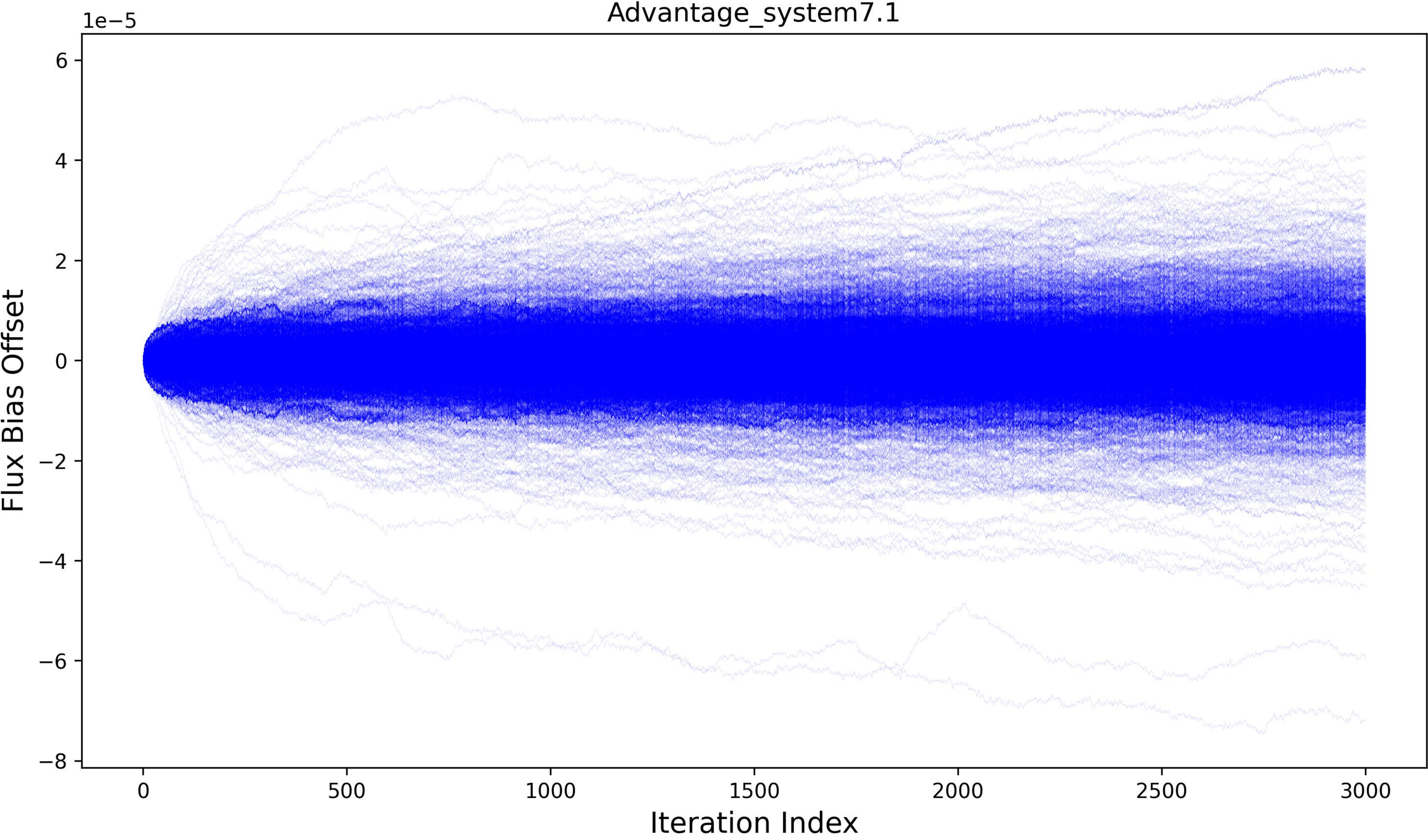}
    \includegraphics[width=0.496\linewidth]{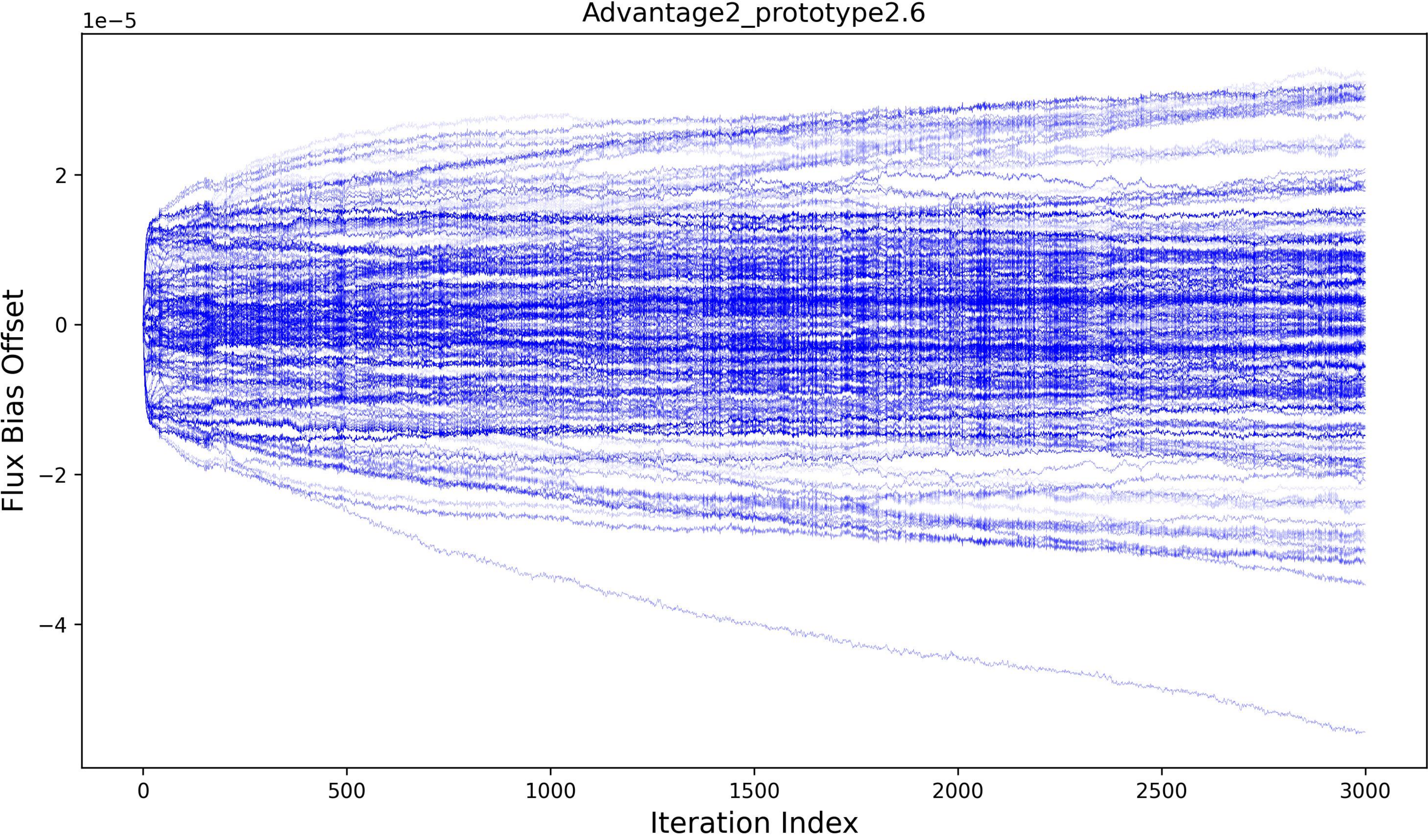}
    \caption{Evolution of FBO values, for each single qubit in the hardware-lattice, as the calibration process evolves up to $3000$ iterations. The Ising model being sampled during this calibration is a full antiferromagnet defined lattice, using every coupler of each of the D-Wave hardware graphs. The initial flux bias offset starts at all zeros for all qubits, and each iteration has a maximum FBO stepsize of $2e^{-6}$.  }
    \label{fig:FBO_updates_shimming_learning}
\end{figure*}

\section{D-Wave Hardware $A(s)$ and $B(s)$ Functions}
\label{section:A_s_B_s_functions}

Fig.~\ref{fig:hardware_A_B_functions} plots the $A(s)$ and $B(s)$ schedules that define the D-Wave hardware Hamiltonian in eq.~\eqref{equation:QA_Hamiltonian_h_gain}, which allows us to then define physical ratios such as $\Gamma/J$. This is important specifically because all of the programmable units on the D-Wave QPUs are specified in terms of hardware-normalized parameters, and therefore it is necessary to extract the correct energy scales of the D-Wave anneal-schedules to show where certain features of the magnetic hysteresis protocols appear in physical units, for example, at what applied field does magnetization reversal begin. Note that these frequencies are not in angular frequency units.

\section{Flux Bias Offset Balancing Calibration}
\label{section:flux_bias_offset_calibration}

Due to the analog nature of the D-Wave quantum annealing hardware, it can be beneficial to calibrate the statistics of physical quantities that can be measured from the annealing simulations. In particular, analog errors and thermal effects can result in biased quantities. In this study we make use of a particular type of calibration procedure which balances the statistics obtained from the full-hardware-lattice-antiferromagnets. The update procedure used follows ref.~\cite{Chern_2023}, and in particular is effectively a type of parallel-update steepest gradient descent machine learning process, and is very much an \emph{autonomous} analog control correction procedure. This is reminiscent of gradient descent approaches in the training of neural networks~\cite{lecun2002gradient, rumelhart1986learning}. The hardware control mechanism that is used here is the flux bias offset (FBO) control of each individual qubit~\cite{PhysRevB.80.052506} which can force the qubit to be biased towards spin up or down, and in this case we very slightly update the FBO proportionally to the size of the bias of the single-qubit magnetization (relative to the mean of the distribution across all qubits). The statistics balancing method in particular makes use of a correction towards the mean sampled observable ($M_z$) across all spins; in this case, we expect the net average magnetization to be zero, and the non-calibrated distributions also have a mean of approximately zero, which makes this correction method work quite well in this case because its primary action is to reduce the variance of the measured single-spin magnetization on the lattice. The magnetization FBO update is relatively easy to perform because all of the qubits are in the same observable ``orbit''~\cite{Chern_2023}, meaning we expect on average the magnetization of each qubit to be the same as all of the other qubits. Note, however, that this balancing did not adjust the $J$ antiferromagnetic couplers; refining the programmed coupler energy values is possible to adjust as well, however boundaries in the lattice mean that different observable orbits must be constructed -- which for the specific case of the highly complex hardware graphs would make the calibration significantly more computationally intensive in terms of total iterations and also parameter tuning. Therefore, we focused on FBO calibration, which yielded improved magnetization curve stability. This calibration procedure can in principle be applied to any programmed Ising model on analog hardware, but it has been especially useful for exploring sensitive phenomena, such as quantum quench dynamics and fluctuations near thermodynamic phase transitions. The FBO update rules applied to each qubit are uniform across the entire lattice, i.e., the bulk and edge qubits are treated identically, unlike in some prior studies~\cite{Kairys_2020}.

Fig.~\ref{fig:magnetization_FBO_calibration} shows the distribution of average magnetization per spin both before and after FBO statistics-balancing calibration. Fig.~\ref{fig:FBO_updates_shimming_learning} plots the values of the flux bias offsets, for every qubit in the hardware graph, as a function of update iteration, which in this case we terminate at a fixed iteration count of $3000$ in order to make the simulations tractable in terms of compute time, and real-time backend usage. The flux bias offset constant step-size used was $2e^{-6}$ (the actual update is proportional to the magnitude of the bias, and this is the constant by which the bias is multiplied by). This specific stepsize we found worked reasonably well from small scale empirical testing; larger stepsizes could in principle correct biases more quickly, but over-correction and thereby numerical instability in the convergence was frequently observed for step sizes larger than $\approx 2e^{-6}$. The calibrated FBOs used in the subsequent hysteresis computations are simply the FBOs in the final learning iteration. Time drift is an important consideration in these D-Wave QPU computations (over long periods of time, D-Wave simulations do change measurably~\cite{Pelofske_2023_noise}), and therefore the complete set of hysteresis cycle data using the calibrated FBO values (shown in Fig.~\ref{fig:FBO_calibrated_hardware_3D_AFM_hysteresis}) were run within $48$ hours of this calibration, with the aim of minimizing the effect of noise drift. At each update step, a total of $3000$ anneal-readout cycles were performed (all in one device job call), and the subsequent update is performed using the (average) magnetization values for every site on the hardware lattice using all $3000$ samples. Importantly, the anneal-schedule used in this calibration procedure \emph{was not} the hysteresis protocol -- there was no anneal-pause, and no h-gain field. Instead, this calibration was performed only on the standard linear-ramp anneal schedule, using the final simulation time in the hysteresis protocol (in this case, always $11.2 \mu$s). This means the same FBO calibration (after the $3000$ learning iterations), for each QPU, was used for each $\Gamma/J$ and each ``slice'' of the longitudinal field ramps. More advanced calibration, which is also more QPU time intensive, could specifically calibrate each $\Gamma/J$, and each longitudinal field sweep step. 
Interestingly, Fig.~\ref{fig:magnetization_FBO_calibration} shows that some of the D-Wave processors are much easier to calibrate, presumably meaning they are less noisy, such as \texttt{Advantage\_system4.1}, whereas \texttt{Advantage\_system7.1} appears to be the most noisy and also exhibited the least de-biasing when this calibration protocol was applied as evidenced by having the widest magnetization distribution variance post-calibration. This is also indicated by the flux bias offset quantities as a function of iteration in Fig.~\ref{fig:FBO_updates_shimming_learning} -- we see that specifically in the case of \texttt{Advantage\_system7.1} and \texttt{Advantage2\_prototype2.6}, some of the qubits did not converge in their FBO values, whereas \texttt{Advantage\_system4.1} and \texttt{Advantage\_system6.4} appear to have converged more quickly.

\begin{figure*}[ht!]
    \centering
    \includegraphics[width=0.999\linewidth]{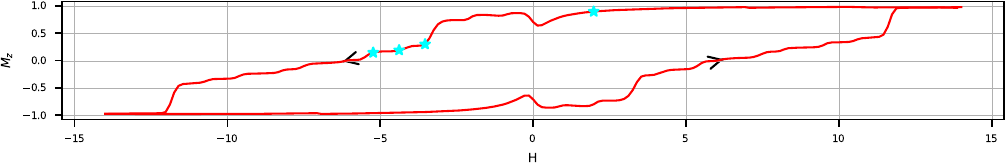}
    \includegraphics[width=0.245\linewidth]{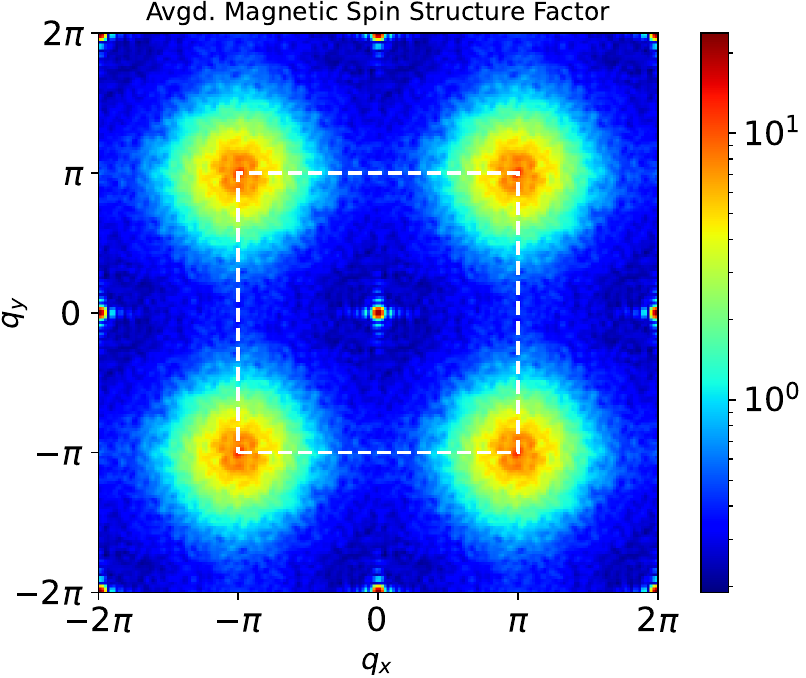}
    \includegraphics[width=0.245\linewidth]{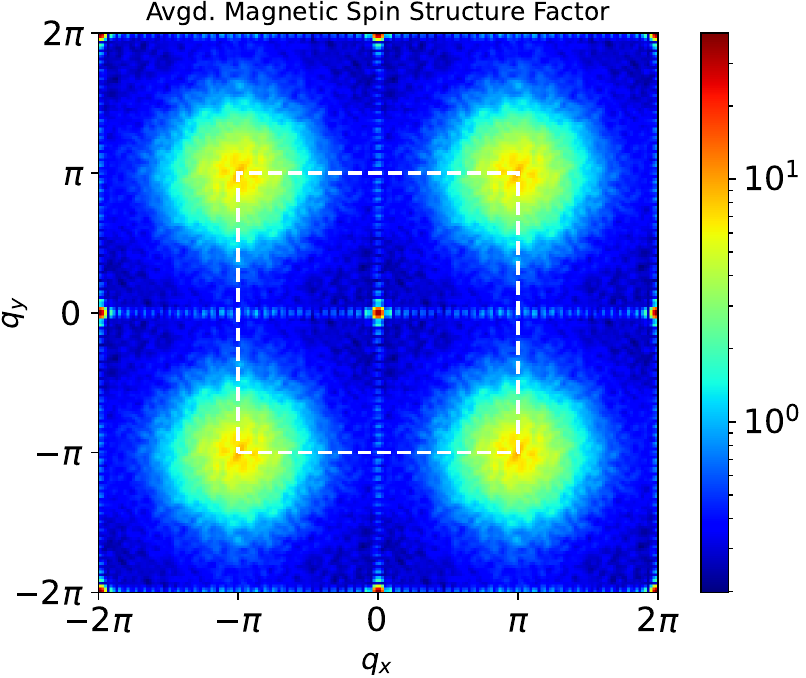}
    \includegraphics[width=0.245\linewidth]{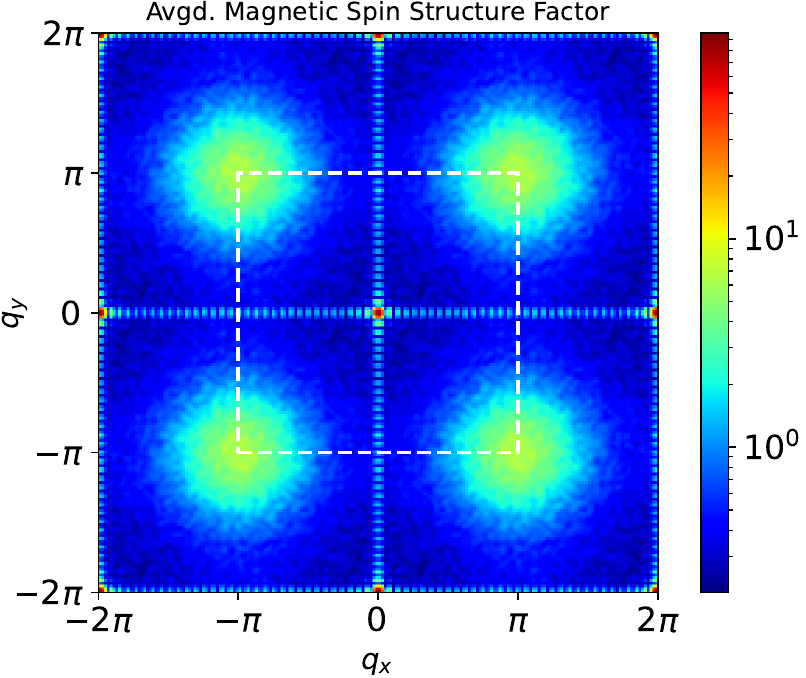}
    \includegraphics[width=0.245\linewidth]{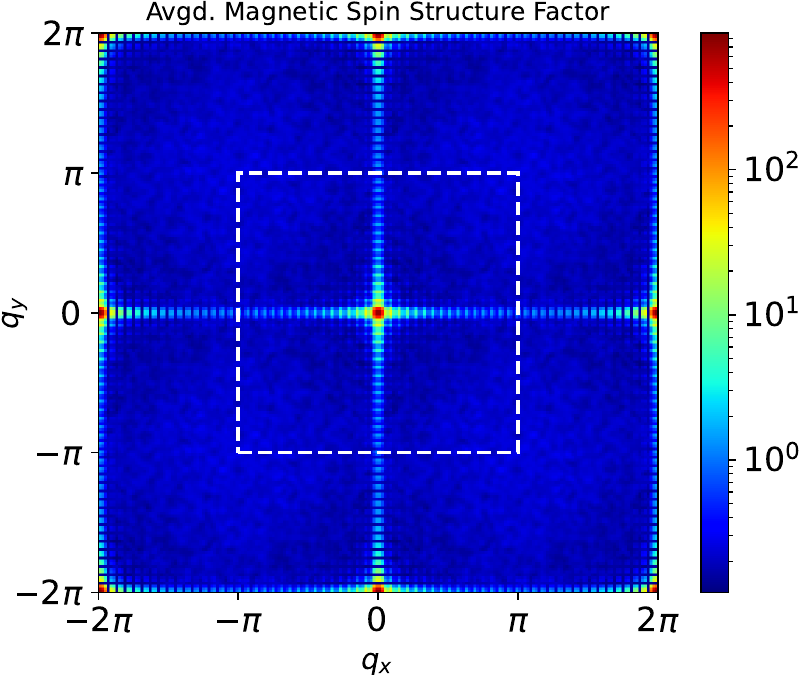}
    \caption{Magnetic structure factors $\abs{S(q)}$ heatmaps (bottom row) within the 2D antiferromagnetic hysteresis cycles (averaged over 100 independent samples), run on \texttt{Advantage\_system7.1} at $s=0.7$. The Ising model being simulated is a $33\times 33$ grid of (antiferromagnetically coupled) spins, with open boundary conditions. The top row shows the single hysteresis cycle in terms of $M_z$ average magnetization, from which we extracted $\abs{S(q)}$ heatmaps for 4 points along the first longitudinal field sweep, denoted by cyan asterisks (the order of the cyan asterisks have the same order as the $\abs{S(q)}$ heatmaps). The dashed white line on the structure factor heat maps outlines the first Brillouin zone. The heat maps are log-scaled and scales are specific to each heatmap subplot. 
    The order of the $\abs{S(q)}$ heatmaps follows the same sweep direction denoted by the top hysteresis cycle plot (the SSF from the fully saturated regime is shown in the right hand subplot). Note the peaks at the corner of the Brillouin zone denoting large antiferromagnetic clusters coexist with a growing peak at the center denoting magnetization. The format of all other magnetic structure plots follow this same figure layout.  }
    \label{fig:SSF2}
\end{figure*}

\begin{figure*}[ht!]
    \centering
    \includegraphics[width=0.999\linewidth]{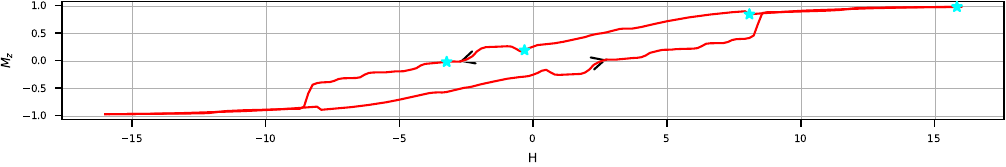}
    \includegraphics[width=0.245\linewidth]{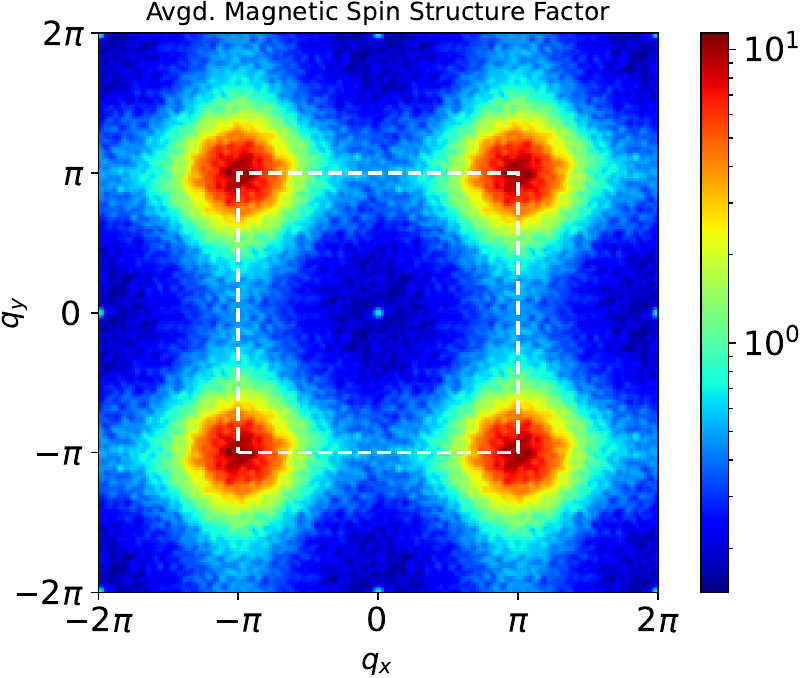}
    \includegraphics[width=0.245\linewidth]{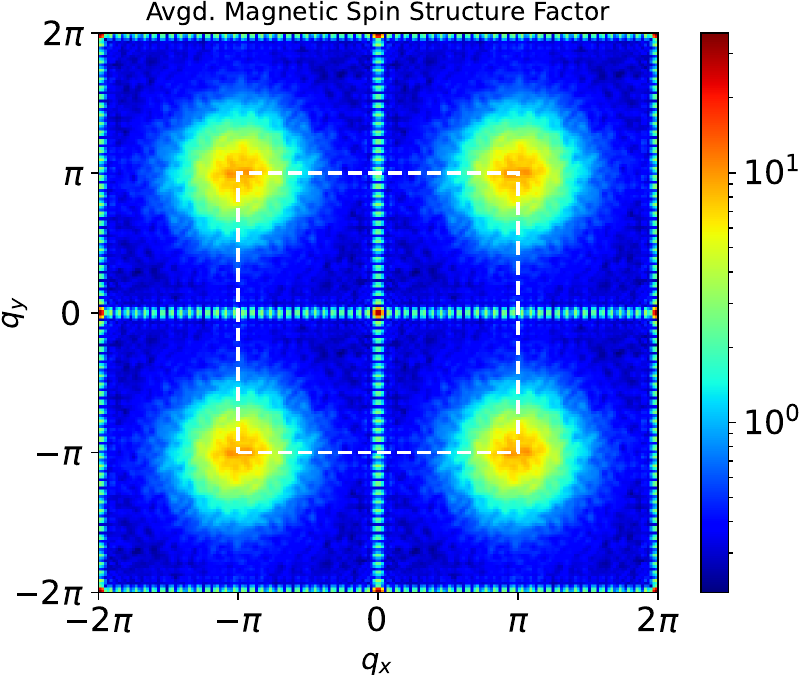}
    \includegraphics[width=0.245\linewidth]{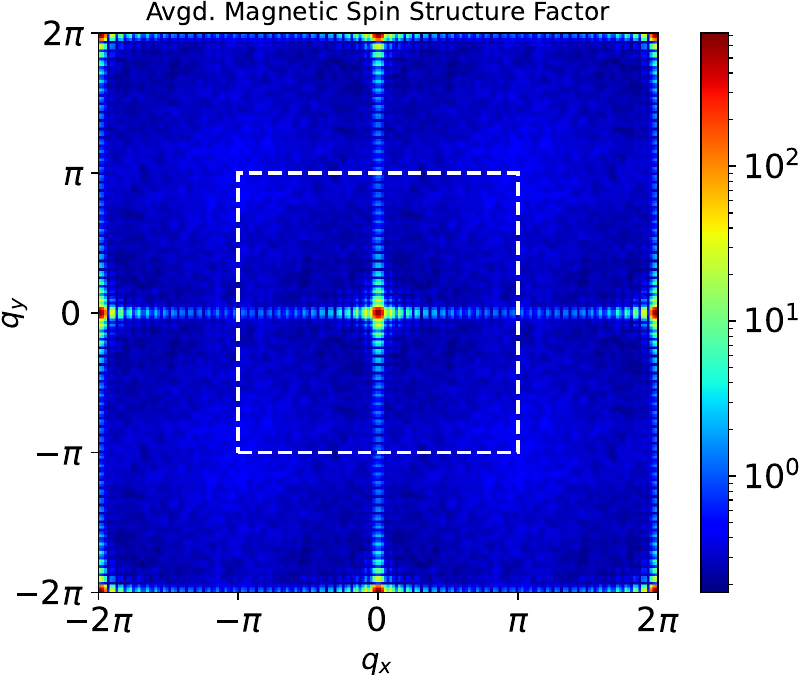}
    \includegraphics[width=0.245\linewidth]{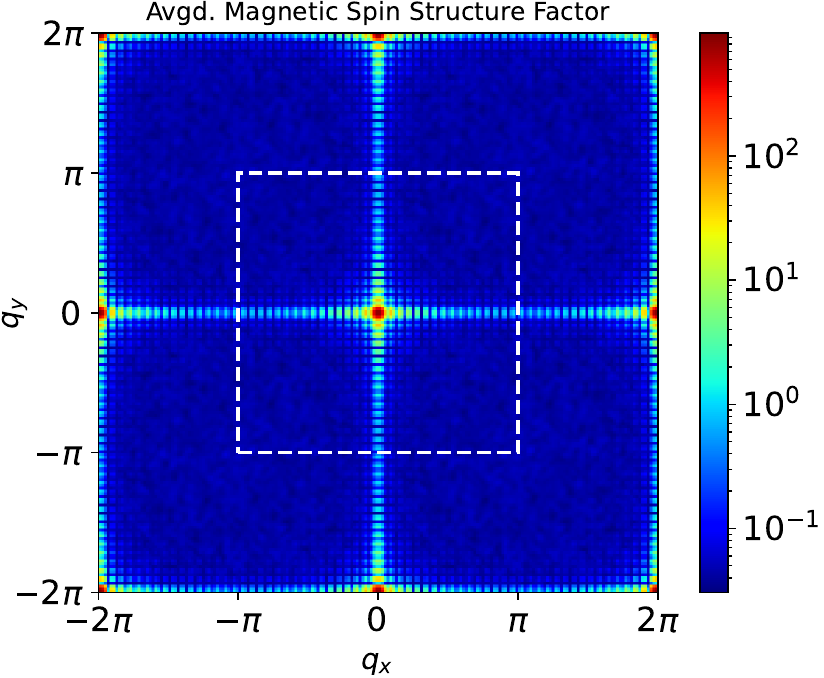}
    \caption{Magnetic structure factors $\abs{S(q)}$ heatmaps (bottom row) within the 2D antiferromagnetic hysteresis cycles, run on \texttt{Advantage\_system6.4} at $s=0.65$. }
    \label{fig:SSF4}
\end{figure*}

\begin{figure*}[ht!]
    \centering
    \includegraphics[width=0.999\linewidth]{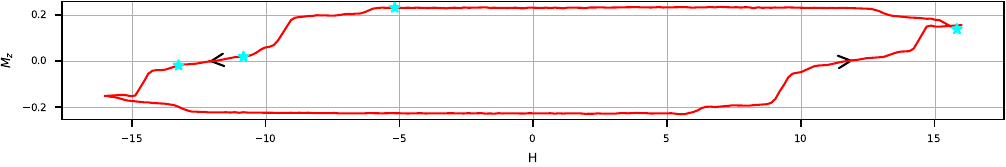}
    \includegraphics[width=0.245\linewidth]{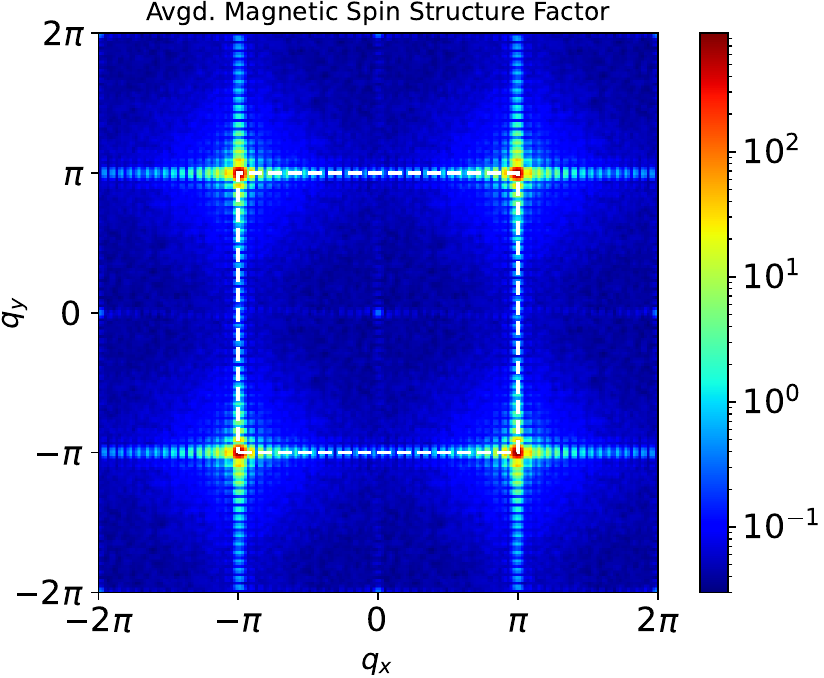}
    \includegraphics[width=0.245\linewidth]{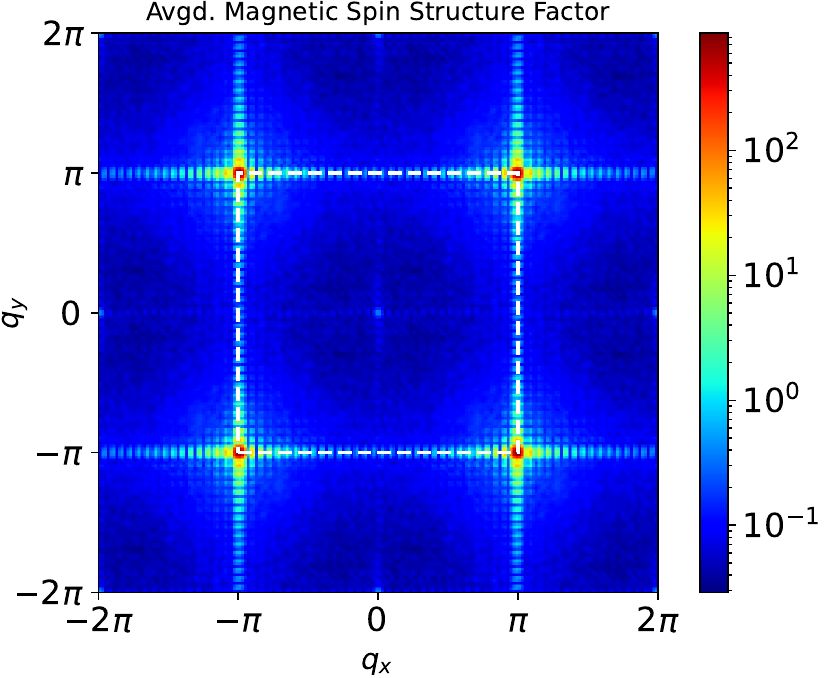}
    \includegraphics[width=0.245\linewidth]{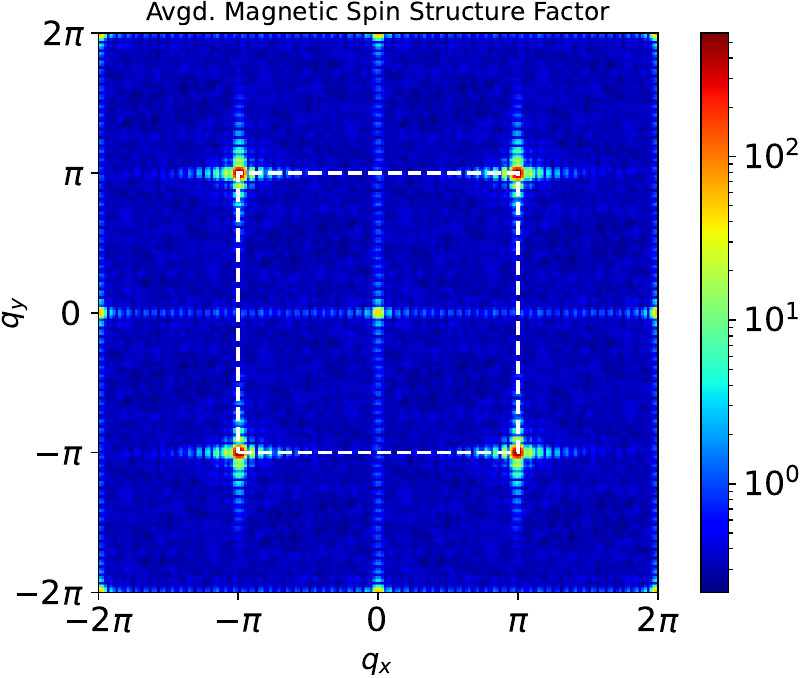}
    \includegraphics[width=0.245\linewidth]{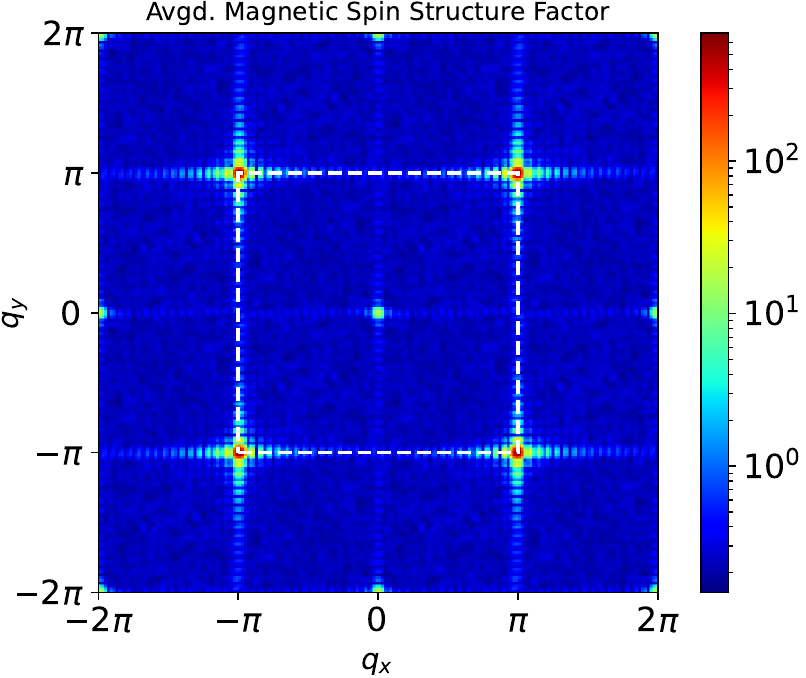}
    \caption{Magnetic structure factors $\abs{S(q)}$ heatmaps (bottom row) within the 2D antiferromagnetic hysteresis cycles, run on \texttt{Advantage\_system6.4} at $s=0.8$. }
    \label{fig:SSF5}
\end{figure*}

\begin{figure*}[ht!]
    \centering
    \includegraphics[width=0.999\linewidth]{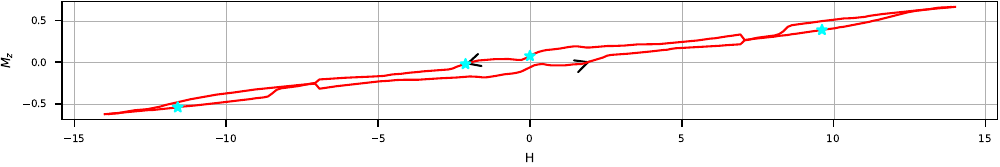}
    \includegraphics[width=0.245\linewidth]{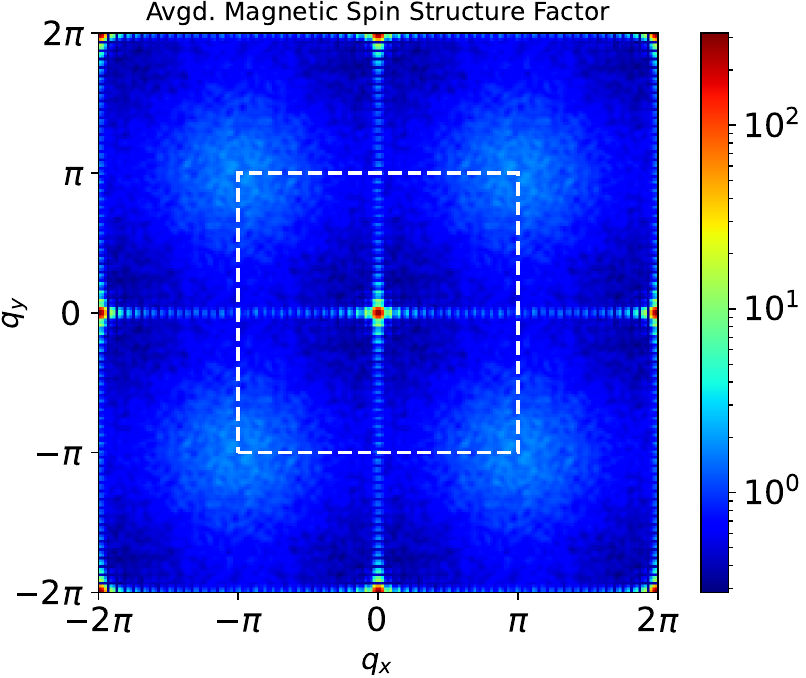}
    \includegraphics[width=0.245\linewidth]{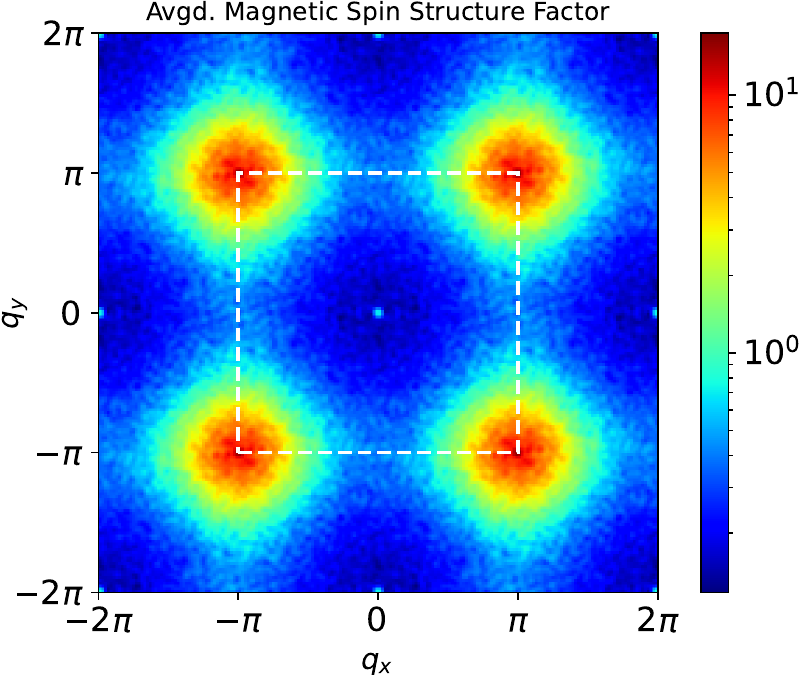}
    \includegraphics[width=0.245\linewidth]{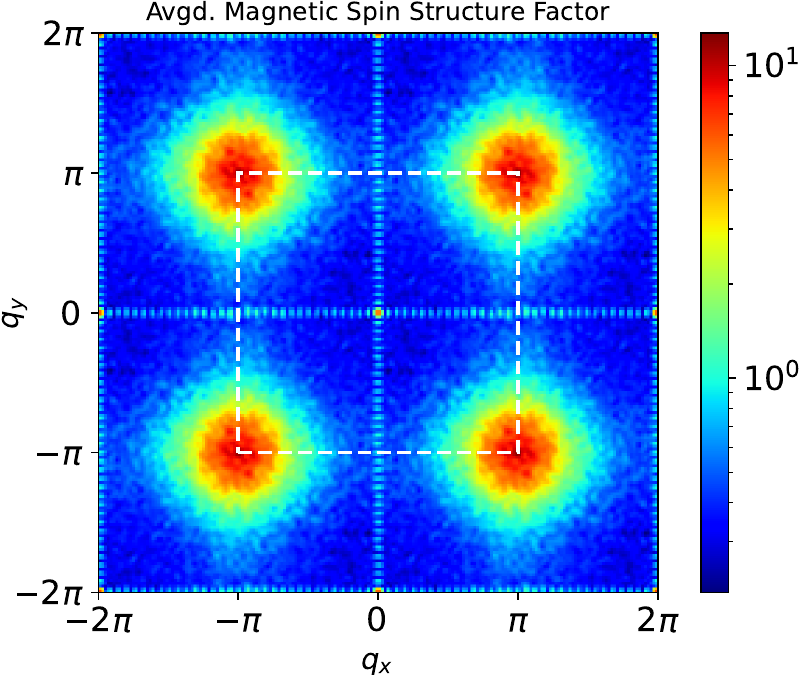}
    \includegraphics[width=0.245\linewidth]{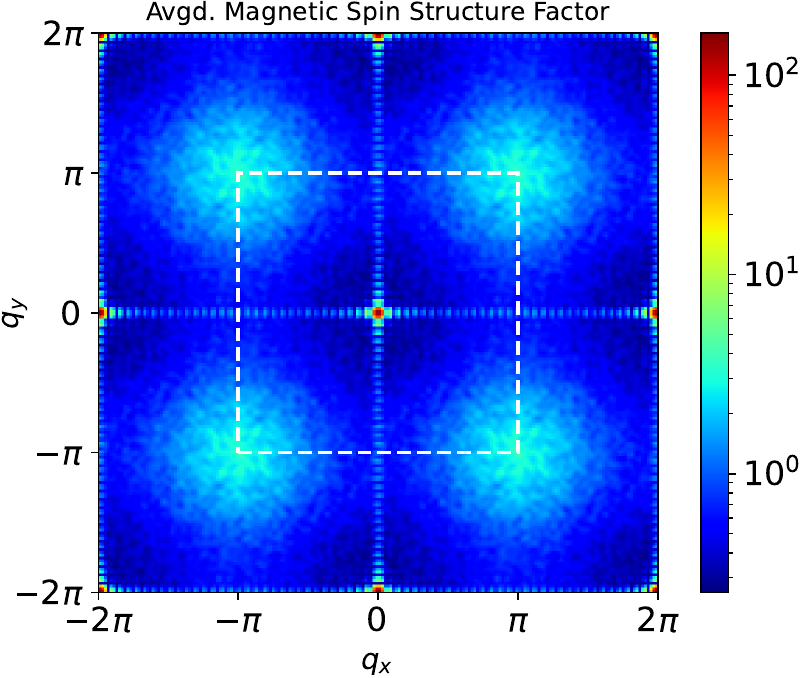}
    \caption{Magnetic structure factors $\abs{S(q)}$ heatmaps (bottom row) within the 2D antiferromagnetic hysteresis cycles, run on \texttt{Advantage\_system7.1} at $s=0.6$. 
    }
    \label{fig:SSF6}
\end{figure*}

\begin{figure*}[ht!]
    \centering
    \includegraphics[width=0.999\linewidth]{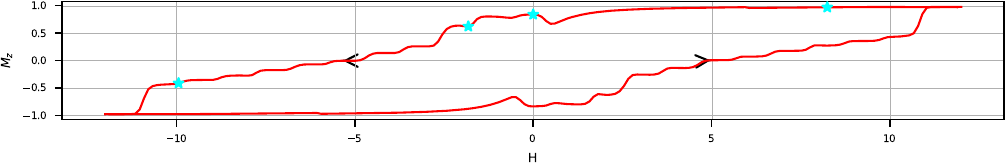}
    \includegraphics[width=0.245\linewidth]{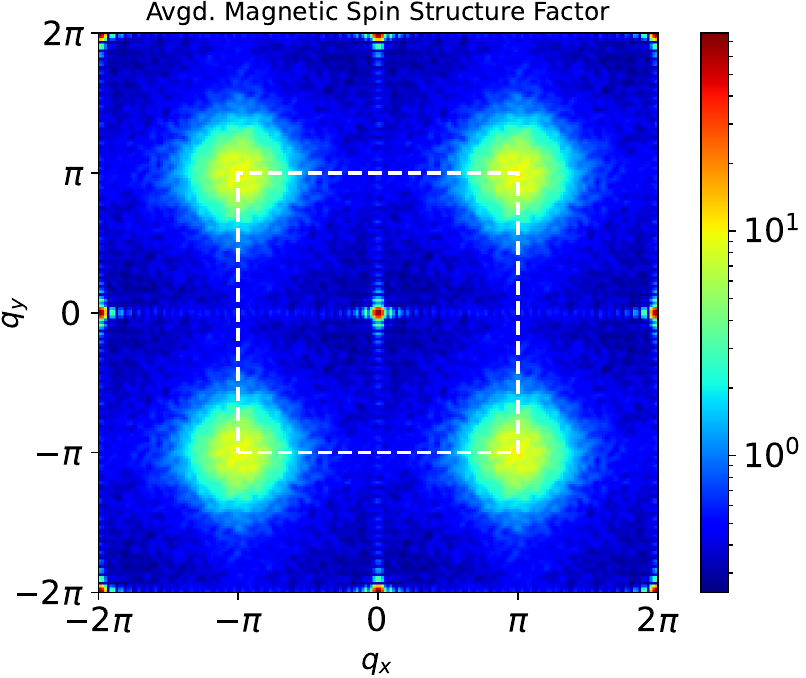}
    \includegraphics[width=0.245\linewidth]{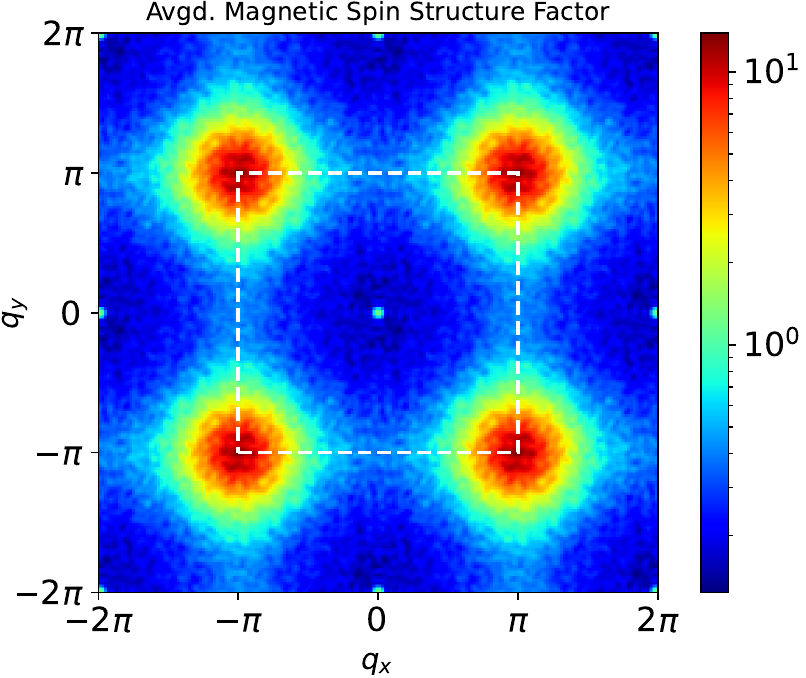}
    \includegraphics[width=0.245\linewidth]{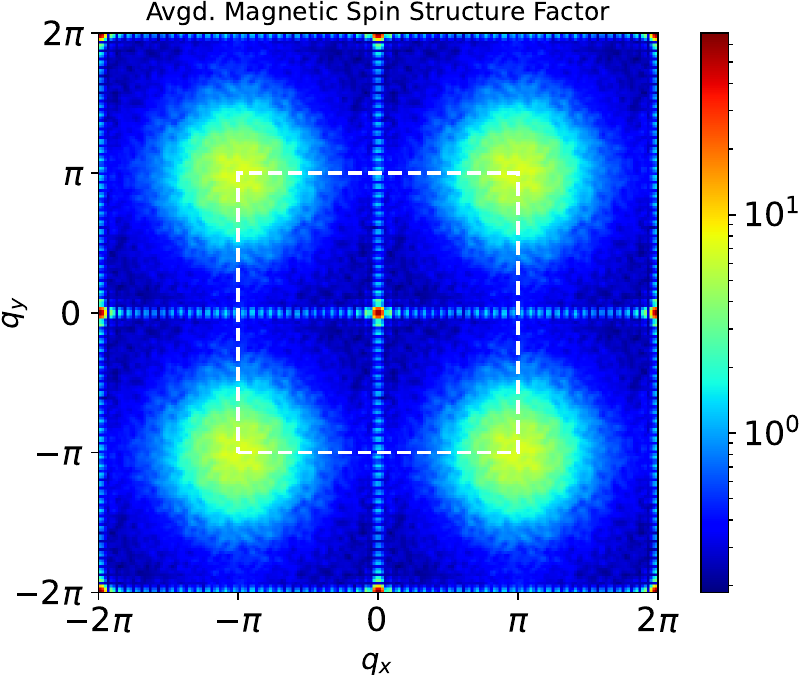}
    \includegraphics[width=0.245\linewidth]{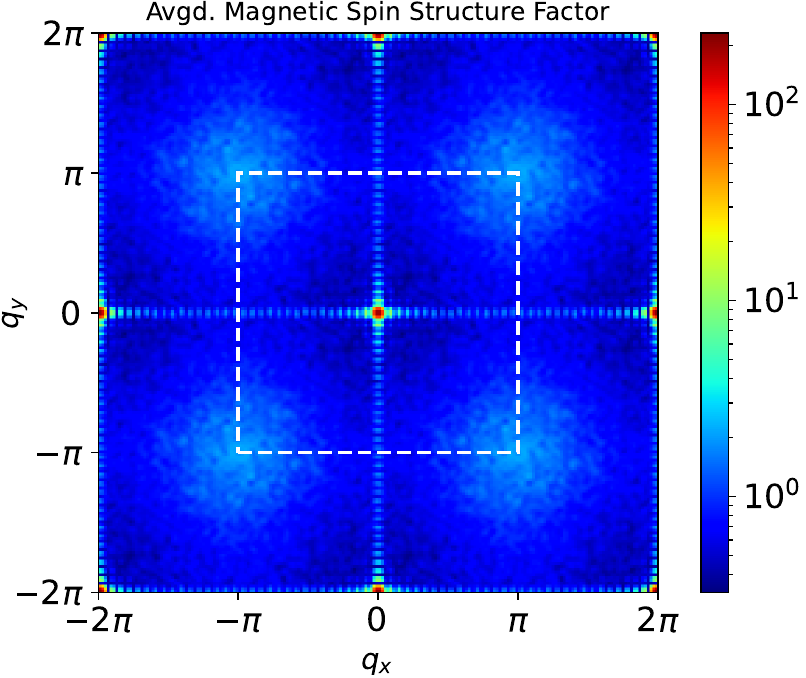}
    \caption{Magnetic structure factors $\abs{S(q)}$ heatmaps (bottom row) within the 2D antiferromagnetic hysteresis cycles, run on \texttt{Advantage\_system4.1} at $s=0.7$.
    }
    \label{fig:SSF7}
\end{figure*}

\begin{figure*}[ht!]
    \centering
    \includegraphics[width=0.999\linewidth]{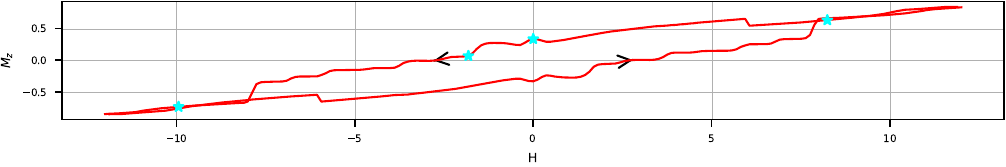}
    \includegraphics[width=0.245\linewidth]{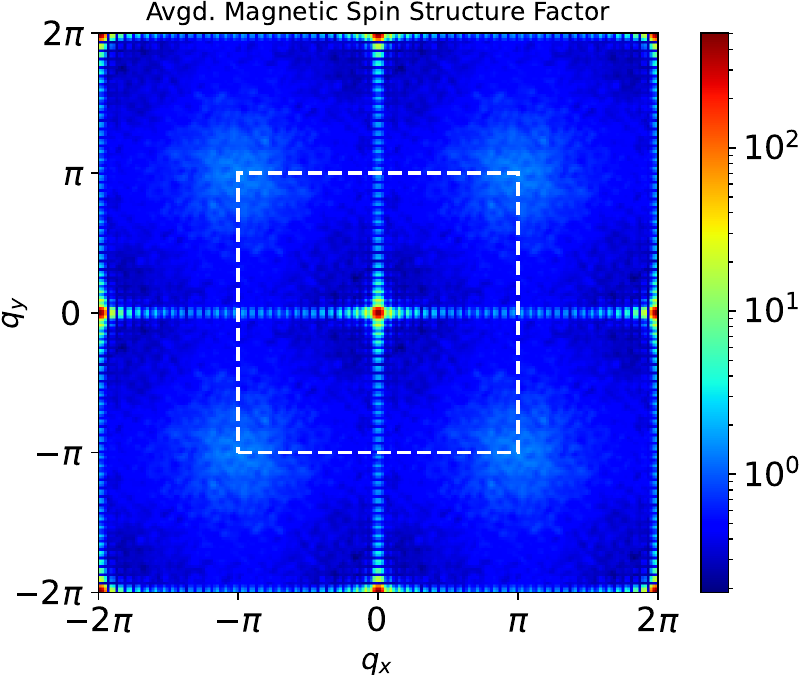}
    \includegraphics[width=0.245\linewidth]{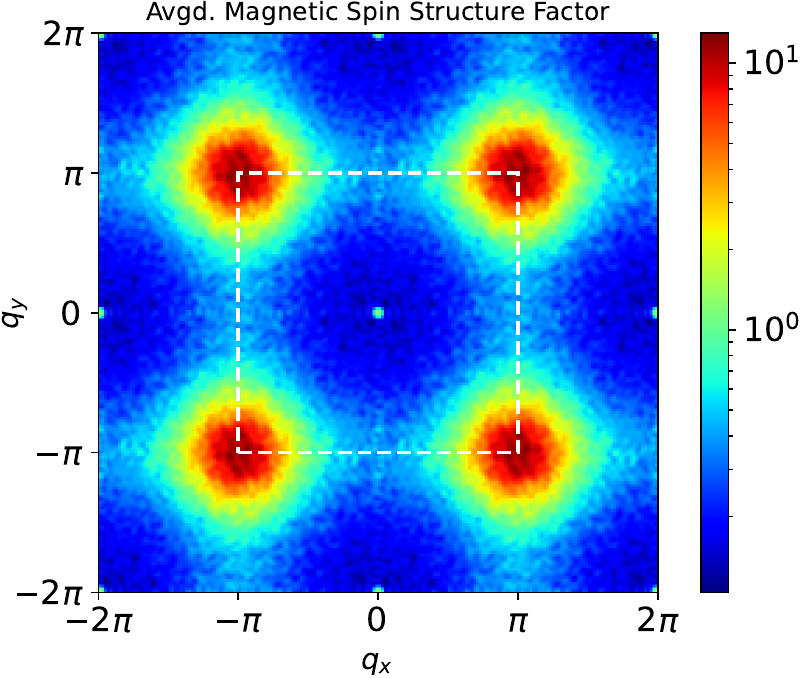}
    \includegraphics[width=0.245\linewidth]{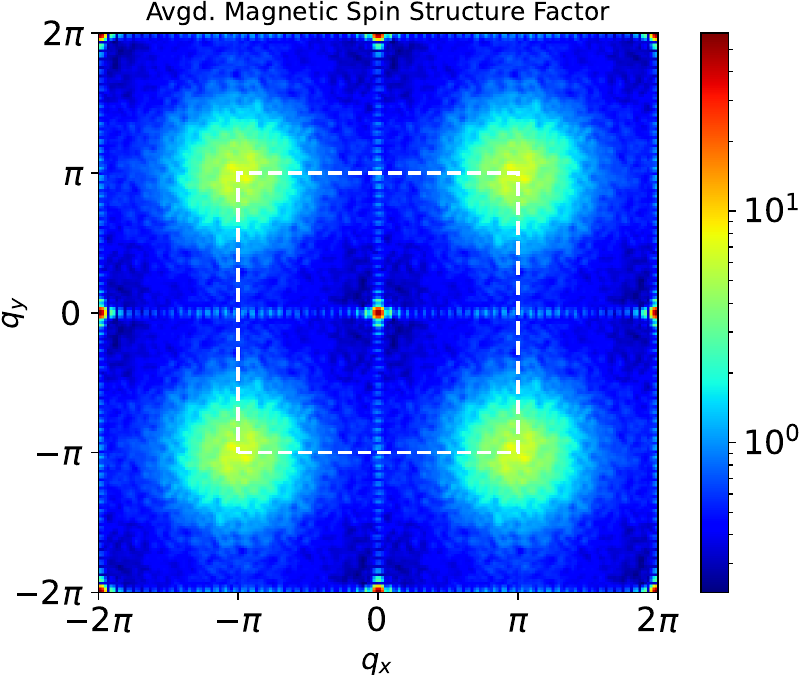}
    \includegraphics[width=0.245\linewidth]{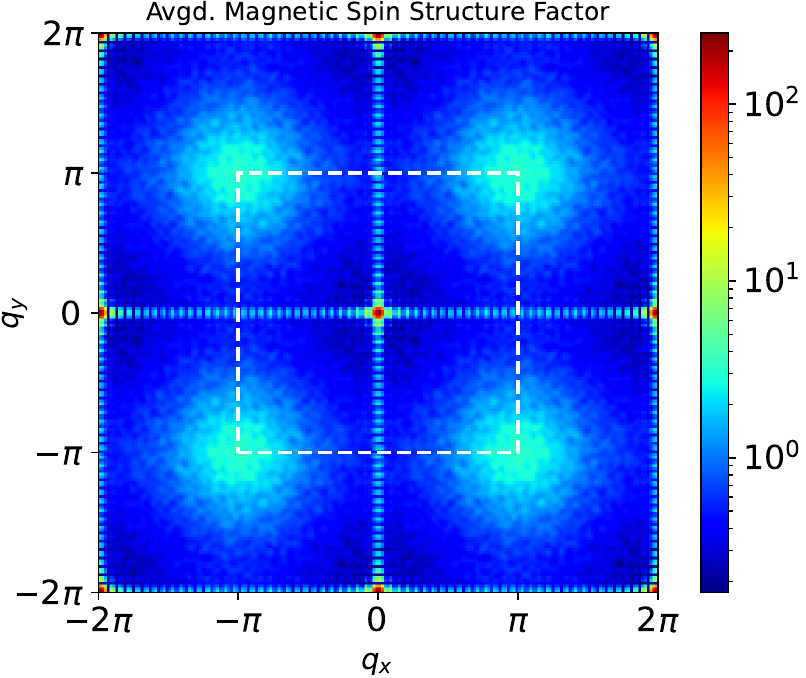}
    \caption{Magnetic structure factors $\abs{S(q)}$ heatmaps (bottom row) within the 2D antiferromagnetic hysteresis cycles, run on \texttt{Advantage\_system4.1} at $s=0.65$. }
    \label{fig:SSF8}
\end{figure*}

\section{Additional Representative Magnetic Spin Structure Factors}
\label{section:additional_SSF_plots}

This section provides several additional sequences of averaged magnetic structure factors at various points along various 2D antiferromagnetic hysteresis simulations. Figs.~\ref{fig:SSF2}, Fig.~\ref{fig:SSF4}, Fig.~\ref{fig:SSF5}, Fig.~\ref{fig:SSF6}, Fig.~\ref{fig:SSF7}, Fig.~\ref{fig:SSF8} show additional sequences, each obtained from different hysteresis cycles at various $s$ values (equivalent to a physical $\Gamma/J$) and run on the several different D-Wave processors. In these structure factors, effects from the open boundary conditions and finite lattice are prominent, particularly at full saturation. In the less saturated regions of the hysteresis cycles, the antiferromagnetic ordering becomes apparent with peaks at $(\pi, \pi)$ in the square lattice. Due to the consistently applied longitudinal field, there is typically a strong peak at $(0, 0)$ which corresponds to ferromagnetic-like ordering, and becomes much stronger at saturation.

The Hamiltonian energy scale ratio $\Gamma/J$ does change the characteristics of the magnetic ordering. Fig.~\ref{fig:SSF4}, left, shows ordering at a low longitudinal field strength with $M_z \approx 0$, where the peak at $(0, 0)$ largely disappears. This peak of ordering disappears because $\Gamma/J$ is strong in Fig.~\ref{fig:SSF4}. Whereas, Fig.~\ref{fig:SSF5} reports magnetic structure factors also near magnetization $M_z \approx 0$, but here $\Gamma/J$ is very weak, and this shows very different magnetic ordering compared to Fig.~\ref{fig:SSF4} with peaks of antiferromagnetic ordering along $q_x=\pm \pi, q_y \pm \pi$.

The general characterization we find is that qualitatively the magnetic ordering, as revealed by $\abs{S(q)}$, is similar across the different physical experiment settings, including different QPUs. In other words, there are no strong outliers that would indicate a particular hardware issue, or an unexpected magnetic ordering. 
The computation of the structure factors is greatly accelerated by the Python libraries Numba and NumPy~\cite{10.1145/2833157.2833162, harris2020array}.

\section{Experiment Standard Deviation Magnetization Measurement Quantities}
\label{section:appendix_standard_deviation_plots}

\begin{figure*}[ht!]
    \centering
    \includegraphics[width=0.49\linewidth]{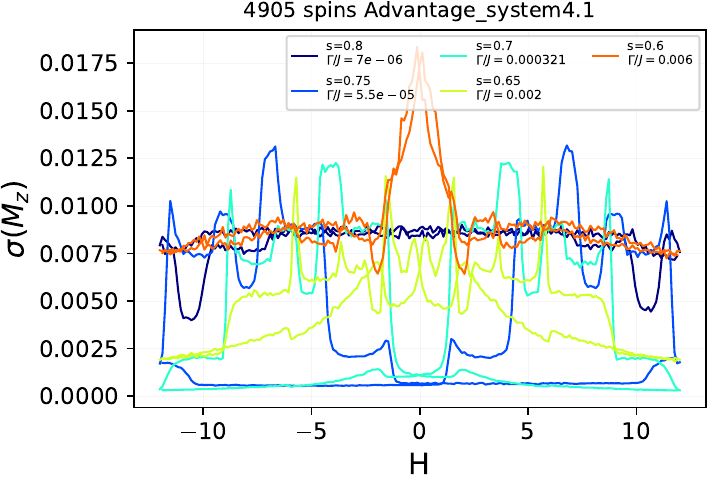}
    \includegraphics[width=0.49\linewidth]{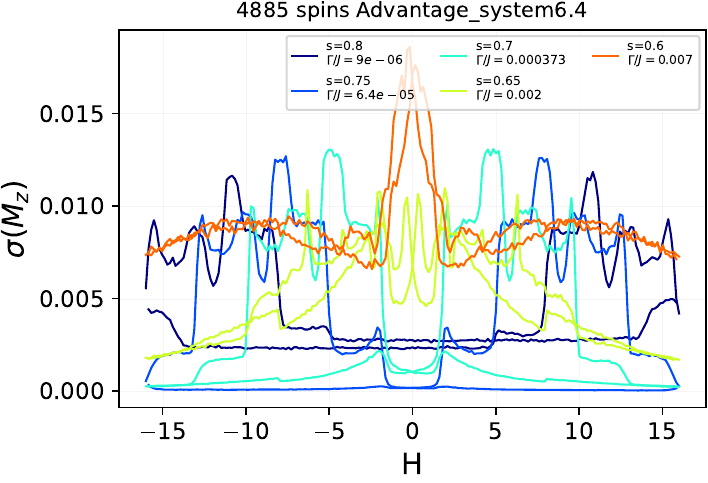}
    \includegraphics[width=0.49\linewidth]{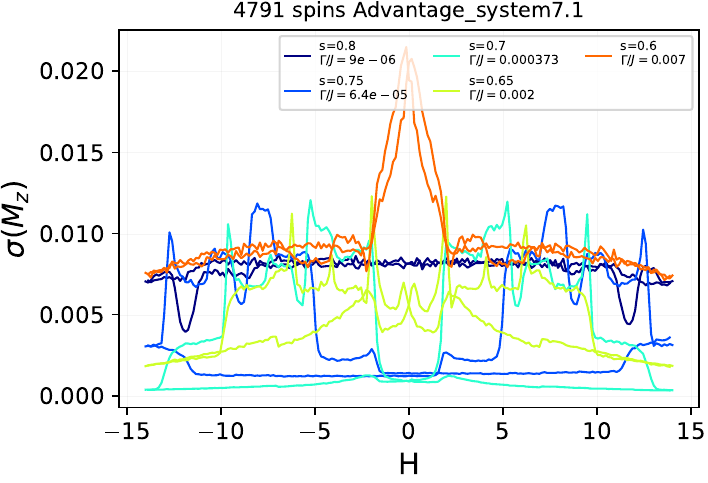}
    \includegraphics[width=0.49\linewidth]{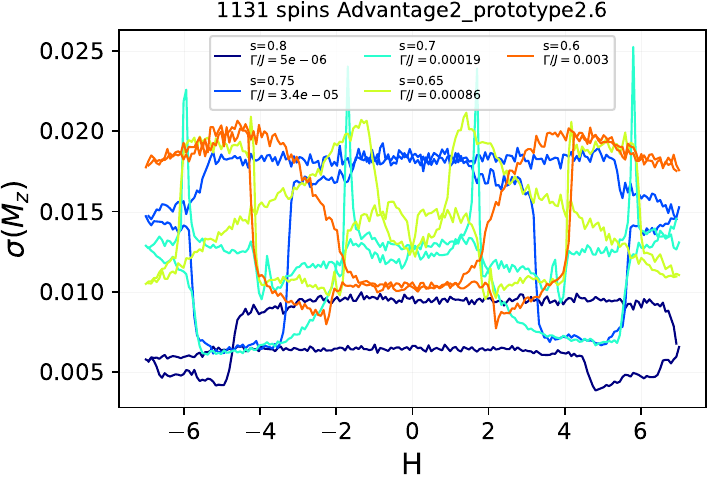}
    \caption{ Standard deviation of the magnetization ($\sigma(M_z)$) measured observable from the hysteresis cycles reported in Fig~\ref{fig:1D_AFM} of the 1D antiferromagnetic rings with periodic boundary conditions.  }
    \label{fig:standard_deviation_mz_1D}
\end{figure*}

\begin{figure*}[ht!]
    \centering
    \includegraphics[width=0.49\linewidth]{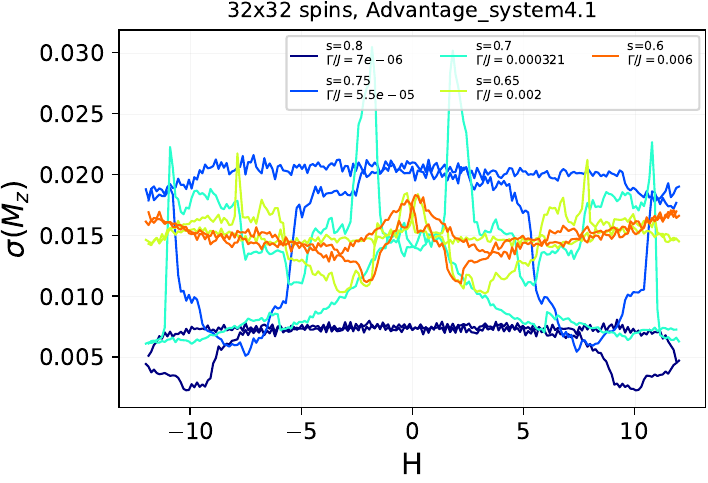}
    \includegraphics[width=0.49\linewidth]{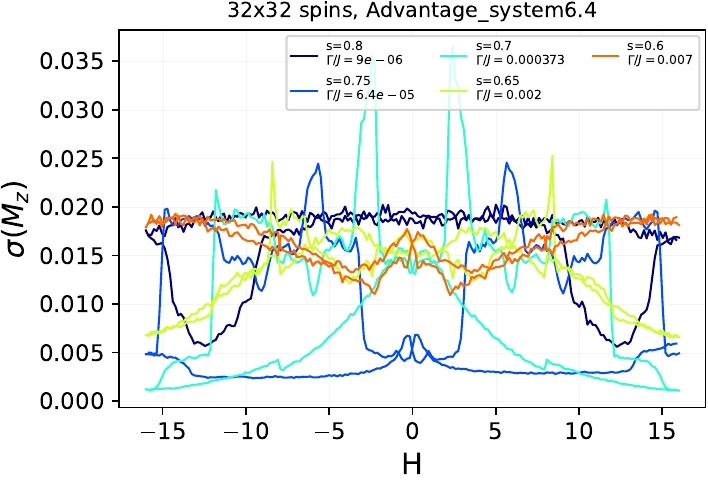}
    \includegraphics[width=0.49\linewidth]{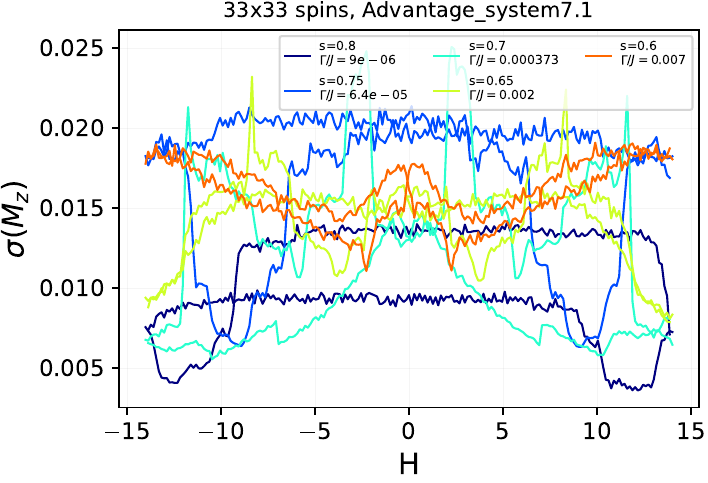}
    \includegraphics[width=0.49\linewidth]{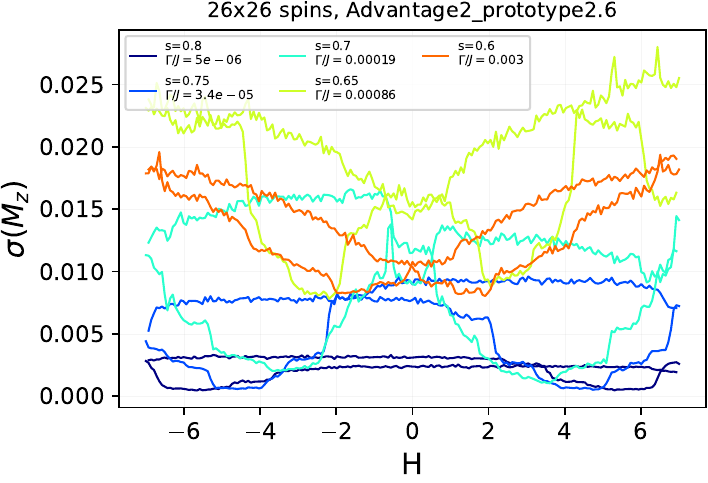}
    \caption{ Standard deviation of the magnetization ($\sigma(M_z)$) measured observable from the hysteresis cycles reported in Fig~\ref{fig:2D_AFM} of 2D antiferromagnetic square lattices with open boundary conditions. }
    \label{fig:standard_deviation_mz_2D}
\end{figure*}

The majority of the statistical uncertainty is due to the finite sampling effect (e.g., shot noise). However, because the protocol is fundamentally a sampling based protocol at intermediate slices of a continuous time computation, there is inherently some additional statistical noise during any magnetic reversal regions of the net magnetization plots. In Figs.~\ref{fig:standard_deviation_mz_1D}, \ref{fig:standard_deviation_mz_2D} we report the standard deviation of several of the antiferromagnetic hysteresis cycles, which shows that overall the largest standard deviations are at most $\approx 0.035$. Comparing the standard deviation measures to the magnetization curves in Fig.~\ref{fig:1D_AFM} and Fig.~\ref{fig:2D_AFM}, typically when $\Gamma/J$ is large (near $s=0.6, s=0.7$) and the system undergoes full reversal, near applied field $H \approx 0$ the standard deviation peaks. This increase in variance at reversal is consistent with the stochastic effects of avalanche-mediated reversal and Barkhausen jumps.

\section{Domain Wall Density on the 3 Other D-Wave QPUs}
\label{section:appendix_Domain_wall_density}

\begin{figure*}[ht!]
    \centering
    \includegraphics[width=0.494\linewidth]{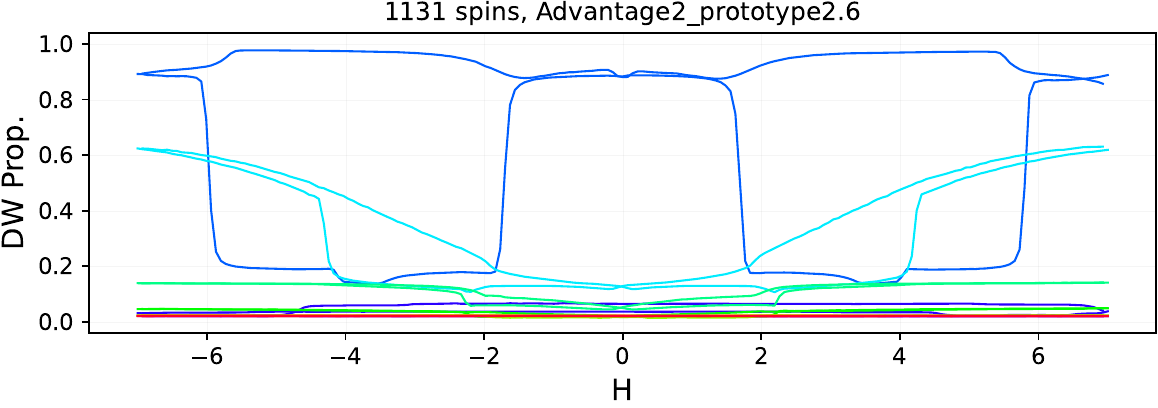}
    \includegraphics[width=0.494\linewidth]{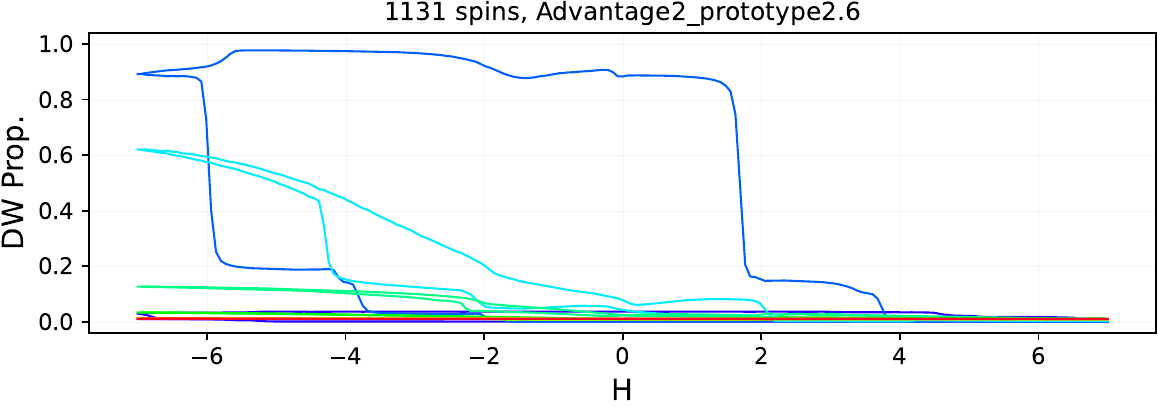}
    \includegraphics[width=0.6\linewidth]{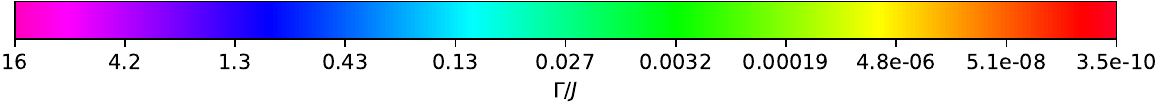}
    \includegraphics[width=0.494\linewidth]{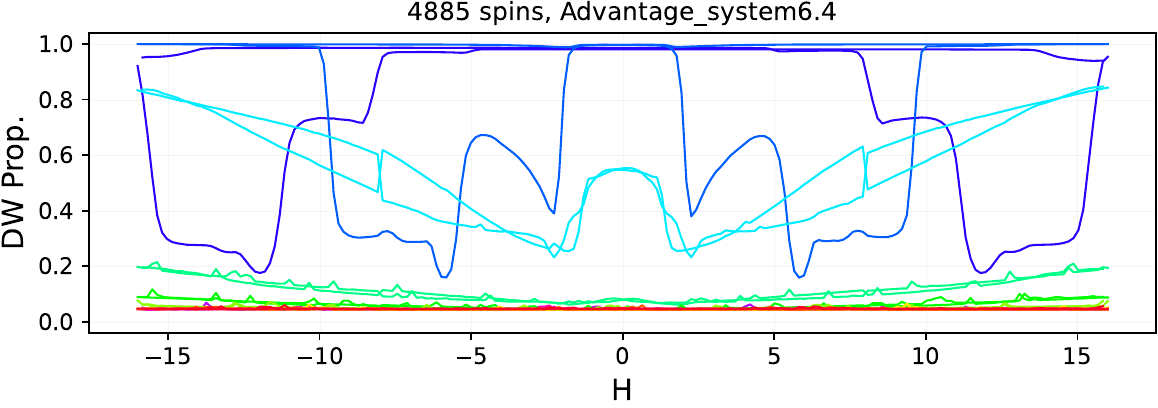}
    \includegraphics[width=0.494\linewidth]{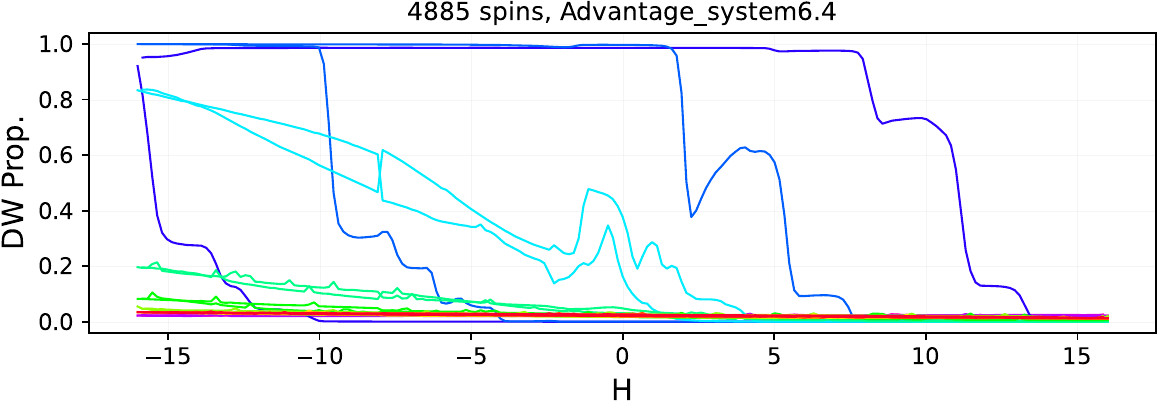}
    \includegraphics[width=0.6\linewidth]{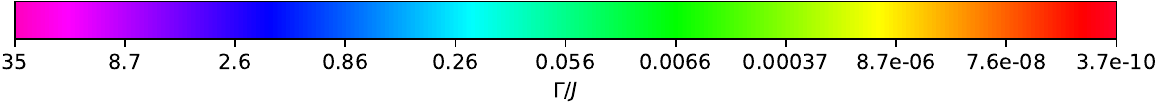}
    \includegraphics[width=0.494\linewidth]{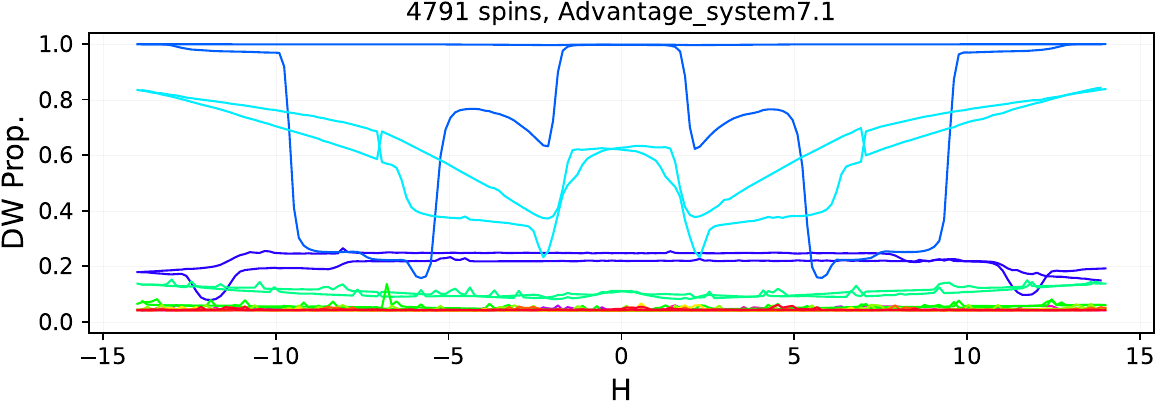}
    \includegraphics[width=0.494\linewidth]{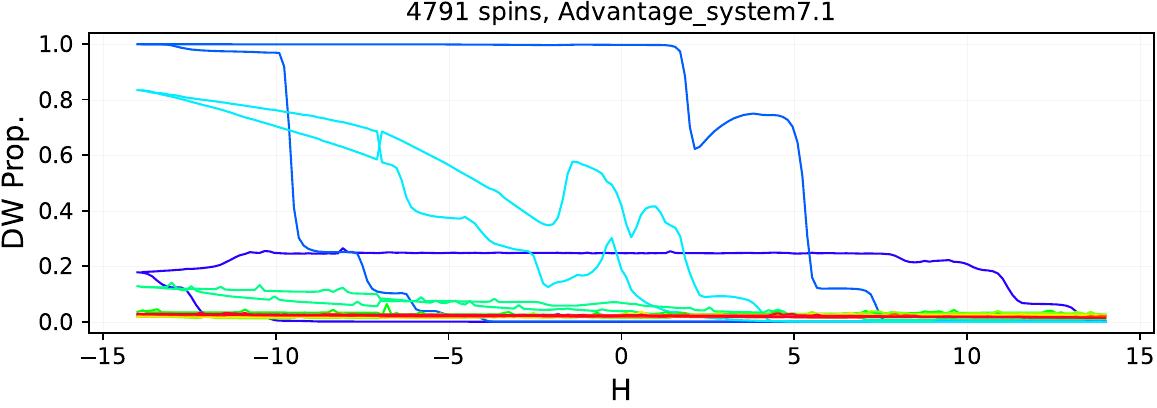}
    \includegraphics[width=0.6\linewidth]{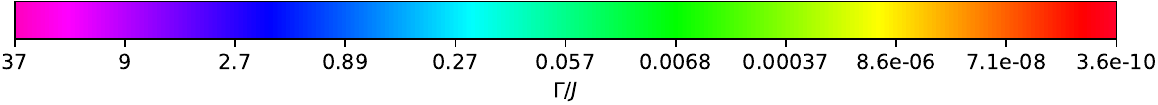}
    \caption{\textbf{Domain wall density on the 1D antiferromagnetic rings during the hysteresis cycles as a function of applied longitudinal field $H$}. Left hand column shows the total, average, domain wall density, regardless of the sign of the domain wall. Right hand column shows the spin-down ($\downarrow \downarrow$) domain wall density. 
    }
    \label{fig:1D_domain_wall_proportions_appendix}
\end{figure*}

Fig.~\ref{fig:1D_domain_wall_proportions_appendix} reports domain wall density during hysteresis cycles in the 1D (odd-spin ring) models from the remaining three QPUs, using the same plot style as Fig.~\ref{fig:1D_domain_wall_proportions}. 

\section{2D Lattice N\'eel Order Parameter Quantities}
\label{section:appendix_2D_Neel_order_param}

We focus on the hysteretic region of $s$, and separate each hysteresis cycle by $\Gamma/J$ in order to make visualization clear. 

\begin{figure*}[ht!]
    \centering
    \includegraphics[width=0.32\linewidth]{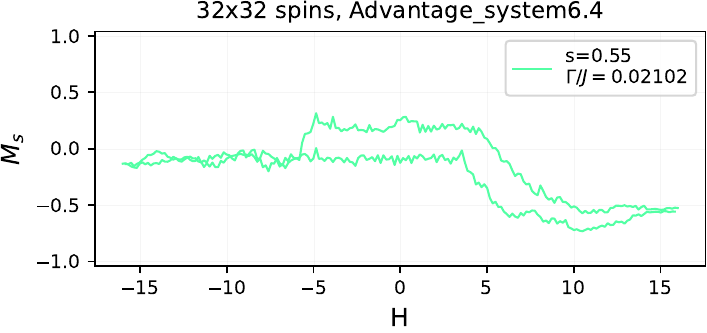}
    \includegraphics[width=0.32\linewidth]{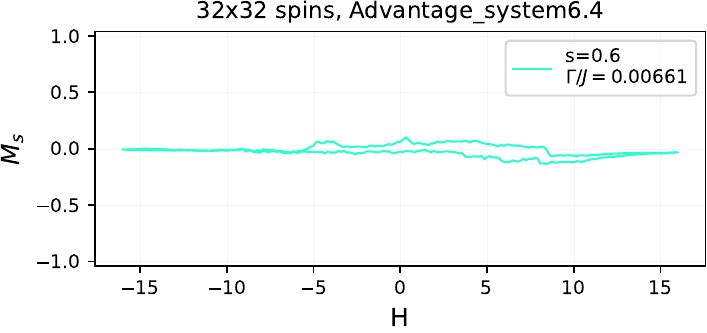}
    \includegraphics[width=0.32\linewidth]{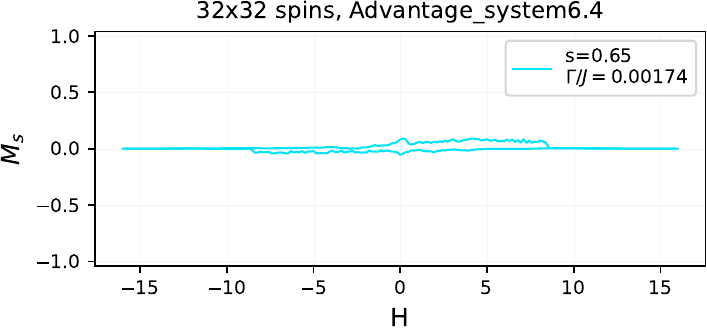}
    \includegraphics[width=0.32\linewidth]{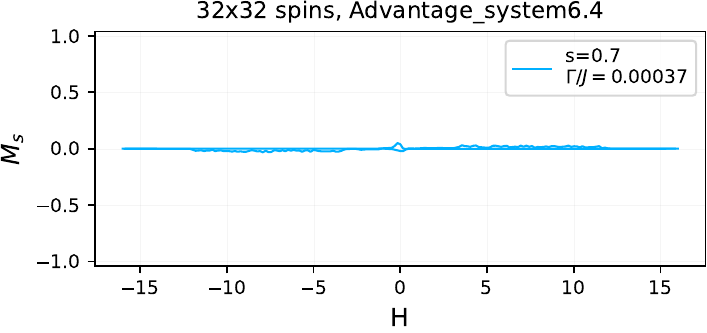}
    \includegraphics[width=0.32\linewidth]{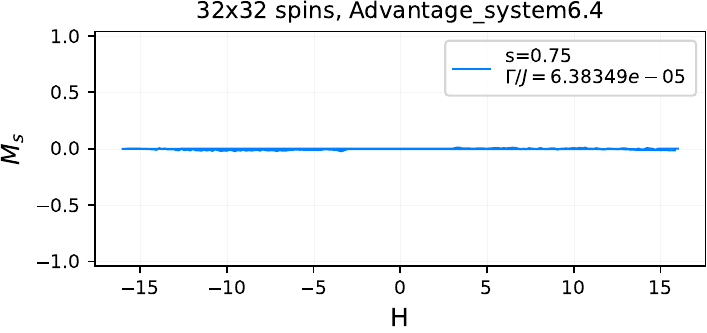}
    \includegraphics[width=0.32\linewidth]{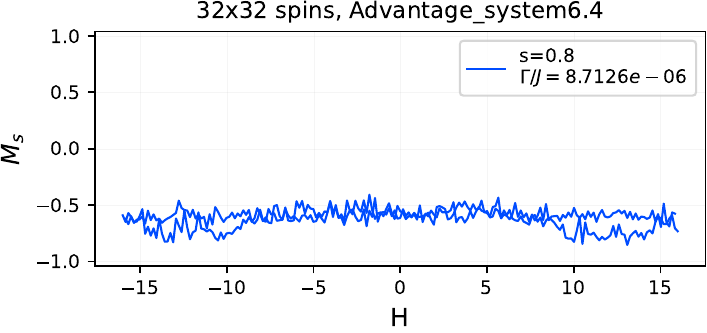}
    \caption{2D Hysteresis cycles in terms of the antiferromagnetic order parameter $M_s$. Data from \texttt{Advantage\_system6.4}.  }
    \label{fig:2D_AFM_order_parameter_Pegasus6.4}
\end{figure*}

\begin{figure*}[ht!]
    \centering
    \includegraphics[width=0.32\linewidth]{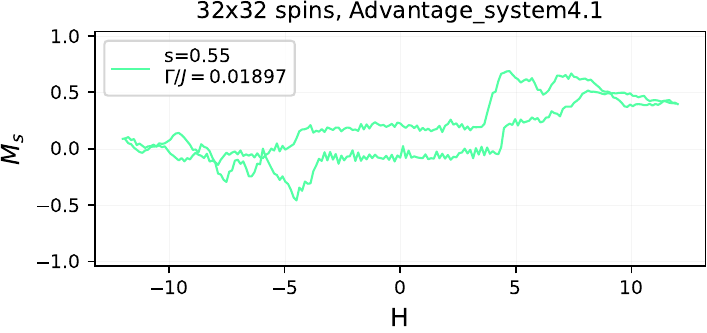}
    \includegraphics[width=0.32\linewidth]{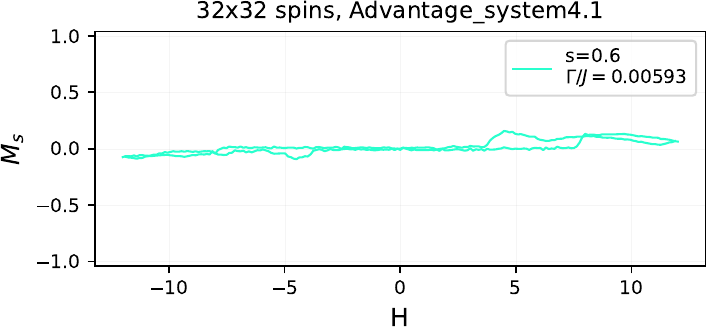}
    \includegraphics[width=0.32\linewidth]{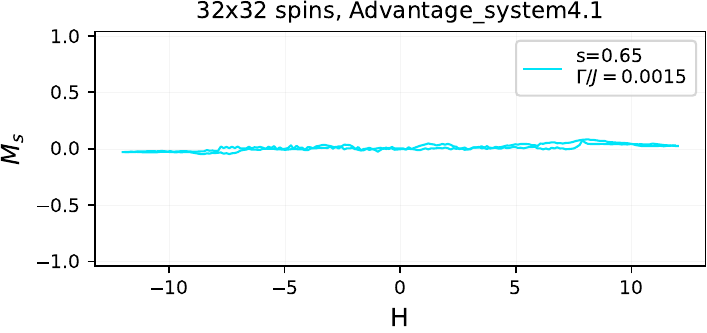}
    \includegraphics[width=0.32\linewidth]{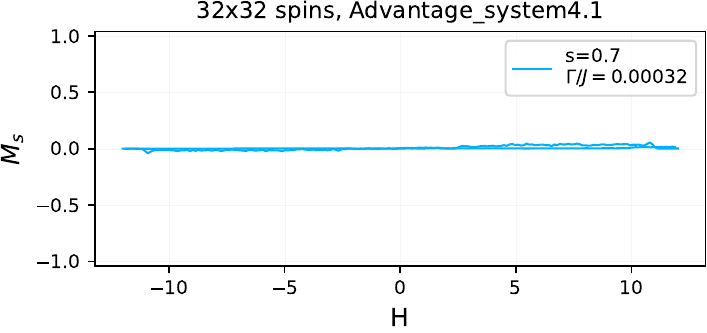}
    \includegraphics[width=0.32\linewidth]{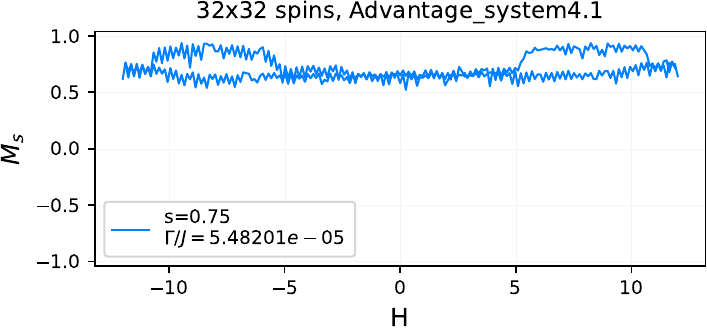}
    \includegraphics[width=0.32\linewidth]{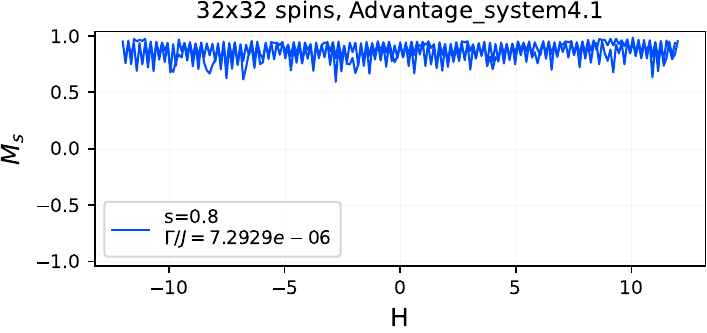}
    \caption{2D Hysteresis cycles in terms of the antiferromagnetic order parameter $M_s$. Data from \texttt{Advantage\_system4.1}.  }
    \label{fig:2D_AFM_order_parameter_Pegasus4.1}
\end{figure*}

\begin{figure*}[ht!]
    \centering
    \includegraphics[width=0.32\linewidth]{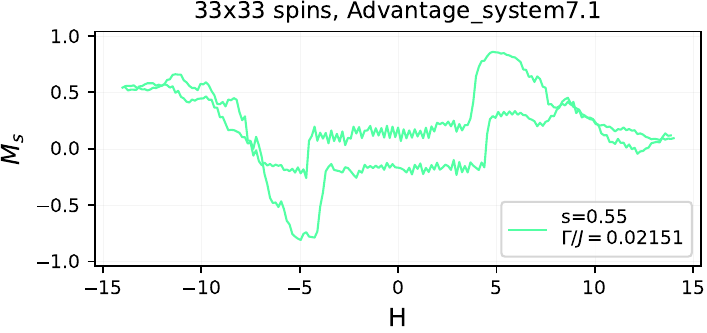}
    \includegraphics[width=0.32\linewidth]{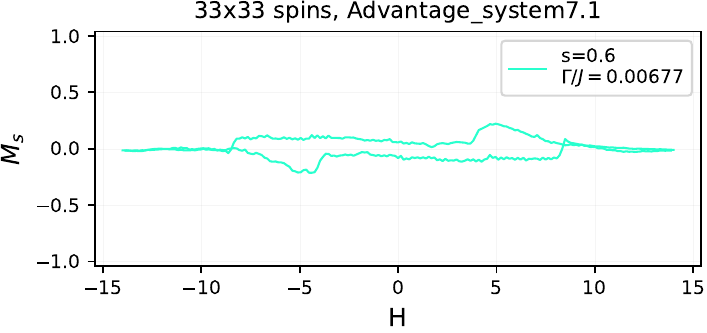}
    \includegraphics[width=0.32\linewidth]{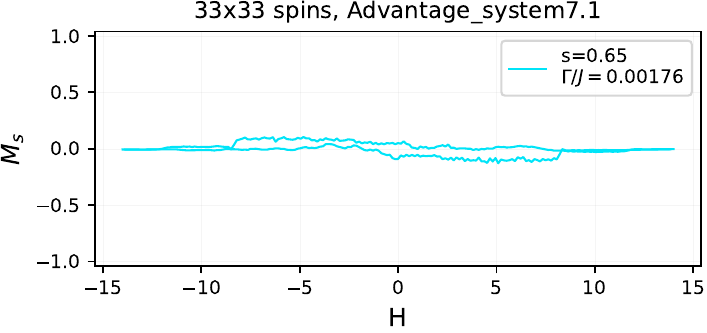}
    \includegraphics[width=0.32\linewidth]{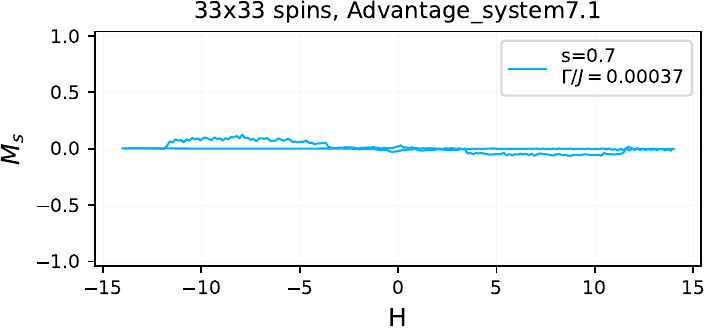}
    \includegraphics[width=0.32\linewidth]{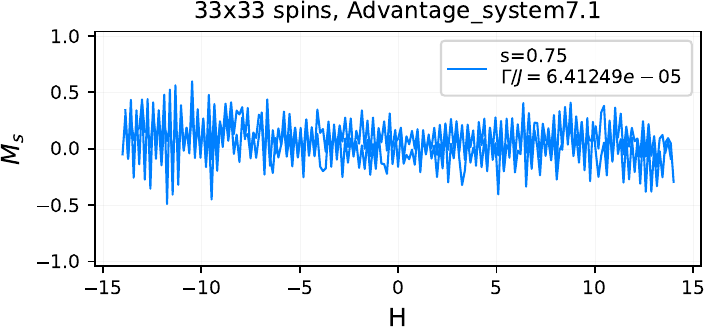}
    \includegraphics[width=0.32\linewidth]{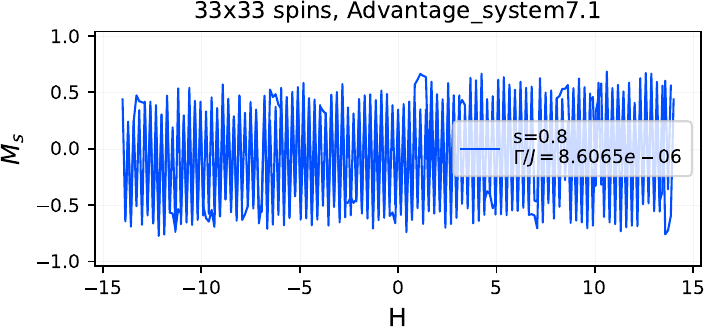}
    \caption{2D Hysteresis cycles in terms of the antiferromagnetic order parameter $M_s$. Data from \texttt{Advantage\_system7.1}.  }
    \label{fig:2D_AFM_order_parameter_Pegasus7.1}
\end{figure*}

\begin{figure*}[ht!]
    \centering
    \includegraphics[width=0.32\linewidth]{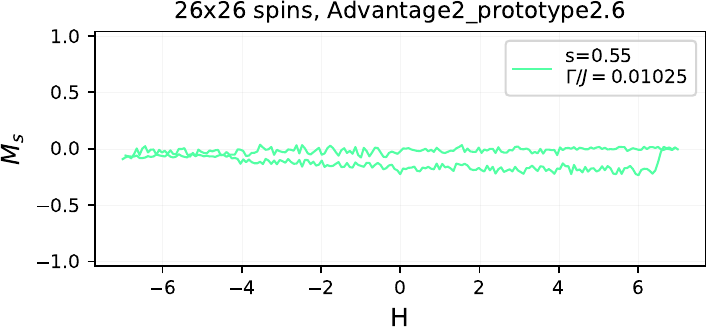}
    \includegraphics[width=0.32\linewidth]{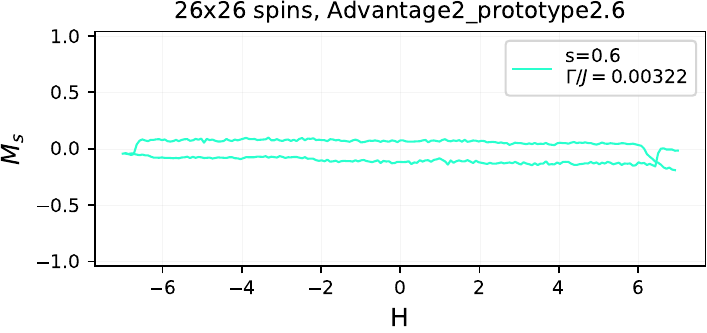}
    \includegraphics[width=0.32\linewidth]{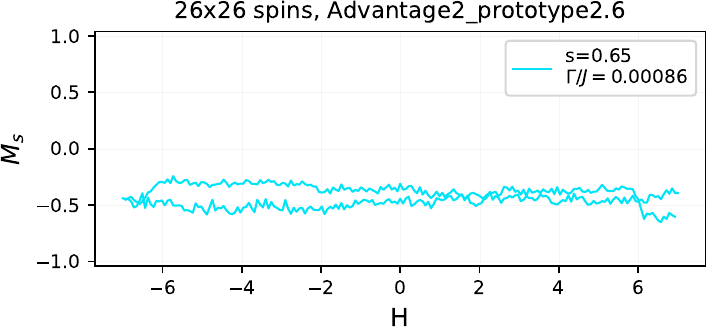}
    \includegraphics[width=0.32\linewidth]{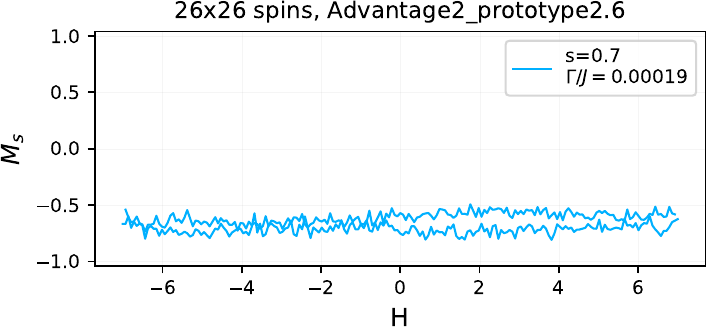}
    \includegraphics[width=0.32\linewidth]{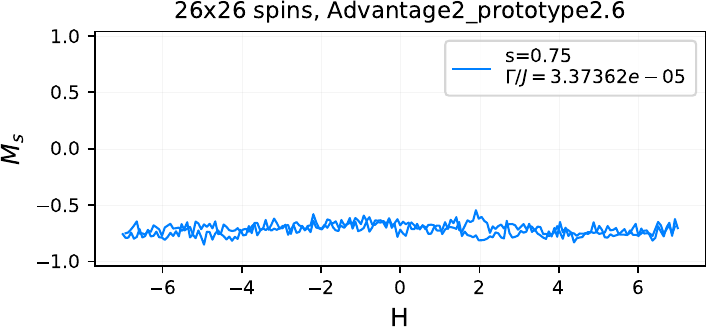}
    \includegraphics[width=0.32\linewidth]{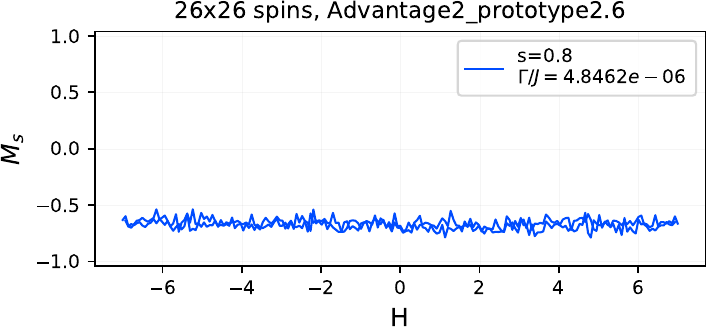}
    \caption{2D Hysteresis cycles in terms of the antiferromagnetic order parameter $M_s$. Data from \texttt{Advantage2\_prototype2.6}.  }
    \label{fig:2D_AFM_order_parameter_Zephyr2.6}
\end{figure*}

Fig.~\ref{fig:2D_AFM_order_parameter_Pegasus6.4} shows that the staggered antiferromagnetic order parameter is non-zero, and that it also forms hysteresis cycles. Data are from experiments on \texttt{Advantage\_system6.4}. Figs.~\ref{fig:2D_AFM_order_parameter_Pegasus4.1}, \ref{fig:2D_AFM_order_parameter_Pegasus7.1}, and \ref{fig:2D_AFM_order_parameter_Zephyr2.6} show the same set of data but from the other three quantum annealing processors. The occasional hysteresis is very distinct from the ferromagnetic-like magnetization curves seen in Fig.~\ref{fig:2D_AFM}; for example, we see non-zero area between sweeps at $s \approx 0.75, 0.55$ in Fig.~\ref{fig:2D_AFM_order_parameter_Pegasus6.4}. Interestingly, at very small $\Gamma/J$ we do see long-range antiferromagnetic ordering. This is expected because state transitions are much harder when the transverse field is weak. However, the data for the staggered N\'eel parameter are significantly noisier than average net magnetization in Fig.~\ref{fig:2D_AFM}. This effect could in part originate in shot noise; also, the N\'eel order parameter could be more sensitive to hardware biases. In future studies we will apply statistics-balancing calibration refinement~\cite{Chern_2023} on the local fields and on the couplers to reduce the noise. For the three devices where full magnetic $M_z$ saturation was reached, $M_s$ converges entirely to zero during the hysteresis cycle for the values of $s$ corresponding to maximal hysteretic area in $M_z$ (see Fig.~\ref{fig:2D_AFM}).

\section{Additional Representative 2D Sampled Configurations During Hysteresis Cycles}
\label{section:appendix_additional_2D_samples}

\begin{figure*}[ht!]
    \centering
    \includegraphics[width=0.15\linewidth]{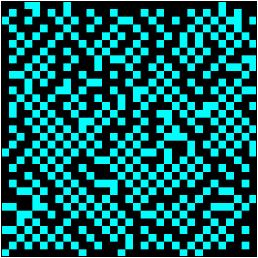}
    \includegraphics[width=0.15\linewidth]{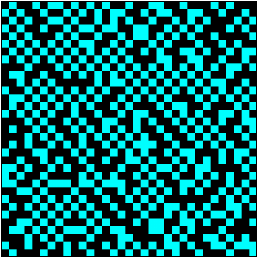}
    \includegraphics[width=0.15\linewidth]{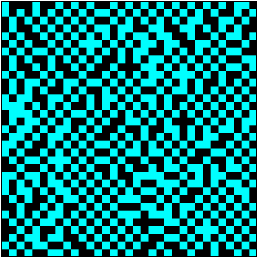}
    \includegraphics[width=0.15\linewidth]{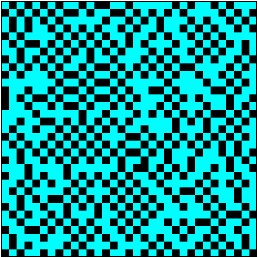}
    \includegraphics[width=0.15\linewidth]{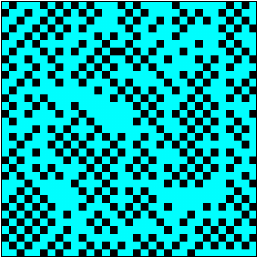}
    \includegraphics[width=0.15\linewidth]{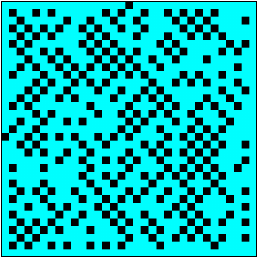}
    \includegraphics[width=0.15\linewidth]{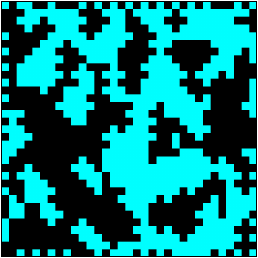}
    \includegraphics[width=0.15\linewidth]{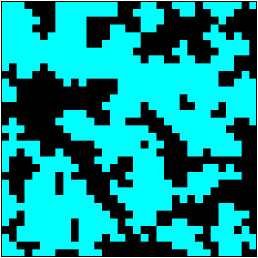}
    \includegraphics[width=0.15\linewidth]{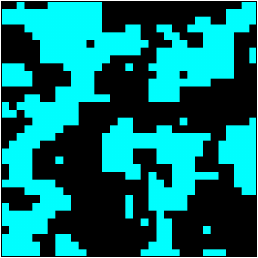}
    \includegraphics[width=0.15\linewidth]{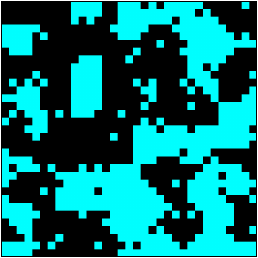}
    \includegraphics[width=0.15\linewidth]{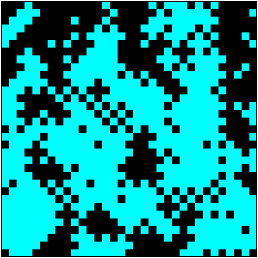}
    \includegraphics[width=0.15\linewidth]{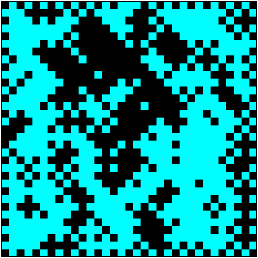}
    \caption{\textbf{Selected real-space spin configurations sampled on the \texttt{Advantage\_system7.1} QPU,} illustrated as pixel plots where black pixels denote spin-down qubit measurements and cyan pixels denote spin up qubit measurements (top row), from the middle of the magnetization reversal at various longitudinal field strengths of the first magnetic hysteresis sweep of a $33\times 33$ 2D square grid antiferromagnet at $s=0.7$. Notice that the checkerboard antiferromagnetic ground-state emerges in sections of the lattices. Also notice that the boundary effects from the open boundary conditions are noticeable in many of the measured spin configurations. Bottom row: staggered domain height function $\tau$ from Eq.~(\ref{eq:tau}), corresponding to the same spin configurations on top, making the antiferromagnetic domains apparent. }
    \label{fig:2D_spin_configs_Advantage_system7.1_appendix}
\end{figure*}

Fig.~\ref{fig:2D_spin_configs_Advantage_system7.1_appendix} shows additional representative sampled configurations of the 2D antiferromagnetic model during a hysteresis cycle.

\section{D-Wave Quantum Annealer Hardware Embeddings}
\label{section:appendix_embeddings}

\begin{figure*}[ht!]
    \centering
    \includegraphics[width=0.49\linewidth]{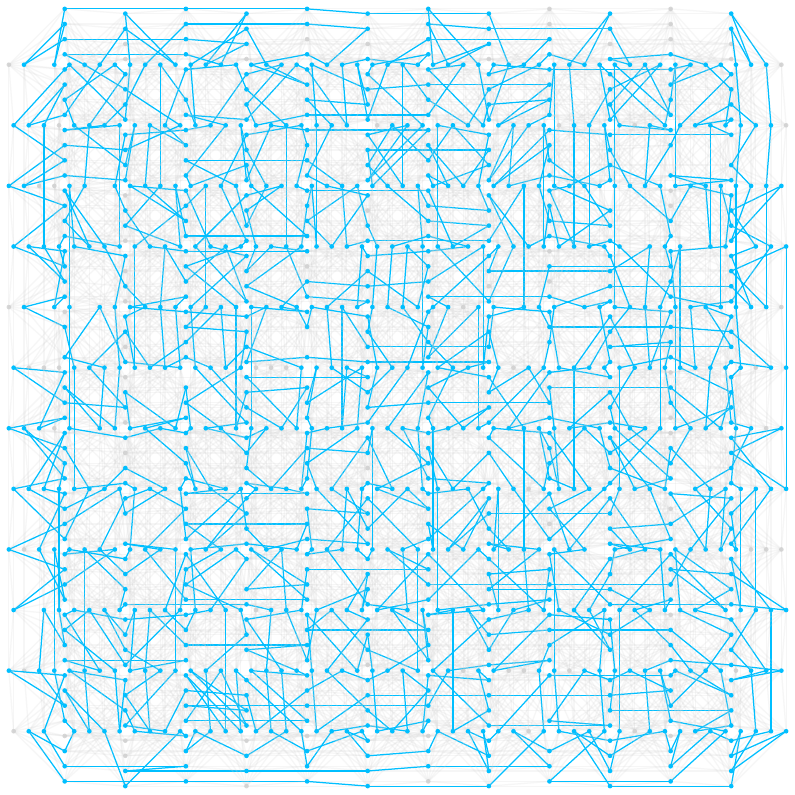}
    \includegraphics[width=0.49\linewidth]{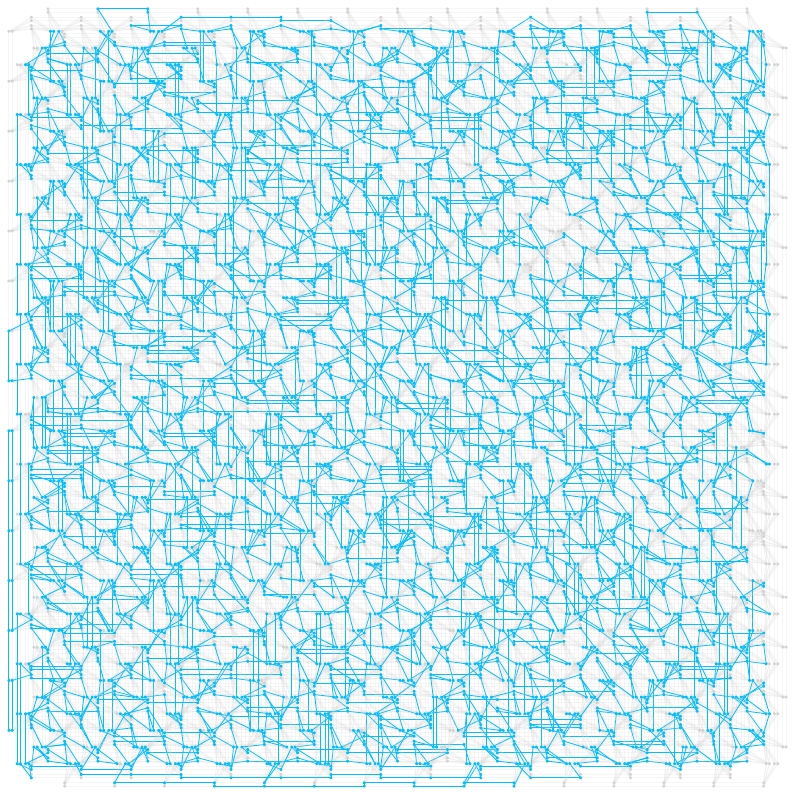}
    \includegraphics[width=0.49\linewidth]{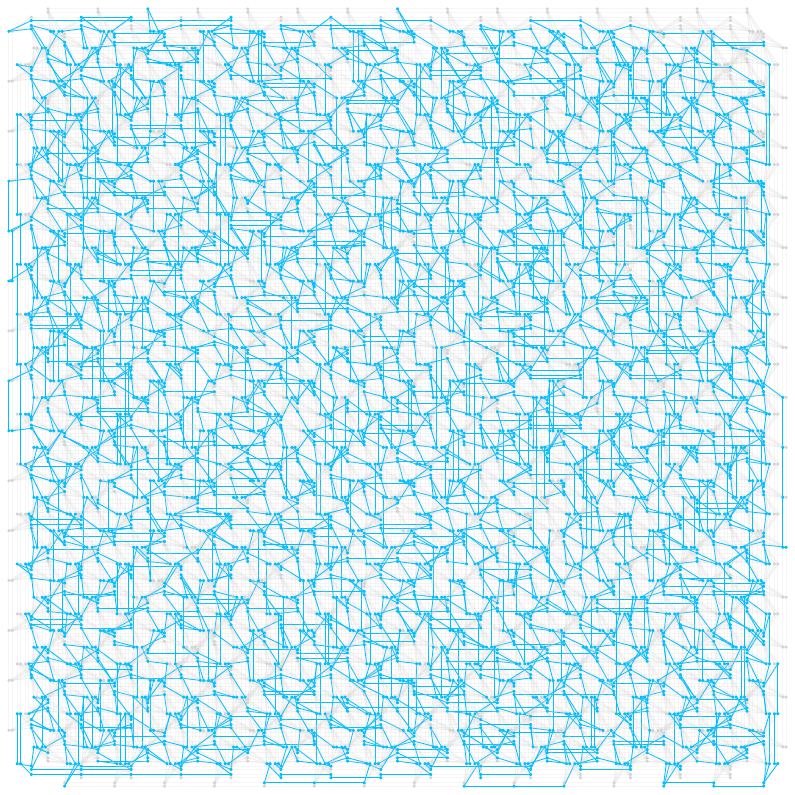}
    \includegraphics[width=0.49\linewidth]{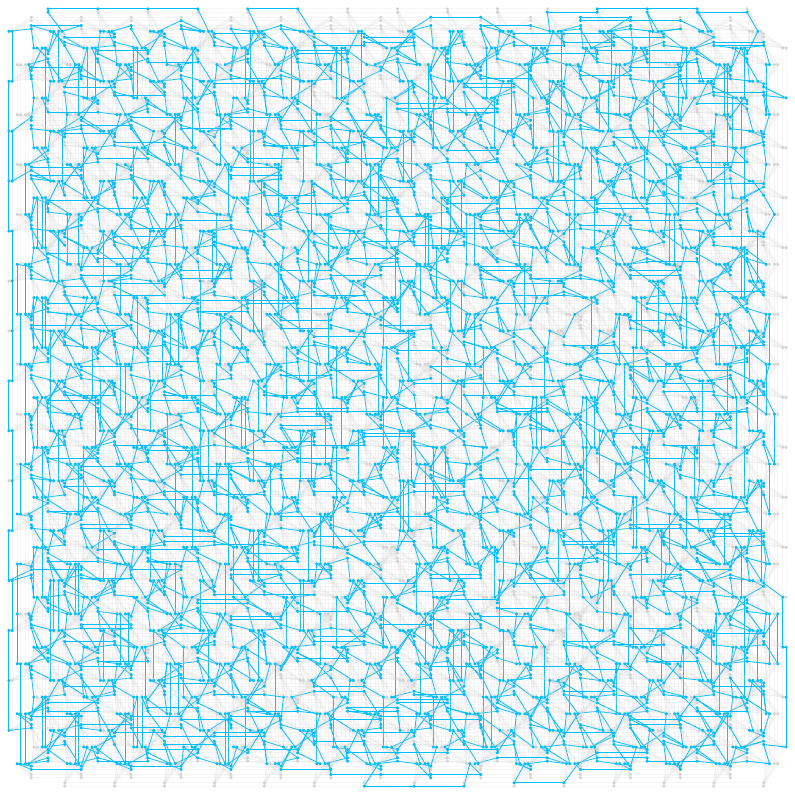}
    \caption{ 1D ring embeddings on the four hardware graphs of the D-Wave QPUs (\texttt{Advantage2\_prototype2.6} top left, \texttt{Advantage\_system4.1} top right, \texttt{Advantage\_system6.4} lower left, and \texttt{Advantage\_system7.1} lower right). Blue nodes and edges denote qubits and couplers used in the model embedding. Grey nodes and edges denote hardware not used in the embedding. Missing couplers or nodes from the hardware graph are inactive qubits/couplers that the user can not program (and in the case of \texttt{Advantage\_system7.1}, the full Pegasus $P_{16}$ graph was used for the visual rendering). }
    \label{fig:1D_embeddings}
\end{figure*}

\begin{figure*}[ht!]
    \centering
    \includegraphics[width=0.49\linewidth]{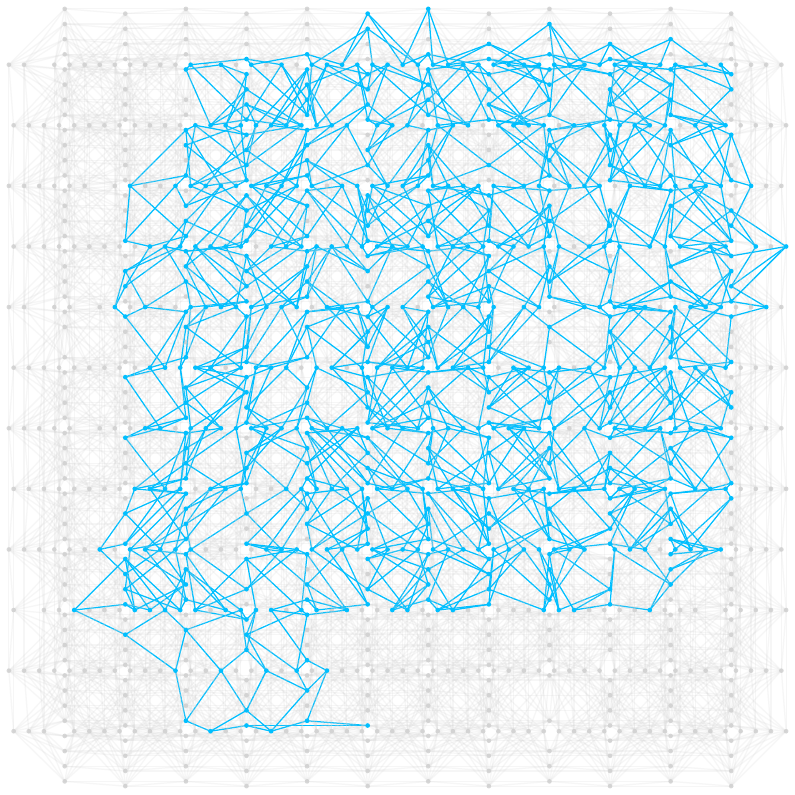}
    \includegraphics[width=0.49\linewidth]{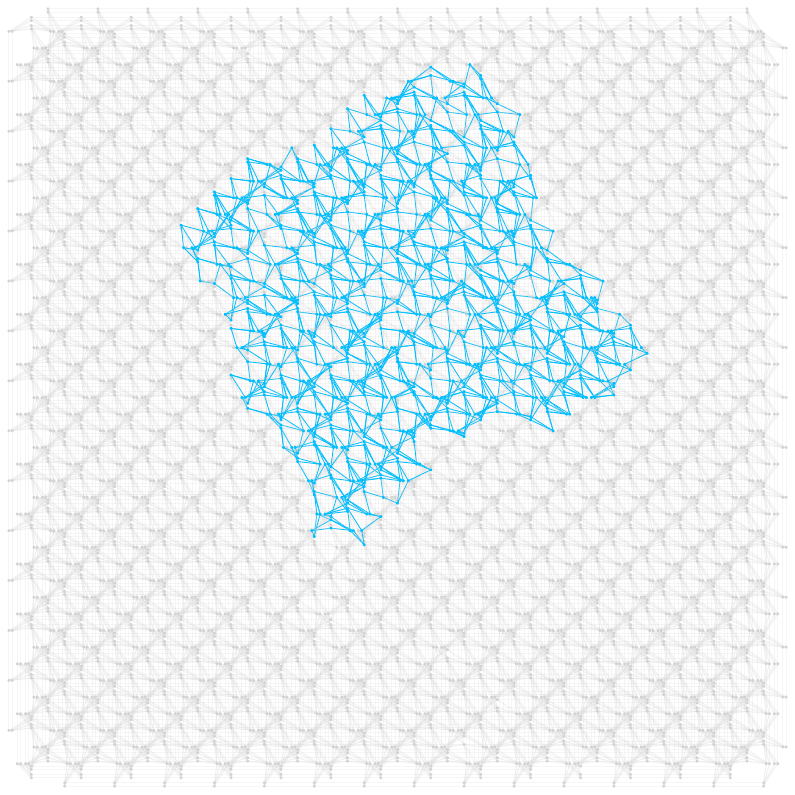}
    \includegraphics[width=0.49\linewidth]{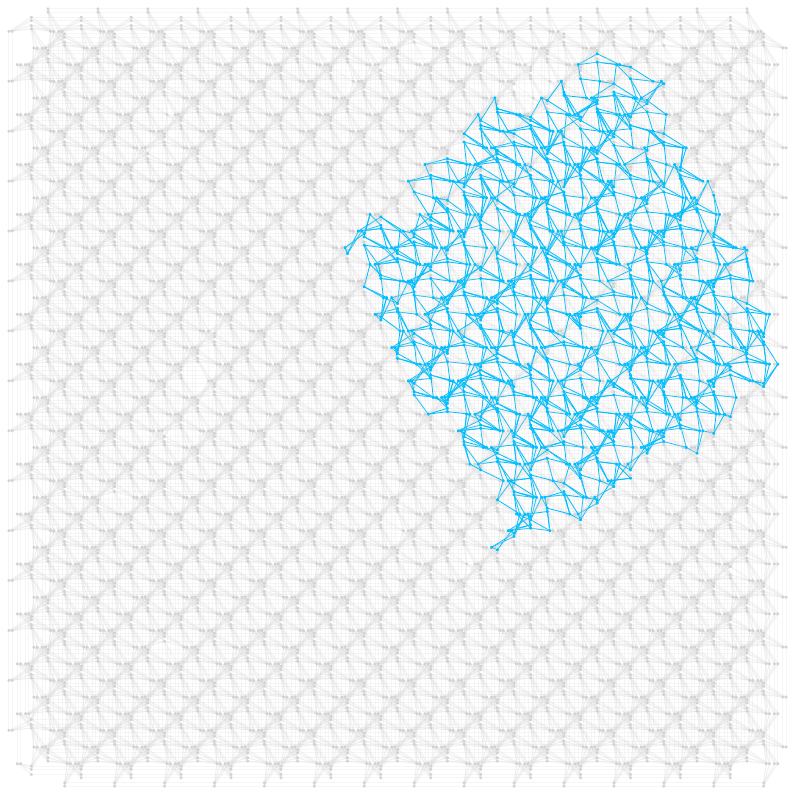}
    \includegraphics[width=0.49\linewidth]{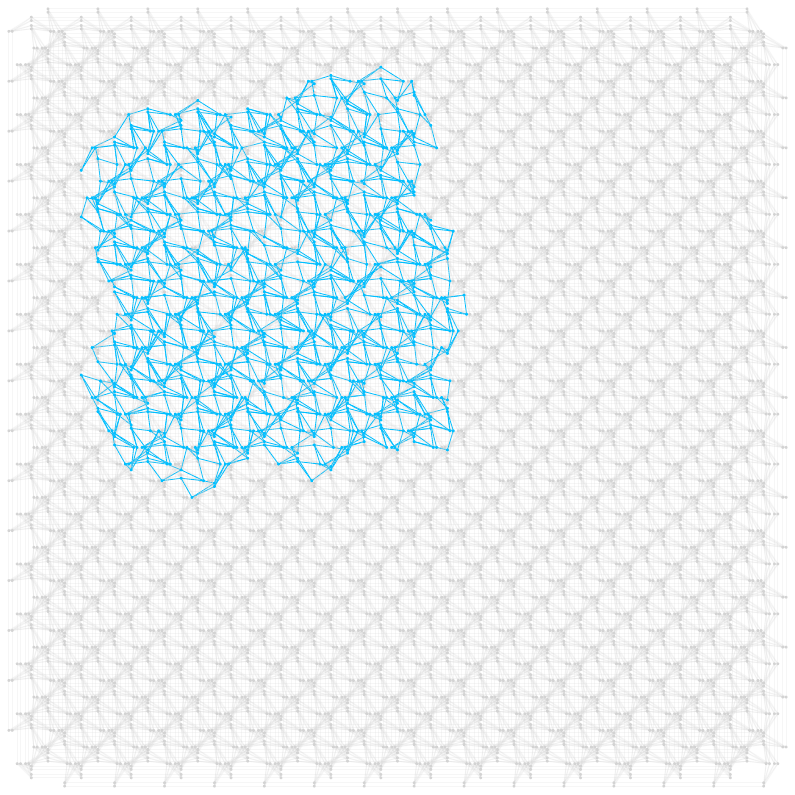}
    \caption{ 2D square grid, with open boundary conditions, embeddings on the four hardware graphs of the D-Wave QPUs (\texttt{Advantage2\_prototype2.6} top left, \texttt{Advantage\_system4.1} top right, \texttt{Advantage\_system6.4} lower left, and \texttt{Advantage\_system7.1} lower right). Blue nodes and edges denote qubits and couplers used in the model embedding. Grey nodes and edges denote hardware not used in the embedding. Missing couplers or nodes from the hardware graph are inactive qubits/couplers that the user can not program (and in the case of \texttt{Advantage\_system7.1}, the full Pegasus $P_{16}$ graph was used for the visual rendering).  }
    \label{fig:2D_embeddings}
\end{figure*}

All Ising models that were used in this study were hardware-native embeddings, meaning a spin-to-qubit mapping was uniquely defined for each model, for each D-Wave hardware graph. Fig.~\ref{fig:1D_embeddings} renders the 1-dimensional model embeddings, and Fig.~\ref{fig:2D_embeddings} renders the 2-dimensional model embeddings. Note that the hardware graph defined antiferromagnetic models do not require an isomorphism embedding because they, by definition, use the entire hardware graph.

\clearpage

\bibliographystyle{apsrev4-2-titles}
\bibliography{references}
\end{document}